\documentclass[10pt,twocolumn,twoside]{IEEEtran}

\usepackage[cmex10]{amsmath}
\usepackage{cite}

\usepackage{latexsym}
\usepackage{graphics}
\usepackage{amssymb}
\usepackage{amsthm}
\usepackage{color}

\usepackage{cite}
\usepackage{array}
\usepackage{graphicx}
\usepackage{caption}
\usepackage{subcaption}

\usepackage{url}

\def\MSE{\mathrm{MSE}}

\def\tr{\mathrm{tr}}

\def\rank{\mathrm{rank}}

\def\diag{\mathrm{diag}}

\def\p{d}

\newcounter{MYtempeqncnt}

\newcommand{\fracSumtwo}[2]{\overset{#2}{\underset{#1}{\sum}}}
\newcommand{\vect}[1]{\mathbf{#1}}

\theoremstyle{plain}
\newtheorem{remark}{Remark}

\newtheorem{theorem}{Theorem}
\newtheorem{corollary}{Corollary}
\newtheorem{lemma}{Lemma}
\newtheorem{definition}{Definition}

\begin{document}

\title{Massive MIMO Systems with Non-Ideal Hardware: \\ Energy Efficiency, Estimation, and Capacity Limits}

\IEEEoverridecommandlockouts

\author{Emil~Bj\"ornson,~\IEEEmembership{Member,~IEEE,}
        Jakob Hoydis,~\IEEEmembership{Member,~IEEE,}
        Marios Kountouris,~\IEEEmembership{Member,~IEEE,}
        and~M\'erouane~Debbah,~\IEEEmembership{Senior~Member,~IEEE}%
\thanks{Copyright (c) 2014 IEEE. Personal use of this material is permitted.  However, permission to use this material for any other purposes must be obtained from the IEEE by sending a request to pubs-permissions@ieee.org.}%
\thanks{The Matlab code that reproduces all simulation results is available online, see https://github.com/emilbjornson/massive-MIMO-hardware-impairments/}%
\thanks{E.~Bj\"ornson was with the Alcatel-Lucent Chair on Flexible Radio, Sup\'elec, Gif-sur-Yvette, France, and with the Department of Signal Processing, KTH Royal Institute of Technology, Stockholm, Sweden. He is currently with the Department of Electrical Engineering (ISY), Link\"{o}ping University, Sweden (email: emil.bjornson@liu.se).}%
\thanks{J.~Hoydis was with Bell Laboratories, Alcatel-Lucent, Germany. He is now with Spraed SAS, Orsay, France (email: hoydis@ieee.org).}
\thanks{M.~Kountouris and M.~Debbah are SUPELEC, Gif-sur-Yvette, France (e-mail: marios.kountouris@supelec.fr, merouane.debbah@supelec.fr).}%
\thanks{This paper was presented in part at the International Conference on Digital Signal Processing (DSP), Santorini, Greece, July 2013 \cite{Bjornson2013f}.}%
\thanks{The work of E.~Bj\"ornson was funded by the International Postdoc Grant 2012-228 from The Swedish Research Council. This research has been supported by the ERC Starting Grant 305123 MORE (Advanced Mathematical Tools for Complex Network Engineering). Parts of this work have been performed in the framework of the FP7 project ICT-317669 METIS. This work was supported by the Future and Emerging Technologies (FET) project HIATUS within the Seventh Framework Programme for Research of the European Commission under FET-Open grant number 265578.}
}

\maketitle

\begin{abstract}
The use of large-scale antenna arrays can bring substantial improvements in energy and/or spectral efficiency to wireless systems due to the greatly improved spatial resolution and array gain. Recent works in the field of massive multiple-input multiple-output (MIMO) show that the user channels decorrelate when the number of antennas at the base stations (BSs) increases, thus strong signal gains are achievable with little inter-user interference. Since these results rely on asymptotics, it is important to investigate whether the conventional system models are reasonable in this asymptotic regime. This paper considers a new system model that incorporates general transceiver hardware impairments at both the BSs (equipped with large antenna arrays) and the single-antenna user equipments (UEs). As opposed to the conventional case of ideal hardware, we show that hardware impairments create finite ceilings on the channel estimation accuracy and on the downlink/uplink capacity of each UE. Surprisingly, the capacity is mainly limited by the hardware at the UE, while the impact of impairments in the large-scale arrays vanishes asymptotically and inter-user interference (in particular, pilot contamination) becomes negligible. Furthermore, we prove that the huge degrees of freedom offered by massive MIMO can be used to reduce the transmit power and/or to tolerate larger hardware impairments, which allows for the use of inexpensive and energy-efficient antenna elements.
\end{abstract}


\IEEEpeerreviewmaketitle

\section{Introduction}

The spectral efficiency of a wireless link is limited by the information-theoretic capacity \cite{Telatar1999a}, which depends not only on the signal-to-noise ratio (SNR) but also on  spatial correlation in the propagation environment \cite{Gao2011a,Hoydis2012a}, channel estimation accuracy \cite{Medard2000a}, transceiver hardware impairments \cite{Bjornson2013c,Zhang2012a}, and signal processing resources \cite{Mueller2014a,Rusek2013a}. It is of profound importance to increase the spectral efficiency of future networks, to keep up with the increasing demand for wireless services. However, this is a challenging task and usually comes at the price of having stricter hardware and overhead requirements.

A new network architecture has recently been proposed with the remarkable potential of both increasing the spectral efficiency and relaxing the aforementioned implementation issues. It is known as massive MIMO, or large-scale MIMO, and is based on having a very large number of antennas at each BS and exploiting channel reciprocity in time-division duplex (TDD) mode \cite{Marzetta2010a,Jose2011b,Hoydis2013a,Ngo2013a,Rusek2013a}. Some key features are: 1) propagation losses are mitigated by a large array gain due to coherent beamforming/combining; 2) interference-leakage due to channel estimation errors vanish asymptotically in the large-dimensional vector space; 3) low-complexity signal processing algorithms are asymptotically optimal; and 4) inter-user interference is easily mitigated by the high beamforming resolution.

The amount of research on massive MIMO increases rapidly, but
the impact of transceiver hardware impairments on these systems has received little attention so far---although large arrays might only be attractive for network deployment if
each antenna element consists of inexpensive hardware. Cheap hardware components are particularly prone to the impairments that exist in any transceiver (e.g., amplifier non-linearities, I/Q-imbalance, phase noise, and quantization errors \cite{Schenk2008a,Studer2010a,Zetterberg2011a,Wenk2010a,Goransson2008a,Bjornson2012b,Bjornson2013d,Mezghani2010a,Qi2012a,Maham2012a}). The influence of hardware impairments is usually mitigated by compensation algorithms \cite{Schenk2008a}, which can be implemented by analog and digital signal processing. These techniques cannot remove the impairments completely, but there remain residual impairments since
the time-varying hardware characteristics cannot be fully parameterized and estimated, and because there is randomness induced by different types of noise. Transceiver impairments are known to fundamentally limit the capacity in the high-power regime \cite{Bjornson2013c,Koch2009a}, while there are only a few publications that analyze the behavior in the large number of antenna regime. Lower bounds on the achievable uplink sum rate in massive single-cell systems with phase noise from free-running oscillators were derived in \cite{Pitarokoilis2014a}. The impact of amplifier non-linearities in a transmitter can be reduced by having a low peak-to-average power ratio (PAPR). The excess degrees of freedom offered by massive MIMO were used in \cite{Studer2013a} to optimize the downlink precoding for low PAPR, while \cite{Mohammed2013a} considered a constant-envelope precoding scheme designed for very low PAPR.

This paper analyzes the \emph{aggregate} impact of different hardware impairments on systems with large antenna arrays, in contrast to the ideal hardware considered in \cite{Marzetta2010a,Jose2011b,Hoydis2013a,Ngo2013a} and the single type of impairments considered in \cite{Pitarokoilis2014a,Studer2013a,Mohammed2013a}. We assume that appropriate compensation algorithms have been applied and focus on the residual hardware impairments. Motivated by the analytic analysis and experimental results in \cite{Schenk2008a,Goransson2008a,Studer2010a,Wenk2010a,Zetterberg2011a}, the residual hardware impairments at the transmitter and receiver are modeled as additive distortion noises with certain important properties. The system model with hardware impairments is defined and motivated in Section \ref{sec:system-model}. Section \ref{sec:channel-estimation} derives a new pilot-based channel estimator and shows that the estimation accuracy is limited by the levels of impairments. The focus of Section \ref{sec:downlink-uplink-capacity} is on a single link in the system where we derive lower and upper bounds on the downlink and uplink capacities. Our results reveal the existence of finite capacity ceilings due to hardware impairments. Despite these discouraging results, Section \ref{sec:energy-efficiency} shows that a high energy efficiency and resilience towards hardware impairments at the BS can be achieved. Section \ref{sec:multi-cell-scenario} puts these results in a multi-cell context and shows that inter-user interference (including pilot contamination) basically drowns in the distortion noise from hardware impairments. Section \ref{sec:advanced-impairment-models} describes the impact of various refinements of the system model, while Section \ref{sec:conclusion} summarizes the contributions and insights of the paper.

To encourage reproducibility and extensions to this paper, all the simulation results can be generated by the Matlab code that is available at https://github.com/emilbjornson/massive-MIMO-hardware-impairments/

\begin{figure}
\begin{center}
\includegraphics[width=0.82\columnwidth]{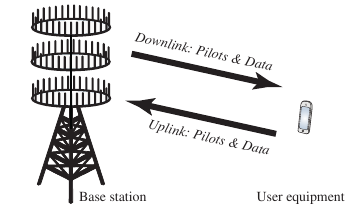}
\end{center} \vskip-3mm
\caption{Illustration of the reciprocal channel between a BS equipped with a large antenna array and a single-antenna UE.} \label{figure:system-model} \vskip-4mm
\end{figure}

\textbf{Notation:} Boldface (lower case) is used for column vectors, $\vect{x}$, and (upper case) for matrices, $\vect{X}$. Let $\vect{X}^T$, $\vect{X}^*$, and $\vect{X}^H$ denote the transpose, conjugate, and conjugate transpose of $\vect{X}$, respectively. $\vect{X}_1 \succeq \vect{X}_2$ means that $\vect{X}_1 - \vect{X}_2$ is positive semi-definite.
A diagonal matrix with $a_{1},\ldots,a_{N}$ on the main diagonal is denoted $\diag(a_{1},\ldots,a_{N})$ and $\vect{I}$ denotes an identity matrix (of appropriate dimensions).
The Frobenius and spectral norms of a matrix $\vect{X}$ are denoted by $\|\vect{X}\|_F$ and $\|\vect{X}\|_2$, respectively, while $\| \vect{x} \|_k$ denotes the $L_k$ norm of a vector $\vect{x}$. { A stochastic variable $\vect{x}$ and its realization is denoted in the same way, for brevity. The expectation operator with respect to a stochastic variable $\vect{x}$ is denoted $\mathbb{E}\{\vect{x}\}$, while $\mathbb{E}\{\vect{x} | \vect{y} \}$ is the conditional expectation when $\vect{y}$ is given.} A Gaussian stochastic variable $x$ is denoted $x \sim\mathcal{N}(\bar{x},q)$, where $\bar{x}$ is the mean and $q$ is the variance. A circularly symmetric complex Gaussian stochastic vector $\vect{x}$ is denoted $\vect{x} \sim \mathcal{CN}(\bar{\vect{x}},\vect{Q})$, where $\bar{\vect{x}}$ is the mean and $\vect{Q}$ is the covariance matrix. { The empty set is denoted by $\emptyset$.} The big $\mathcal{O}$ notation $f(x) = \mathcal{O}(g(x))$ means that $\left| \frac{f(x)}{g(x)}\right|$ is bounded as $x\rightarrow \infty$.

\section{Channel and System Model}
\label{sec:system-model}

For analytical clarity, the major part of this paper analyzes the fundamental spectral and energy efficiency limits of a single link, which operates under arbitrary interference conditions. The link is established between an $N$-antenna BS and a single-antenna UE. A main characteristic in the analysis is that the number of antennas $N$ can be very large. We consider a TDD protocol that toggles between uplink (UL) and downlink (DL) transmission on the same flat-fading subcarrier. This enables efficient channel estimation even when $N$ is large, because the estimation accuracy and overhead in the UL is independent of $N$ \cite{Rusek2013a}. The acquired instantaneous channel state information (CSI) is utilized for UL data detection as well as DL data transmission, by exploiting channel reciprocity;\footnote{The physical channels are always reciprocal, but different transceiver chains are typically used in the UL and DL. Careful calibration is therefore necessary to utilize the reciprocity for transmission; see Section \ref{subsec:imperfect-reciprocity}.} see Fig.~\ref{figure:system-model}.
In Section \ref{sec:multi-cell-scenario}, we put our results in a multi-cell context with many users, inter-cell interference, and pilot contamination.

\begin{figure}
\begin{center}
\includegraphics[width=\columnwidth]{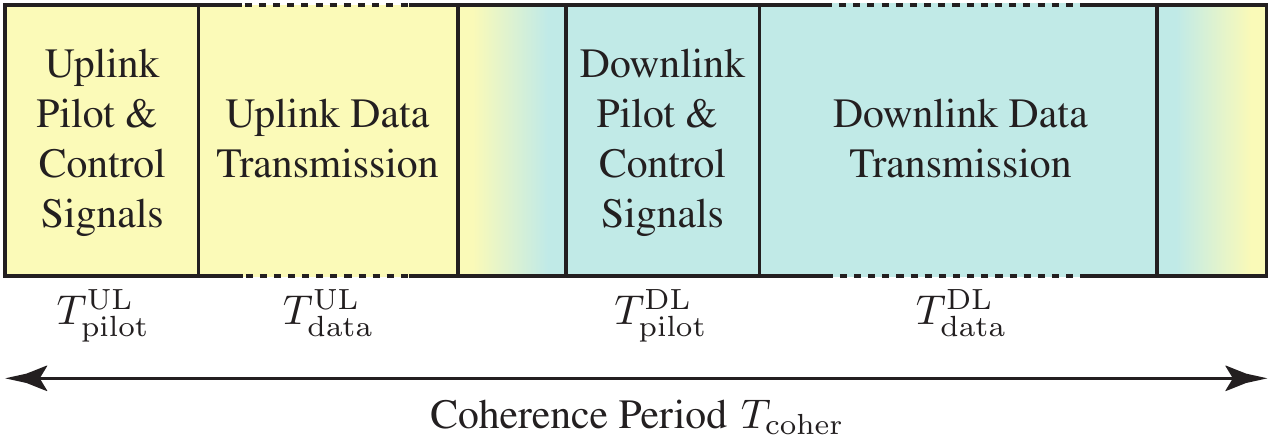}
\end{center} \vskip-3mm
\caption{Cyclic operation of a block-fading TDD system, where the coherence period $T_{\mathrm{coher}}$ is divided into phases for UL/DL pilot and data transmission.} \label{figure:tdd_operation} \vskip-4mm
\end{figure}

We assume a block fading structure where each channel is static for a \emph{coherence period} of $T_{\mathrm{coher}}$ channel uses. The channel realizations are generated randomly and are independent between blocks. { For simplicity, $T_{\mathrm{coher}}$ is the same for the useful channel and any interfering channels, and the coherence periods are synchronized. We consider the conventional TDD protocol in Fig.~\ref{figure:tdd_operation}, which can be found in many previous works; see for example \cite{Caire2010a} and \cite{Bjornson2013b}}. Each block begins with UL pilot/control signaling for $T^{\mathrm{UL}}_{\mathrm{pilot}}$ channel uses, followed by UL data transmission for $T^{\mathrm{UL}}_{\mathrm{data}}$ channel uses. Next, the system toggles to the DL. This part begins with $T^{\mathrm{DL}}_{\mathrm{pilot}}$ channel uses of DL pilot/control signaling. These pilots are typically used by the UEs to estimate their effective channel (with precoding) and the current interference conditions, which enables coherent DL reception. Note that these quantities are scalars irrespective of $N$, thus the DL pilot signaling need not scale with $N$. The coherence period ends with DL data transmission for $T^{\mathrm{DL}}_{\mathrm{data}}$ channel uses. The four parameters satisfy $T^{\mathrm{UL}}_{\mathrm{pilot}} + T^{\mathrm{UL}}_{\mathrm{data}} + T^{\mathrm{DL}}_{\mathrm{pilot}} + T^{\mathrm{DL}}_{\mathrm{data}} = T_{\mathrm{coher}}$. The analysis of this paper is valid for arbitrary fixed values of those parameters, but we note that these can also be optimized dynamically based on $T_{\mathrm{coher}}$, user load, user conditions, ratio of UL/DL traffic, etc.

The { stochastic block-fading} channel between the BS and the UE is denoted as $\vect{h} \in \mathbb{C}^{N \times 1}$. It is modeled as an ergodic process with a fixed independent realization $\vect{h} \sim \mathcal{CN}(\vect{0},\vect{R})$ in each coherence period. This is known as Rayleigh block fading and $\vect{R} = \mathbb{E} \{ \vect{h} \vect{h}^H\} \in \mathbb{C}^{N \times N}$ is the positive semi-definite covariance matrix. The statistical distribution is assumed to be known at the BS. In the asymptotic analysis, we make the following technical assumptions:

\begin{itemize}
\item The spectral norm of $\vect{R}$ is uniformly bounded, irrespective of the number of antennas $N$ (i.e., $\| \vect{R} \|_2 = \mathcal{O}(1)$);

\item The trace of $\vect{R}$ scales linearly with $N$ (i.e., $0 < \lim \inf_N \frac{1}{N} \tr(\vect{R}) \leq \lim \sup_N \frac{1}{N} \tr(\vect{R}) < \infty$) and $\vect{R}$ has strictly positive diagonal elements.

\end{itemize}

The first assumption is a necessary physical property that originates from the law of energy conservation. It is also a common enabler for asymptotic analysis (cf.~\cite{Hoydis2013a}).
The second assumption is a typical consequence of increasing the array size with $N$ and thereby improving the spatial resolution and aperture \cite{Rusek2013a}.\footnote{Although these assumptions make sense for practically large $N$ \cite{Hoydis2012a}, we cannot physically let $N \rightarrow \infty$ since the propagation environment is enclosed by a finite volume \cite{Rusek2013a}. Nevertheless, our simulations reveal that the asymptotic analysis enabled by the technical assumptions is accurate at quite small $N$.} These assumptions imply $0 <\lim \inf_N \frac{1}{N} \rank(\vect{R}) \leq 1$, which means that $\vect{R}$ can be rank deficient but the rank increases with $N$ such that $c N \leq \rank(\vect{R}) \leq N$ for some $c>0$. We stress that $\vect{R}$ is generally \emph{not} a scaled identity matrix, but describes the spatial propagation environment { and array geometry}. It might be rank-deficient (e.g., have a large conditional number) for large arrays due to insufficient richness of the scattering \cite{Gao2011a,Hoydis2012a}.

\subsection{Transceiver Hardware Impairments}

The majority of papers on massive MIMO systems considers channels with ideal transceiver hardware. However, practical transceivers suffer from hardware impairments that 1) create a mismatch between the intended transmit signal and what is actually generated and emitted; and 2) distort the received signal in the reception processing. In this paper, we analyze how these impairments impact the performance and key asymptotic properties of massive MIMO systems.

Physical transceiver implementations consist of many different hardware components (e.g., amplifiers, converters, mixers, filters, and oscillators \cite{Cui2005a}) and each one distorts the signals in its own way. The hardware imperfections are unavoidable, but the severity of the impairments depends on engineering decisions---larger distortions can be deliberately introduced to decrease the hardware cost and/or the power consumption \cite{Zhang2012a}. The non-ideal behavior of each component can be modeled in detail for the purpose of designing compensation algorithms, but even after compensation there remain residual transceiver impairments \cite{Studer2010a,Wenk2010a}; for example, due to insufficient modeling accuracy, imperfect estimation of model parameters, and time varying characteristics induced by noise.

From a system performance perspective, it is the aggregate effect of all the residual transceiver impairments that is important, not the individual behavior of each hardware component. Recently, a new system model has been proposed in \cite{Schenk2008a,Goransson2008a,Studer2010a,Wenk2010a,Zetterberg2011a,Bjornson2012b} where the aggregate residual hardware impairments are modeled by { independent} additive distortion noises at the BS as well as at the UE. We adopt this model herein due its analytical tractability and the experimental verifications in \cite{Studer2010a,Wenk2010a,Zetterberg2011a}.
The details of the DL and UL system models are given in the next subsections, and these are then used in Sections \ref{sec:channel-estimation}--\ref{sec:multi-cell-scenario} to analyze different aspects of massive MIMO systems. Possible model refinements are then provided in Section \ref{sec:advanced-impairment-models}, along with discussions on how these might impact the main results of this paper.

\subsection{Downlink System Model}
\label{subsec:generalized-downlink-model}

The downlink channel is used for data transmission and pilot-based channel estimation; see Fig.~\ref{figure:system-model}. The received DL signal $y \in \mathbb{C}$ in a flat-fading multiple-input single-output (MISO) channel is conventionally modeled as
\begin{equation} \label{eq:downlink_channel-classic}
y = \vect{h}^T \vect{s} + n
\end{equation}
where $\vect{s} \in \mathbb{C}^{N \times 1}$ is either a deterministic pilot signal (during channel estimation) or a stochastic zero-mean data signal; in any case, the covariance matrix is denoted $\vect{W} = \mathbb{E} \{  \vect{s} \vect{s}^H \}$ and the average power is $p^{\mathrm{BS}} = \tr ( \vect{W} )$. { $\vect{W}$ is a design parameter that might be a function of the channel realization $\vect{h}$ and the realizations of any other channel in the system (e.g., due to precoding); we let $ \mathcal{H}$ denote the set of channel realizations for all useful and interfering channels (i.e., $\vect{h} \in \mathcal{H}$). Hence, $\vect{W}$ is constant within each coherence period but changes between coherence periods since $\mathcal{H}$ changes. The additive term $n= n_{\mathrm{noise}}+n_{\mathrm{interf}}$ is an ergodic stochastic process that consists of independent receiver noise $n_{\mathrm{noise}} \sim \mathcal{CN}(0,\sigma_{\mathrm{UE}}^2)$ and interference $n_{\mathrm{interf}}$ from simultaneous transmissions (e.g., to other UEs). The interference has zero mean and is independent of the data signal, but might depend on any channel in the system (e.g., such that carry interference). Hence, the conditional interference variance is $\mathbb{E}  \{ | n_{\mathrm{interf}}|^2 | \mathcal{H} \} = I_{\mathcal{H}}^{\mathrm{UE}} \geq 0$ in the coherence period where the channel realizations are $\mathcal{H}$. The long-term interference variance is denoted $\mathbb{E}\{ I_{\mathcal{H}}^{\mathrm{UE}} \}$.} It is only for brevity that we use a common notation $n$ for interference and receiver noise---it does not mean that the interference must be treated as noise at the UE. A detailed interference model is provided in Section \ref{sec:multi-cell-scenario}.

To model systems with non-ideal hardware more accurately, we consider the new system model from \cite{Schenk2008a,Studer2010a,Wenk2010a,Goransson2008a,Zetterberg2011a,Bjornson2012b} where the received signal at the UE is
\begin{equation} \label{eq:downlink_channel}
y = \vect{h}^T (\vect{s} + \boldsymbol{\eta}_{t}^{\mathrm{BS}}) + \eta_{r}^{\mathrm{UE}} + n.
\end{equation}
{ The difference from the conventional model in \eqref{eq:downlink_channel-classic} is the additive distortion noise terms $\boldsymbol{\eta}_{t}^{\mathrm{BS}} \in \mathbb{C}^{N \times 1}$ and $\eta_{r}^{\mathrm{UE}} \in \mathbb{C}$, which are ergodic stochastic processes that describe the residual transceiver impairments of the transmitter hardware at the BS and the receiver hardware at UE, respectively. We assume that these are independent of the signal $\vect{s}$, but depend on the channel $\vect{h}$ and thus are stationary only within each coherence period.\footnote{{ These are model assumptions that originate from the experimental works of \cite{Studer2010a,Wenk2010a,Zetterberg2011a}. An analytic motivation of the assumptions (which should not be misinterpret as a proof) can be obtained from the Bussgang theorem; see Section \ref{sec:advanced-impairment-models}.}} In particular, we consider the conditional distributions $\boldsymbol{\eta}_{t}^{\mathrm{BS}} \sim \mathcal{CN}(\vect{0},\vect{\Upsilon}_t^{\mathrm{BS}})$ and $\eta_{r}^{\mathrm{UE}} \sim \mathcal{CN}(0,\upsilon_r^{\mathrm{UE}})$ for given channel realizations $\mathcal{H}$.}
The Gaussian distributions of $\boldsymbol{\eta}_{t}^{\mathrm{BS}}$ and $\eta_{r}^{\mathrm{UE}}$ have been verified experimentally (see e.g.,~\cite[Fig.~4.13]{Wenk2010a}) and can be motivated analytically by the central limit theorem---the distortion noises describe the aggregate effect of many residual hardware impairments. A key property is that the distortion noise caused at an antenna is proportional to the signal power at this antenna (see \cite{Studer2010a,Wenk2010a,Zetterberg2011a} for experimental verifications), thus we have
\begin{align} \label{eq:distortion-statistics-DL-BS}
\vect{\Upsilon}_t^{\mathrm{BS}} &= \kappa_t^{\mathrm{BS}} \, \diag(W_{11},\ldots,W_{NN}) \\
\upsilon_r^{\mathrm{UE}} &= \kappa_r^{\mathrm{UE}} \, \vect{h}^T \vect{W} \vect{h}^* \label{eq:distortion-statistics-DL-UE}
\end{align}
where $W_{ii}$ is the $i$th diagonal element of $\vect{W}$ and $\kappa_t^{\mathrm{BS}},\kappa_r^{\mathrm{UE}} \geq 0$ are the proportionality coefficients. The intuition is that a fixed portion of the signal is turned into distortion; for example, due to quantization errors in { automatic-gain-controlled analog-to-digital conversion (ADC)}, inter-carrier interference induced by phase noise, leakage from the mirror subcarrier under I/Q imbalance, and amplitude-amplitude nonlinearities in the power amplifier \cite{Mezghani2010a,Schenk2008a,Holma2011a}. The proportionality coefficients are treated as constants in the analysis, but can generally increase with the signal power; see Section \ref{subsec:power-increases-EVM} for details.

\begin{remark}[Distortion Noise and EVM] \label{remark:distortion-noise}
Distortion noise is an alteration of the useful signal, while the classical receiver noise models random fluctuations in the electronic circuits at the receiver. A main difference is thus that the distortion noise power { is non-stationary since it is} proportional to the signal power $p^{\mathrm{BS}}$ and the current channel gain $\|\vect{h}\|_2^2$. The proportionality coefficients $\kappa_t^{\mathrm{BS}}$ and $\kappa_r^{\mathrm{UE}}$ characterize the levels of impairments and are related to the error vector magnitude (EVM) \cite{Studer2010a}; for example, the EVM at the BS is defined as
\begin{equation} \label{eq:EVM-definition}
\mathrm{EVM}_t^{\mathrm{BS}} = \sqrt{ \frac{
\mathbb{E}  \{ \| \boldsymbol{\eta}_{t}^{\mathrm{BS}} \|^2_2 | \mathcal{H} \}
}{\mathbb{E} \{ \|\vect{s}\|^2_2 | \mathcal{H}\}}  } = \sqrt{ \frac{
\tr(  \vect{\Upsilon}_t^{\mathrm{BS}} ) }{ \tr( \vect{W})}  } = \sqrt{\kappa_t^{\mathrm{BS}}}.
\end{equation}
The EVM is a common quality measure of transceivers and the 3GPP LTE standard specifies total EVM requirements in the range $[0.08,0.175]$, where higher spectral efficiencies (modulations) are supported if the EVM is smaller \cite[Sec.~14.3.4]{Holma2011a}. LTE transceivers typically support all the standardized modulations, thus the EVM is below 0.08. Larger EVMs are, however, of interest in massive MIMO systems since such relaxed hardware constraints enable the use of low-cost equipment. Therefore, the simulations in this paper consider $\kappa$-parameters in the range $[0,0.15^2]$, where small values represent accurate and expensive transceiver hardware.
\end{remark}

The system model in \eqref{eq:downlink_channel} captures the main characteristics of non-ideal hardware, in the sense that it allows us to identify some fundamental differences in the behavior of massive MIMO systems as compared to the case of ideal hardware. However, it cannot capture all practical characteristics of residual transceiver hardware impairments. Possible refinements, and their respective implications on our analytical results and observations, are outlined in Section \ref{sec:advanced-impairment-models}.

\subsection{Uplink System Model}

The reciprocal UL channel is used for pilot-based channel estimation and data transmission; see Fig.~\ref{figure:system-model} and Sections \ref{sec:channel-estimation}--\ref{sec:downlink-uplink-capacity}.
Similar to \eqref{eq:downlink_channel}, we consider a system model with the received signal $\vect{z} \in \mathbb{C}^{N}$ at the BS being
\begin{equation} \label{eq:uplink_channel}
\vect{z} = \vect{h} (\p + \eta_{t}^{\mathrm{UE}})  + \boldsymbol{\eta}_{r}^{\mathrm{BS}} + \boldsymbol{\nu}
\end{equation}
where $\p \in \mathbb{C}$ is either a deterministic pilot signal (used for channel estimation) or a stochastic data signal; in any case, the average power is
$p^{\mathrm{UE}} = \mathbb{E}\{ |\p|^2 \}$. { The additive term $\boldsymbol{\nu} = \boldsymbol{\nu}_{\mathrm{noise}} + \boldsymbol{\nu}_{\mathrm{interf}} \in \mathbb{C}^{N \times 1}$ is an ergodic process that consists of independent receiver noise $\boldsymbol{\nu}_{\mathrm{noise}} \sim \mathcal{CN}(\vect{0},\sigma_{\mathrm{BS}}^2 \vect{I})$ as well as potential interference $\boldsymbol{\nu}_{\mathrm{interf}}$ from other simultaneous transmissions. The interference is independent of $\p$ but might depend on the channel realizations in $\mathcal{H}$.} Moreover, the interference statistics can be different in the pilot and data transmission phases; for example, it is common to assume that each cell uses time-division multiple access (TDMA) for pilot transmission, since this can provide sufficient CSI accuracy to enable spatial-division multiple access (SDMA) for data transmission \cite{Marzetta2010a,Jose2011b,Hoydis2013a,Ngo2013a,Rusek2013a}. { Therefore, we assume that $\boldsymbol{\nu}_{\mathrm{interf}}$ has zero mean and $\vect{S} = \mathbb{E}\{ \boldsymbol{\nu}_{\mathrm{interf}} \boldsymbol{\nu}_{\mathrm{interf}}^H \}$ is that the covariance matrix during pilot transmission. We assume that $\vect{S}$ has a uniformly bounded spectral norm, $\| \vect{S} \|_2 = \mathcal{O}(1)$, for the same physical reasons as for $\vect{R}$. For data transmission, we define the conditional covariance matrix $\vect{Q}_\mathcal{H} = \mathbb{E}\{ \boldsymbol{\nu}_{\mathrm{interf}} \boldsymbol{\nu}_{\mathrm{interf}}^H | \mathcal{H} \}$, in a coherence period with channel realizations $\mathcal{H}$, and the corresponding long-term covariance matrix $\mathbb{E}\{ \vect{Q}_\mathcal{H} \}$. The covariance matrices $\vect{S},\vect{Q}_{\mathcal{H}}  \in \mathbb{C}^{N \times N}$ are positive semi-definite. The spectral norm of $\vect{Q}_{\mathcal{H}}$ might  grow unboundedly with $N$ due to pilot contamination in multi-cell scenarios \cite{Marzetta2010a,Jose2011b,Hoydis2013a,Ngo2013a,Rusek2013a}; see Section \ref{sec:multi-cell-scenario} for further details.}

{ Similar to the DL, the aggregate residual transceiver impairments in the hardware used for UL transmission are modeled by the independent distortion noises
$\eta_{t}^{\mathrm{UE}} \in \mathbb{C}$ and $\boldsymbol{\eta}_{r}^{\mathrm{BS}} \in \mathbb{C}^{N \times 1}$ at the transmitter and receiver, respectively. These ergodic stochastic processes are independent of $\p$, but depend on the channel realizations $\mathcal{H}$. The conditional distribution for a given $\mathcal{H}$ are $\eta_{t}^{\mathrm{UE}} \sim \mathcal{CN}(0,\upsilon_t^{\mathrm{UE}})$ and $\boldsymbol{\eta}_{r}^{\mathrm{BS}} \sim \mathcal{CN}(\vect{0},\vect{\Upsilon}_r^{\mathrm{BS}})$. Similar to \eqref{eq:distortion-statistics-DL-BS} and \eqref{eq:distortion-statistics-DL-UE}, the conditional covariance matrices are modeled as}
\begin{align} \label{eq:distortion-statistics-UL-UE}
\upsilon_t^{\mathrm{UE}} &= \kappa_t^{\mathrm{UE}} p^{\mathrm{UE}} \\
\vect{\Upsilon}_r^{\mathrm{BS}} &= \kappa_r^{\mathrm{BS}} p^{\mathrm{UE}} \, \diag(|h_1|^2,\ldots,|h_{N}|^2). \label{eq:distortion-statistics-UL-BS}
\end{align}

Note that the hardware quality is characterized by $\kappa_t^{\mathrm{BS}},\kappa_r^{\mathrm{BS}}$ at the BS and by $\kappa_t^{\mathrm{UE}},\kappa_r^{\mathrm{UE}}$ at the UE. We can have  $\kappa_t^{\mathrm{BS}} \neq \kappa_r^{\mathrm{BS}}$ and $\kappa_t^{\mathrm{UE}} \neq \kappa_r^{\mathrm{UE}}$ since different transceiver chains are used for transmission and reception at a device.

Generally speaking, we would like to achieve high performance using cheap hardware. This is particularly evident in massive MIMO systems since the deployment cost of large antenna arrays might scale linearly with $N$ unless we can accept higher levels of impairments, $\kappa_t^{\mathrm{BS}},\kappa_r^{\mathrm{BS}}$, at the BSs than in conventional systems. This aspect is analyzed in Section \ref{sec:energy-efficiency}.

\section{Uplink Channel Estimation}
\label{sec:channel-estimation}

This section considers estimation of the current channel realization $\vect{h}$ by comparing the received UL signal $\vect{z}$ in \eqref{eq:uplink_channel} with the predefined UL pilot signal $\p$ (recall: $p^{\mathrm{UE}} = |\p|^2$).
The classic results on pilot-based channel estimation consider Rayleigh fading channels that are observed in independent complex Gaussian noise with known statistics \cite{Kay1993a,Kotecha2004a,Bjornson2010a,Hassibi2003a}. However, this is not the case herein because the distortion noises $\eta_{t}^{\mathrm{UE}}$ and $\boldsymbol{\eta}_{r}^{\mathrm{BS}}$ effectively depend on the unknown stochastic channel $\vect{h}$. The dependence is either through the multiplication $\vect{h} \eta_{t}^{\mathrm{UE}}$ or the conditional variance of $\boldsymbol{\eta}_{r}^{\mathrm{BS}}$ in  \eqref{eq:distortion-statistics-UL-BS}, which is essentially the same type of relation. Although the distortion noises are Gaussian when conditioned on a channel realization, the effective distortion is the product of Gaussian variables and, thus, has a \emph{complex double Gaussian} distribution \cite{Donoughue2012a}.\footnote{For example, the $i$th element of $\boldsymbol{\eta}_{r}^{\mathrm{BS}}$ can be expressed as $x_i = h_i \xi_i$, which is the product of the $i$th channel coefficient $h_i \sim \mathcal{CN}(0,r_{ii})$ and an independent variable $\xi_i \sim \mathcal{CN}(0, \kappa_r^{\mathrm{BS}} p^{\mathrm{UE}})$. The joint product distribution is complex double Gaussian with the PDF $f(x_i) = \frac{2}{\pi \mu_i} K_0 \left( 2 \frac{|x_i|}{\sqrt{\mu_i}}\right)$, where $\mu_i = r_{ii} \kappa_r^{\mathrm{BS}} p^{\mathrm{UE}}$ is the variance and $K_0(\cdot)$ denotes the zeroth-order modified Bessel function of the second kind \cite{Donoughue2012a}.} Consequently, an optimal channel estimator cannot be deduced from the standard results provided in \cite{Kay1993a,Kotecha2004a,Bjornson2010a,Hassibi2003a}.

We now derive the \emph{linear} minimum mean square error (LMMSE) estimator of $\vect{h}$ under hardware impairments.

\begin{theorem} \label{theorem:LMMSE-estimator}
The LMMSE estimator of $\vect{h}$ from the observation of $\vect{z}$ in \eqref{eq:uplink_channel} is
\begin{equation} \label{eq:LMMSE-estimator}
\hat{\vect{h}} = \underbrace{\p^* \vect{R} \bar{\vect{Z}}^{-1}}_{\triangleq \vect{A}} \vect{z}
\end{equation}
where $\vect{R}_{\diag} = \diag(r_{11},\ldots,r_{NN})$ consists of the diagonal elements of $\vect{R}$ and the covariance matrix of $\vect{z}$ is denoted as
\begin{equation}
\bar{\vect{Z}} = \mathbb{E}\{ \vect{z} \vect{z}^H \} =  p^{\mathrm{UE}} (1+\kappa_t^{\mathrm{UE}}) \vect{R} + p^{\mathrm{UE}} \kappa_r^{\mathrm{BS}}  \vect{R}_{\diag} + \vect{S} + \sigma_{\mathrm{BS}}^2 \vect{I}.
\end{equation}

The total MSE is $\MSE = \mathbb{E}\{ \|\hat{\vect{h}}  - \vect{h}\|_2^2 \} = \tr( \vect{C} )$, where the error covariance matrix is
\begin{equation} \label{eq:LMMSE-error-covariance}
\begin{split}
\vect{C} =  \mathbb{E}\{ (\hat{\vect{h}}  - \vect{h})(\hat{\vect{h}}  - \vect{h})^H \} =\vect{R} - p^{\mathrm{UE}} \vect{R} \bar{\vect{Z}} ^{-1} \vect{R}.
\end{split}
\end{equation}
\end{theorem}
\begin{IEEEproof}
The LMMSE estimator has the form $\hat{\vect{h}} = \vect{A} \vect{z}$ where $\vect{A}$ minimizes the MSE.
The MSE definition gives
\begin{equation} \label{eq:MSE_expression_general}
\begin{split}
\MSE = \tr  \bigg( \vect{R} - \p \vect{A} \vect{R}  - \p^* \vect{R} \vect{A}^H + \vect{A} \bar{\vect{Z}}  \vect{A}^H \bigg)
\end{split}
\end{equation}
where the expectations that involve $\eta_t^{\mathrm{UE}},\boldsymbol{\eta}_r^{\mathrm{BS}}$ in $\MSE = \mathbb{E}\{ \|\hat{\vect{h}}  - \vect{h}\|_2^2 \}$ are computed by first having a fixed value of $\vect{h}$ and then average over $\vect{h}$.
The LMMSE estimator in \eqref{eq:LMMSE-estimator} is achieved by differentiation of \eqref{eq:MSE_expression_general} with respect to $\vect{A}$ and equating to zero. This vector minimizes the MSE since the Hessian is always positive definite. The error covariance matrix and the MSE are obtained by plugging \eqref{eq:LMMSE-estimator} into the respective definitions.
\end{IEEEproof}

Based on Theorem \ref{theorem:LMMSE-estimator}, the channel can be decomposed as $\vect{h} = \hat{\vect{h}} + \boldsymbol{\epsilon}$ where $\hat{\vect{h}}$ is the LMMSE estimate in \eqref{eq:LMMSE-estimator} and $\boldsymbol{\epsilon} \in \mathbb{C}^{N \times 1}$ denotes the unknown estimation error. Contrary to conventional estimation with independent Gaussian noise (cf.~\cite[Chapter 15.8]{Kay1993a}), $\hat{\vect{h}}$ and $\boldsymbol{\epsilon}$ are neither independent nor jointly complex Gaussian, but only uncorrelated and have zero mean. The covariance matrices are $\mathbb{E}\{ \hat{\vect{h}} \hat{\vect{h}}^H\} = \vect{R} - \vect{C}$ and $\mathbb{E}\{ \boldsymbol{\epsilon} \boldsymbol{\epsilon}^H\} = \vect{C}$ where $\vect{C}$ was given in \eqref{eq:LMMSE-error-covariance}.

We remark that there might exist non-linear estimators that achieve smaller MSEs than the LMMSE estimator in Theorem~\ref{theorem:LMMSE-estimator}. This stands in contrast to conventional channel estimation with independent Gaussian noise, where the LMMSE estimator is also the MMSE estimator \cite{Bjornson2010a}. However, the difference in MSE performance should be small, since the dependent distortion noises are relatively weak.

\begin{corollary} \label{cor:asymptotic-estimation-performance}
Consider the special case of $\vect{R} = \lambda \vect{I}$ and $\vect{S} = \vect{0}$. The error covariance matrix in \eqref{eq:LMMSE-error-covariance} becomes
\begin{equation} \label{eq:LMMSE-error-covariance-iid}
\vect{C} =  \lambda \left( 1  - \frac{ p^{\mathrm{UE}} \lambda }{  p^{\mathrm{UE}} \lambda (1 + \kappa_t^{\mathrm{UE}}  +\kappa_r^{\mathrm{BS}}  ) + \sigma_{\mathrm{BS}}^2 }  \right) \vect{I}.
\end{equation}
In the high UL power regime, we have
\begin{equation} \label{eq:LMMSE-error-covariance-iid-asympt}
\lim_{p^{\mathrm{UE}} \rightarrow \infty } \vect{C} = \lambda \left( 1 - \frac{1}{1 + \kappa_t^{\mathrm{UE}}  +\kappa_r^{\mathrm{BS}}} \right) \vect{I}.
\end{equation}
\end{corollary}

This corollary brings important insights on the average estimation error per element in $\vect{h}$. The error variance is given by the factor in front of the identity matrix in \eqref{eq:LMMSE-error-covariance-iid}. It is independent of the number of antennas $N$, thus letting $N$ grow large neither increases nor decreases the estimation error \emph{per element}.\footnote{The MSE per element is finite, i.e.~$\frac{1}{N} \tr(\vect{C}) < \infty$, but the sum MSE behaves as $\tr( \vect{C} ) \rightarrow \infty$ when $N \rightarrow \infty$ since the number of elements grows.} The estimation error is clearly a decreasing function of the pilot power $p^{\mathrm{UE}} = |\p|^2$, but contrary to the ideal hardware case the error variance is \emph{not} converging to zero as $p^{\mathrm{UE}}  \rightarrow \infty$. As seen in \eqref{eq:LMMSE-error-covariance-iid-asympt}, there is a strictly positive error floor of $\lambda ( 1 -\frac{1}{1 + \kappa_t^{\mathrm{UE}}  +\kappa_r^{\mathrm{BS}}})$ due to the transceiver hardware impairments. Thus, perfect estimation accuracy cannot be achieved in practice, not even asymptotically. The error floor is characterized by the sum of the levels of impairments $\kappa_t^{\mathrm{UE}}$ and $\kappa_r^{\mathrm{BS}}$ in the transmitter and receiver hardware, respectively. In terms of estimation accuracy, it is thus equally important to have high-quality hardware at the BS and at the UE.

Non-ideal hardware exhibits an error floor also when $\vect{R}$  is non-diagonal and when there is interference such that $\vect{S} \neq \vect{0}$; the general high-power limit is easily computed from \eqref{eq:LMMSE-error-covariance}. In fact, the results hold for any zero-mean channel and interference distributions with covariance matrices $\vect{R}$ and $\vect{S}$, because the LMMSE estimator and its MSE are computed using only the first two moments of the statistical distributions \cite{Kay1993a,Bjornson2010a}.

\subsection{Impact of the Pilot Length}
\label{subsec:pilot-length}

The LMMSE estimator in Theorem \ref{theorem:LMMSE-estimator} considers a scalar pilot signal $\p$, which is sufficient to excite all $N$ channel dimensions in the UL and is used in Section \ref{subsec:lower-bounds-capacity} to derive lower bounds on the UL and DL capacities.  With ideal hardware and a total pilot energy constraint, a scalar pilot signal is also sufficient to minimize the MSE \cite{Bjornson2010a}.  In contrast, we have non-ideal hardware and per-symbol energy constraints in this paper. In this case we can improve the MSE by increasing the pilot length.

Suppose we use a pilot signal $\vect{\p} \in \mathbb{C}^{1 \times B}$ that spans $1 \leq B \leq T^{\mathrm{UL}}_{\mathrm{pilot}}$ channel uses { and where each element of $\vect{\p}$ has squared norm $p^{\mathrm{UE}}$.} A simple estimation approach would be to compute $B$ separate LMMSE estimates,
$\hat{\vect{h}}_i = \vect{h} - \boldsymbol{\epsilon}_i$ for $i=1,\ldots,B$, using Theorem  \ref{theorem:LMMSE-estimator}. By averaging, we obtain
\begin{equation}
\widehat{\hat{\vect{h}}} = \frac{1}{B} \sum_{i=1}^B \hat{\vect{h}}_i = \vect{h} - \frac{1}{B} \sum_{i=1}^B \boldsymbol{\epsilon}_i.
\end{equation}
If the distortion noises are temporally uncorrelated and identically distributed, the MSE of the estimate $\widehat{\hat{\vect{h}}}$ is
\begin{equation} \label{eq:diminishing-MSE}
\mathbb{E} \left\{ \bigg(\frac{1}{B} \sum_{i=1}^B \boldsymbol{\epsilon}_i \bigg)^H \bigg( \frac{1}{B} \sum_{j=1}^B \boldsymbol{\epsilon}_j \bigg) \right\} = \frac{\tr(\vect{C})}{B}.
\end{equation}
{ Hence, the MSE goes to zero as $1/B$ when we increase the pilot length $B$, although the MSE per pilot channel use is limited by the non-zero error floor demonstrated in Corollary \ref{cor:asymptotic-estimation-performance}. This is interesting because one pilot signal with energy $B p^{\mathrm{UE}}$ exhibits a noise floor, while $B$ pilot signals with energy $p^{\mathrm{UE}}$ per signal does not.\footnote{ Since we have per-symbol energy constraints, what we really compare is one system that has an average symbol energy of $B p^{\mathrm{UE}}$ and one with $p^{\mathrm{UE}}$.} This stands in contrast to the case of ideal hardware, where the MSE is exactly the same in both cases \cite{Bjornson2010a}.
The reason is that we can average out the distortion noise (similar to the law of large numbers) when we have $B$ independent realizations.}

Despite the averaging effect, we stress that $B \leq T_{\mathrm{coher}}$ and thus there is always an estimation error floor for non-ideal hardware---we can, at most, reduce the floor by a factor $1/T_{\mathrm{coher}}$ by increasing the pilot length. Moreover, the derivation above is based on having temporally uncorrelated distortions, but the distortions might be temporally correlated in practice (especially if the same pilot signal $\p$ is transmitted multiple times through the same hardware). In these cases, the benefit of increasing $B$ is smaller and $\widehat{\hat{\vect{h}}}$ should be replaced by an estimator that exploits the temporal correlation by estimating $\vect{h}$ jointly from all the $B$ observations. Finally, we note that it is of great interest to find the $B$ that maximizes some measure of system-wide performance, but this is outside the scope of our current paper. We refer to \cite{Hassibi2003a,Bjornson2010a,Hoydis2011a,Wagner2012a} for some relevant works in the case of ideal hardware.

\subsection{Numerical Illustrations}
\label{subsec:estimation-numerical}

This section exemplifies the impact of transceiver hardware impairments on the channel estimation accuracy.

\begin{figure}
\begin{center}
\includegraphics[width=\columnwidth]{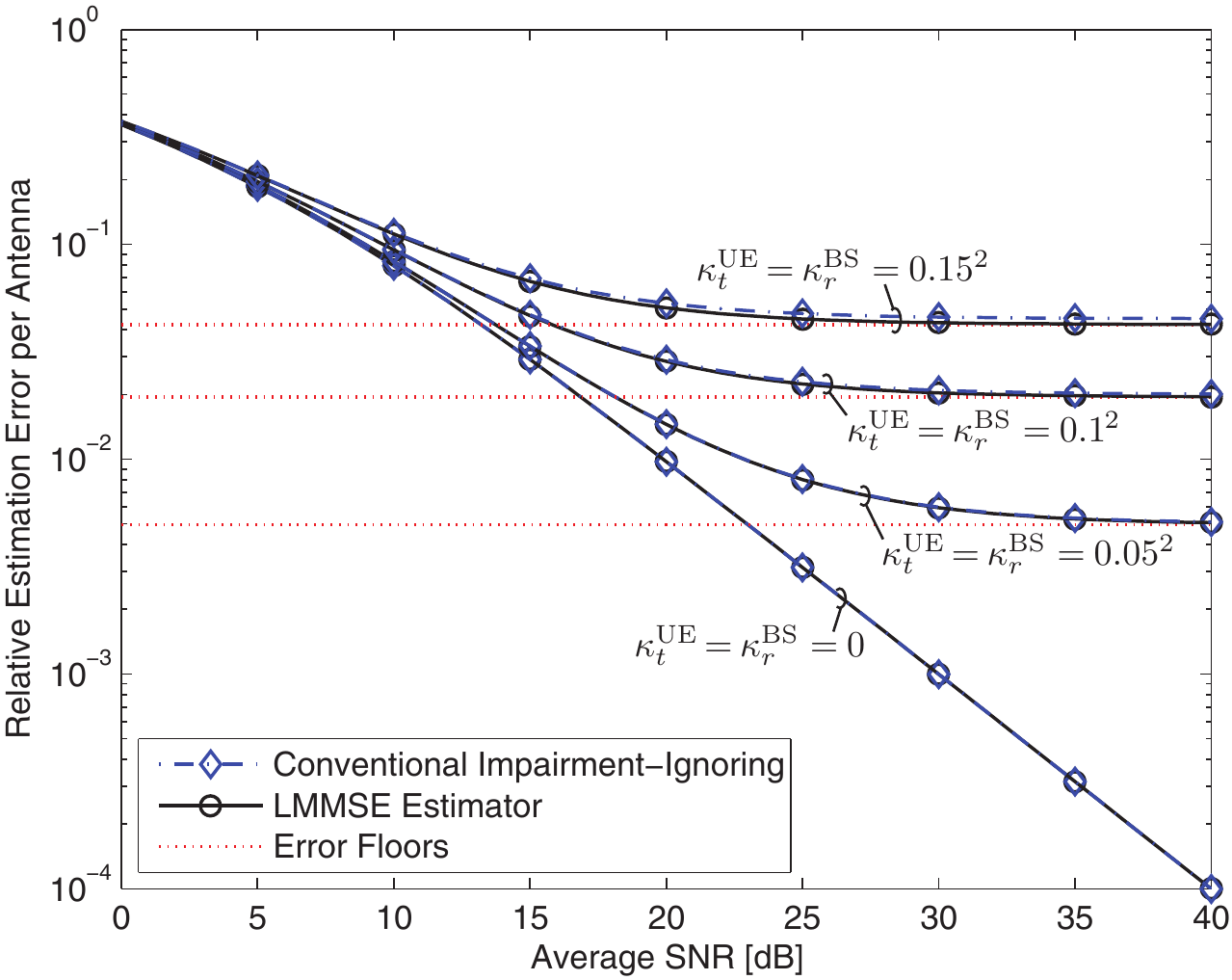} \vskip-2mm
\caption{Estimation error per antenna element for the LMMSE estimator in Theorem \ref{theorem:LMMSE-estimator} and the conventional impairment-ignoring MMSE estimator. Transceiver hardware impairments create non-zero error floors.}\label{figure_estimation_diffimpairments}
\end{center} \vskip-4mm
\end{figure}

In Fig.~\ref{figure_estimation_diffimpairments}, we consider $N=50$ antennas at the BS and  no interference (i.e., $\vect{S} = \vect{0}$). The channel covariance matrix $\vect{R}$ is generated by the exponential correlation model from \cite{Loyka2001a}, which means that the $(i,j)$th element of $\vect{R}$ is
\begin{equation} \label{eq:exponential-model}
[\vect{R}]_{i,j} = \begin{cases} \delta \, r^{j-i}, & i \leq j, \\ \delta \, (r^{i-j})^*, & i > j, \end{cases}
\end{equation}
where $\delta$ is an arbitrary scaling factor. This model basically describes a uniform linear array (ULA) where the correlation factor between adjacent antennas is given by $|r|$ (for $0 \leq |r| \leq 1$) and the phase $\angle r$ describes the angle of arrival/departure as seen from the array. The correlation factor $|r|$ determines the eigenvalue spread in $\vect{R}$, while $\angle r$ determines the corresponding eigenvectors. Since we simulate channel estimation without interference, the angle has no impact on the MSE and we can let $r$ be real-valued without loss of generality. We consider a correlation coefficient of $r=0.7$, which is a modest correlation in the sense of behaving similarly to an array with half-wavelength antenna spacings and a large angular spread of 45 degrees (cf.~\cite[Fig.~1]{Bjornson2009c} which shows how practical angular spreads map non-linearly to $|r|$).

Fig.~\ref{figure_estimation_diffimpairments} shows the relative estimation error per channel element, $\mathrm{MSE}_{\mathrm{rel}} = \frac{\mathrm{MSE}}{\tr(\vect{R})}$, as a function of the average SNR in the UL, defined as
\begin{equation}
{\tt SNR }^{\mathrm{UL}} = p^{\mathrm{UE}} \frac{\tr (\vect{R})}{N \sigma_{\mathrm{BS}}^2}.
\end{equation}
Based on the typical EVM ranges described in Remark \ref{remark:distortion-noise}, we consider four hardware setups with different levels of impairments: $\kappa_t^{\mathrm{UE}} =\kappa_r^{\mathrm{BS}} \in \{ 0, 0.05^2, 0.1^2, 0.15^2 \}$. We compare the LMMSE estimator in Theorem \ref{theorem:LMMSE-estimator} with the conventional impairment-ignoring MMSE estimator from \cite{Kay1993a,Kotecha2004a,Bjornson2010a}.\footnote{Note that the MSE of any linear estimator $\tilde{\vect{A}} \vect{z}$ can be computed by plugging the matrix $\tilde{\vect{A}}$ into the general MSE expression in \eqref{eq:MSE_expression_general}. The difference in MSE is easily quantified by comparing with $\tr(\vect{C})$ using \eqref{eq:LMMSE-error-covariance}.}

Fig.~\ref{figure_estimation_diffimpairments} confirms that there are non-zero error floors at high SNRs, as proved by Corollary \ref{cor:asymptotic-estimation-performance} and the subsequent discussion. The error floor increases with the levels of impairments. The estimation error is very close to the floor when the uplink SNR reaches $20$--$30$ dB, thus further increase in SNR only brings minor improvement. This tells us that we need an uplink SNR of at least 20 dB to fully utilize massive MIMO, because coherent transmission/reception requires accurate CSI. Lower SNRs can be compensated by adding extra antennas (see Fig.~\ref{figure_capacity} in Section \ref{sec:downlink-uplink-capacity}), but the practical performance not as large.
Moreover, Fig.~\ref{figure_estimation_diffimpairments} shows that the conventional impairment-ignoring estimator is only slightly worse than the proposed LMMSE estimator. This indicates that although hardware impairments greatly affect the estimation performance, it only brings minor changes to the structure of the optimal estimator.

The influence of the estimation error floors depend on the anticipated spectral efficiency, the uplink SNR, and the number of antennas. To gain some insight, suppose we have ideal hardware and that the fraction of channel uses allocated for UL data transmission is $T^{\mathrm{UL}}_{\mathrm{data}}  / T_{\mathrm{coher}} = 0.45$. The uplink spectral efficiency can then be approximated as
\begin{equation}
0.45 \log_2 \left(1+ \frac{1-\mathrm{MSE}_{\mathrm{rel}}}{\mathrm{MSE}_{\mathrm{rel}}+ \frac{1}{N {\tt SNR }^{\mathrm{UL}}}} \right)
\end{equation}
by using \cite[Lemma 1]{Yoo2006b}. When the number of antennas is large, such that $N \,{\tt SNR }^{\mathrm{UL}} \rightarrow \infty$, this approximation gives a spectral efficiency of 1.5 [bit/channel use] for $\mathrm{MSE}_{\mathrm{rel}}\!=\!10^{-1}$ and 4.5 [bit/channel use] for $\mathrm{MSE}_{\mathrm{rel}}\!=\!10^{-3}$. The impact of the estimation errors on systems with non-ideal hardware is considered in Section \ref{sec:downlink-uplink-capacity}, where we derive lower and upper capacity bounds and analyze these for different SNRs and number of antennas.

Next, we illustrate the possible improvement in estimation accuracy by increasing the pilot length to comprise $B \geq 1$ channel uses.
As discussed in Section \ref{subsec:pilot-length}, it is not clear whether the distortion noise is temporally uncorrelated or correlated in practice. Therefore, we fix the levels of impairments at $\kappa_t^{\mathrm{UE}} =\kappa_r^{\mathrm{BS}} =0.05^2$ and consider the two extremes: temporally uncorrelated and fully correlated distortion noises. The latter means that the distortion noise realizations are the same for all $B$ channel uses, since the same pilot signal is always distorted in the same way. The channel and interference statistics are as in the previous figure (i.e., $N=50$, $\vect{S}=\vect{0}$, and $\vect{R}$ is given by the exponential correlation model with $r=0.7$).

\begin{figure}
\begin{center}
\includegraphics[width=\columnwidth]{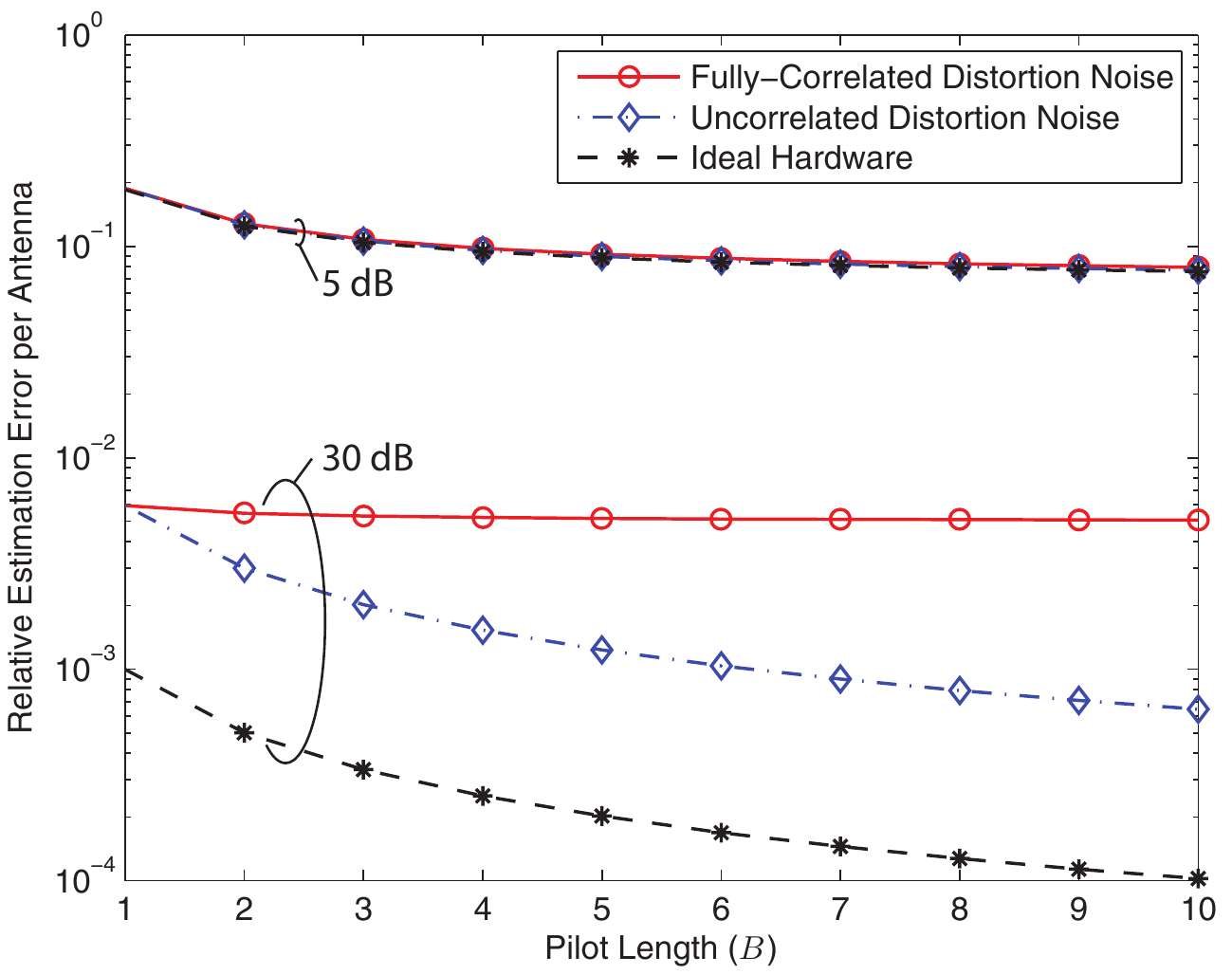} \vskip-2mm
\caption{Estimation error per antenna element for the LMMSE estimator in Theorem \ref{theorem:LMMSE-estimator} as a function of the pilot length $B$. The levels of impairments are $\kappa_t^{\mathrm{UE}} =\kappa_r^{\mathrm{BS}} =0.05^2$ and different temporal correlations are considered.}\label{figure_estimation_diffpilotlengths}
\end{center} \vskip-4mm
\end{figure}

The relative estimation error per antenna element is shown in Fig.~\ref{figure_estimation_diffpilotlengths} as a function of the pilot length. We also show the performance with ideal hardware as a reference. At a low SNR of 5 dB, hardware impairments have little impact and there is a small but clear gain from increasing the pilot length because the total pilot energy increases as $B p^{\mathrm{UE}}$. At a high SNR of 30 dB, the temporal correlation has a large impact. Only small improvements are possible in the fully correlated case, since only the receiver noise can be mitigated by increasing $B$. In the uncorrelated case the distortion noise can be also mitigated by increasing $B$. This gives a logarithmic slope similar to the case of ideal hardware. We stress that the actual performance lies somewhere in between the two extremes.

Next, we consider different channel covariance models:
\begin{enumerate}
\item Uncorrelated antennas $\vect{R}=\vect{I}$. (Equivalent to the exponential correlation model in \eqref{eq:exponential-model} with $r=0$.)
\item Exponential correlation model with $r=0.7$.
\item One-ring model with 20 degrees angular spread \cite{Shiu2000a}.
\item One-ring model with 10 degrees angular spread \cite{Shiu2000a}.
\end{enumerate}
The exponential correlation model was defined in \eqref{eq:exponential-model}. The classic one-ring model assumes a ring of scatterers around the UE, while there is no scattering close to the BS \cite{Shiu2000a}. From the BS perspective, the multipath components arrive from a main angle of arrival (here: $30$ degrees) and a small angular spread around it (here: $10$ or $20$ degrees). The BS is assumed to have a ULA with half-wavelength antenna spacings. An important property of this model is that $\vect{R}$ might not have full rank as $N$ grows large \cite{Yin2013a,Adhikary2013a}, due to insufficient scattering.

\begin{figure}
\begin{center}
\includegraphics[width=\columnwidth]{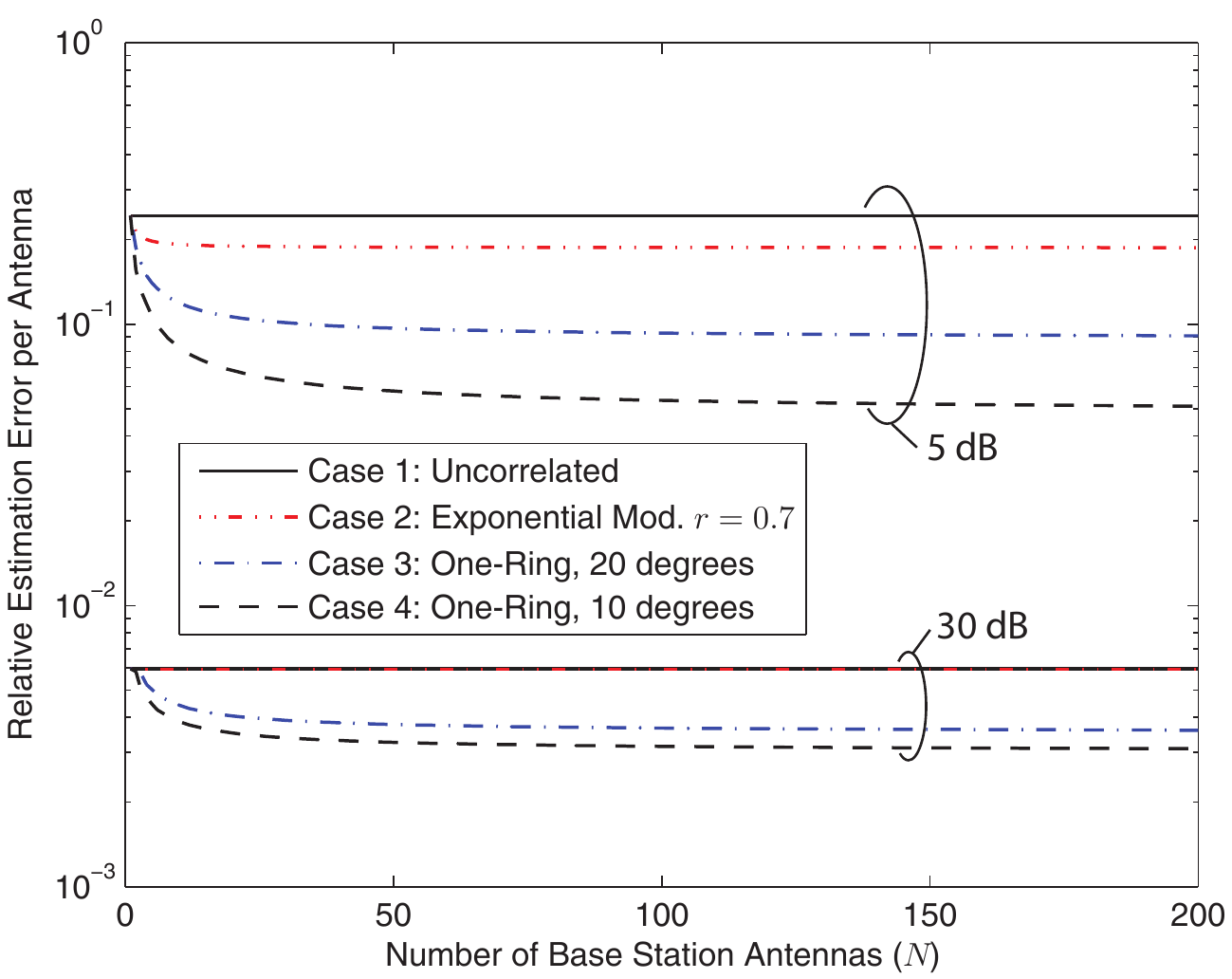} \vskip-2mm
\caption{Estimation error per antenna element for the LMMSE estimator in Theorem \ref{theorem:LMMSE-estimator} as a function of the number of BS antennas. Four different channel covariance models are considered and $\kappa_t^{\mathrm{UE}} =\kappa_r^{\mathrm{BS}} =0.05^2$.}\label{figure_estimation_diffchannelmodels}
\end{center} \vskip-4mm
\end{figure}

The relative estimation error per channel element is shown in Fig.~\ref{figure_estimation_diffchannelmodels} for these four channel covariance models. We consider two SNRs (5 and 30 dB), hardware impairments with $\kappa_t^{\mathrm{UE}} =\kappa_r^{\mathrm{BS}} =0.05^2$, and show the estimation errors as a function of the number of BS antennas. The main observation from Fig.~\ref{figure_estimation_diffchannelmodels} is that the choice of covariance model has a large impact on the estimation accuracy. It was proved in \cite{Bjornson2010a} that spatially correlated channels are easier to estimate and this is consistent with our results; increasing the coefficient $r$ in the exponential correlation model and decreasing the angular spread in the one-ring model lead to higher spatial correlation and smaller errors in Fig.~\ref{figure_estimation_diffchannelmodels}.
However, the error floors due to hardware impairments make the difference between the models reduce with the SNR.
Moreover, the estimation error \emph{per} antenna is virtually independent of $N$ in the exponential correlation model, while increasing $N$ improves the error \emph{per} antenna in the one-ring model. This is explained by the limited richness of the propagation environment in the one-ring model, which is a physical property that we can expect in practice.

\begin{remark}[Acquiring Large Covariance Matrices]
The proposed channel estimator requires knowledge of the $N \times N$ covariance matrices $\vect{R}$ and $\vect{S}$. It becomes increasingly difficult to acquire consistent estimates of covariance matrices as their dimensions grow large \cite{Couillet2011a}. Fortunately, the channel statistics have a much larger coherence time and coherence bandwidth than the channel realization itself; thus, one can obtain many more observations in the covariance estimation than in channel vector estimation. Robust covariance estimators for large matrices were recently considered in \cite{Couillet2014a}. The impact of imperfect covariance information on the channel estimation accuracy was analyzed in \cite{Shariati2014a}. The authors observe that the usual improvement in MSE from having spatial correlation vanishes if the covariance information cannot be trusted, but the MSE degradation is otherwise small (if the estimated covariance matrices are robustified). Another problem is that the large-dimensional matrix inversion in \eqref{eq:LMMSE-estimator} is very computationally expensive, but \cite{Shariati2014a} proposed low-complexity approximations based on polynomial expansions.

Instead of acquiring the covariance matrix of a user directly, the coverage area can be divided into ``location bins'' with approximately the same channel statistics within each bin \cite{Huh2012a}. By acquiring and storing the covariance matrices for each bin in advance, it is sufficient to estimate the location of a user and then associate the user with the corresponding bin.
\end{remark}

\section{Downlink and Uplink Data Transmission}
\label{sec:downlink-uplink-capacity}

This section analyzes the ergodic channel capacities of the downlink in \eqref{eq:downlink_channel} and the uplink in \eqref{eq:uplink_channel}, under the fixed TDD protocol depicted in Fig.~\ref{figure:tdd_operation}. More precisely, we derive upper and lower capacity bounds that reveal the fundamental impact of non-ideal hardware. These bounds are based on having perfect CSI (i.e., exact knowledge of $\vect{h}$) and imperfect pilot-based CSI estimation (using the LMMSE estimator in Theorem \ref{theorem:LMMSE-estimator}), respectively. Since these are two extremes, the capacity bounds hold when using the channel estimation technique proposed in Section \ref{sec:channel-estimation} and for any better CSI acquisition technique that can be derived in the future. We now define the DL and UL capacities for arbitrary CSI quality at the BS and UE.

We consider the ergodic capacity (in bit/channel use) of the memoryless DL system in \eqref{eq:downlink_channel}. { In each coherence period, the BS has some arbitrary imperfect knowledge $\mathcal{H}^{\mathrm{BS}}$ of the current channel states $\mathcal{H}$ and uses it to select the conditional distribution $f(\vect{s}|\mathcal{H}^{\mathrm{BS}})$ of the data signal $\vect{s}$. The UE has a separate arbitrary imperfect knowledge $\mathcal{H}^{\mathrm{UE}}$ of the current channel states $\mathcal{H}$ and uses it to decode the data.} Based on the well-known capacity expressions in \cite{Caire1999a}, the ergodic DL capacity is
\begin{equation} \label{eq:downlink_capacity}
{\tt C}^{\mathrm{DL}} \!=\! \frac{  T^{\mathrm{DL}}_{\mathrm{data}}   }{ T_{\mathrm{coher}} } \, \mathbb{E} \left\{  \max_{f(\vect{s}|\mathcal{H}^{\mathrm{BS}}) \,: \, \mathbb{E}\{ \| \vect{s} \|_2^2\} \leq p^{\mathrm{BS}}}  \mathcal{I}(\vect{s}; y | \mathcal{H}, \mathcal{H}^{\mathrm{BS}}, \mathcal{H}^{\mathrm{UE}})   \right\}
\end{equation}
{ where $\mathcal{I}(\vect{s}; y | \mathcal{H}, \mathcal{H}^{\mathrm{BS}}, \mathcal{H}^{\mathrm{UE}})$ denotes the conditional mutual information between the received signal $y$ and data signal $\vect{s}$ for a given channel realization $\mathcal{H}$ and given channel knowledge $\mathcal{H}^{\mathrm{BS}}$ and $\mathcal{H}^{\mathrm{UE}}$. The expectation in \eqref{eq:downlink_capacity} is taken over the joint distribution of $\mathcal{H}$, $\mathcal{H}^{\mathrm{BS}}$, and $\mathcal{H}^{\mathrm{UE}}$.} Note that the factor $T^{\mathrm{DL}}_{\mathrm{data}} / T_{\mathrm{coher}}$ is the fixed fraction of channel uses allocated for DL data transmission.

In addition, the ergodic capacity (in bit/channel use) of the memoryless fading UL system in \eqref{eq:uplink_channel} is
\begin{equation} \label{eq:uplink_capacity}
{\tt C}^{\mathrm{UL}} \! = \! \frac{  T^{\mathrm{UL}}_{\mathrm{data}}   }{ T_{\mathrm{coher}} } \, \mathbb{E} \left\{  \max_{f(\p| \mathcal{H}^{\mathrm{UE}}) \,: \, \mathbb{E}\{|\p|^2\} \leq p^{\mathrm{UE}}}  \mathcal{I}(\p ; \vect{z} | \mathcal{H}, \mathcal{H}^{\mathrm{BS}}, \mathcal{H}^{\mathrm{UE}})   \right\}
\end{equation}
{ where $\mathcal{I}(\p ; \vect{z} | \mathcal{H}, \mathcal{H}^{\mathrm{BS}}, \mathcal{H}^{\mathrm{UE}})$ denotes the conditional mutual information between the received signal $\vect{z}$ and data signal $\p$ for a given channel realization $\mathcal{H}$ and given channel knowledge $\mathcal{H}^{\mathrm{BS}}$ and $\mathcal{H}^{\mathrm{UE}}$. The conditional probability distribution of the data signal is denoted $f(\p | \mathcal{H}^{\mathrm{UE}})$ and the expectation in  \eqref{eq:uplink_capacity} is taken over the joint distribution of $\mathcal{H}, \mathcal{H}^{\mathrm{BS}}, \mathcal{H}^{\mathrm{UE}}$.} The fraction of channel uses allocated for UL data transmission is $T^{\mathrm{UL}}_{\mathrm{data}}  / T_{\mathrm{coher}}$.

There are a few implicit properties in the capacity definitions.  Firstly, the interference variance $I_{\mathcal{H}}^{\mathrm{UE}}$ and covariance matrix $\vect{Q}_{\mathcal{H}}$ depend on the channel realizations $\mathcal{H}$ and change between coherence periods. We are not limiting the analysis to any specific interference models but take care of it in the capacity bounds; the lower bounds treat the interference as Gaussian noise, while the upper bounds assume perfect interference suppression.  Section \ref{sec:multi-cell-scenario} describes the interference in multi-cell scenarios in detail. Secondly, we assume that the distortion noises are temporally independent, which is a good model when the data signals are also temporally independent.

The next subsections study the capacity behavior in the limit of infinitely many BS antennas ($N \rightarrow \infty$), which bring insights on how hardware impairments affect channels with large antenna arrays. The DL and UL are analyzed side-by-side since the results follow from similar derivations.

\subsection{Upper Bounds on Channel Capacities}

Upper bounds on the capacities in \eqref{eq:downlink_capacity} and \eqref{eq:uplink_capacity} can be obtained by adding extra channel knowledge and removing all interference (i.e., $I_{\mathcal{H}}^{\mathrm{UE}}=0$ and $\vect{Q}_{\mathcal{H}}=\vect{0}$). { We assume that the UL/DL pilot signals provide the BS and UE with perfect channel knowledge in each coherence period: $ \mathcal{H}^{\mathrm{BS}} = \mathcal{H}^{\mathrm{UE}} = \mathcal{H}$.} Since the receiver noise and distortion noises in \eqref{eq:downlink_channel} and \eqref{eq:uplink_channel} are circularly symmetric complex Gaussian distributed and independent of the useful signals under perfect CSI, { we deduce that Gaussian signaling is optimal in the DL and UL \cite{Telatar1999a} and that single-stream transmission with $\rank(\vect{W})=1$ is sufficient to achieve optimality \cite{Bjornson2013c};} that is, we can set $\vect{s} = \vect{w} s$ for $s \sim \mathcal{CN}(0,p^{\mathrm{BS}})$ and some unit-norm beamforming vector $\vect{w}$ in the DL and $\p \sim \mathcal{CN}(0,p^{\mathrm{UE}})$ in the UL. This gives us the following initial upper bounds.

\begin{lemma} \label{lemma:initial-upper-capacity-bounds}
The downlink and uplink capacities in \eqref{eq:downlink_capacity} and \eqref{eq:uplink_capacity}, respectively, are bounded as
\begin{align} \label{eq:downlink_capacity_upper_first}
&{\tt C}^{\mathrm{DL}} \leq \frac{  T^{\mathrm{DL}}_{\mathrm{data}}   }{ T_{\mathrm{coher}} } \times \\  & \,\, \mathbb{E} \left\{ \log_2(1+ \vect{h}^H \Big( \kappa_t^{\mathrm{BS}} \vect{D}_{|\vect{h}|^2} + \kappa_r^{\mathrm{UE}} \vect{h} \vect{h}^H + \frac{\sigma_{\mathrm{UE}}^2}{p^{\mathrm{BS}}} \vect{I} \Big)^{-1} \vect{h} \right\} \notag \\ \label{eq:uplink_capacity_upper_first}
&{\tt C}^{\mathrm{UL}} \leq  \frac{  T^{\mathrm{UL}}_{\mathrm{data}}   }{ T_{\mathrm{coher}} } \times \\ &  \,\, \mathbb{E} \left\{ \log_2 \left( 1+ \vect{h}^H  \Big( \kappa_t^{\mathrm{UE}} \vect{h} \vect{h}^H + \kappa_r^{\mathrm{BS}} \vect{D}_{|\vect{h}|^2} + \frac{\sigma_{\mathrm{BS}}^2}{p^{\mathrm{UE}}} \vect{I} \Big)^{-1} \vect{h}  \right) \right\} \notag
\end{align}
where $\vect{D}_{|\vect{h}|^2} = \diag(|h_1|^2,\ldots,|h_N|^2)$ with $\vect{h}=[h_1 \, \ldots \, h_N]^T$. These upper bounds are achieved with equality under perfect CSI, using the beamforming vector
\begin{equation} \label{eq:optimal_beamforming_downlink_perfectCSI}
\begin{split}
\vect{w}^{\mathrm{DL}}_{\mathrm{upper}} =
\frac{ ( \kappa_t^{\mathrm{BS}} \vect{D}_{|\vect{h}|^2} + \frac{\sigma_{\mathrm{UE}}^2}{p^{\mathrm{BS}}} \vect{I} )^{-1} \vect{h}^* }{\big\| \big( \kappa_t^{\mathrm{BS}} \vect{D}_{|\vect{h}|^2} + \frac{\sigma_{\mathrm{UE}}^2}{p^{\mathrm{BS}}} \vect{I} \big)^{-1} \vect{h}^* \big\|_2}
\end{split}
\end{equation}
in the downlink and by applying a receive combining vector
\begin{equation} \label{eq:optimal_beamforming_uplink_perfectCSI}
\begin{split}
\vect{w}^{\mathrm{UL}}_{\mathrm{upper}} = \frac{( \kappa_r^{\mathrm{BS}} \vect{D}_{|\vect{h}|^2} + \frac{\sigma_{\mathrm{BS}}^2}{p^{\mathrm{UE}}} \vect{I} )^{-1} \vect{h}}{\big\| ( \kappa_r^{\mathrm{BS}} \vect{D}_{|\vect{h}|^2} + \frac{\sigma_{\mathrm{BS}}^2}{p^{\mathrm{UE}}} \vect{I} )^{-1} \vect{h} \big\|_2 }.
\end{split}
\end{equation}
in the uplink.\footnote{A receive combining vector $\vect{w}$ is a linear filter $\vect{w}^H \vect{z}$ that transforms the system into an effective single-input single-output (SISO) system.}
\end{lemma}
\begin{IEEEproof}
The proof is given in Appendix \ref{proof:lemma:initial-upper-capacity-bounds}.
\end{IEEEproof}

Note that the beamforming vector in \eqref{eq:optimal_beamforming_downlink_perfectCSI} and receive combining vector in \eqref{eq:optimal_beamforming_uplink_perfectCSI} only depend on the channel vector $\vect{h}$, hardware impairments at the BS, and the receiver noise. Hardware impairments at the UE have no impact on $\vect{w}^{\mathrm{DL}}_{\mathrm{upper}}$ and $\vect{w}^{\mathrm{UL}}_{\mathrm{upper}} $ since their distortion noise essentially act as an interferer with the same channel as the data signal; thus filtering cannot reduce it.

The bounds in Lemma \ref{lemma:initial-upper-capacity-bounds} are not amenable to simple analysis, but the lemma enables us to derive further bounds on the channel capacities that are expressed in closed form.

\begin{theorem} \label{theorem:upperbounds-capacities}
The downlink and uplink capacities in \eqref{eq:downlink_capacity} and \eqref{eq:uplink_capacity}, respectively, are bounded as
\begin{align} \label{eq:downlink_capacity_upper2}
{\tt C}^{\mathrm{DL}} &\leq {\tt C}^{\mathrm{DL}}_{\mathrm{upper}} = \frac{  T^{\mathrm{DL}}_{\mathrm{data}}   }{ T_{\mathrm{coher}} } \log_2\left( 1 + \frac{G^{\mathrm{DL}} }{1 +\kappa_r^{\mathrm{UE}} G^{\mathrm{DL}}} \right) \\
{\tt C}^{\mathrm{UL}} &\leq {\tt C}^{\mathrm{UL}}_{\mathrm{upper}} = \frac{  T^{\mathrm{UL}}_{\mathrm{data}}   }{ T_{\mathrm{coher}} } \log_2\left( 1 + \frac{G^{\mathrm{UL}} }{1 +\kappa_t^{\mathrm{UE}} G^{\mathrm{UL}}} \right) \label{eq:uplink_capacity_upper2}
\end{align}
where $r_{11},\ldots,r_{NN}$ are the diagonal elements of $\vect{R}$,
\begin{align} \label{eq:expression_G}
G^{\mathrm{DL}} &= \sum_{i=1}^{N} \frac{1}{\kappa_t^{\mathrm{BS}}} \! \left( 1 - \frac{\sigma_{\mathrm{UE}}^2 e^{\frac{\sigma_{\mathrm{UE}}^2}{p^{\mathrm{BS}} \kappa_t^{\mathrm{BS}} r_{ii}}} }{p^{\mathrm{BS}} \kappa_t^{\mathrm{BS}} r_{ii}} E_1 \bigg( \frac{\sigma_{\mathrm{UE}}^2}{p^{\mathrm{BS}} \kappa_t^{\mathrm{BS}} r_{ii}} \bigg) \right) \!, \\
G^{\mathrm{UL}} & = \sum_{i=1}^{N} \frac{1}{\kappa_r^{\mathrm{BS}}} \! \left( 1 - \frac{\sigma_{\mathrm{BS}}^2 e^{\frac{\sigma_{\mathrm{BS}}^2}{p^{\mathrm{UE}} \kappa_r^{\mathrm{BS}} r_{ii}}} }{p^{\mathrm{UE}} \kappa_r^{\mathrm{BS}} r_{ii}} E_1 \bigg( \frac{\sigma_{\mathrm{BS}}^2}{p^{\mathrm{UE}} \kappa_r^{\mathrm{BS}} r_{ii}} \bigg)  \right) \!, \label{eq:expression_G_uplink}
\end{align}
and $E_1(x) = \int_{1}^{\infty} \frac{e^{-tx} }{t} dt$ denotes the exponential integral.
\end{theorem}
\begin{IEEEproof}
The proof is given in Appendix \ref{proof:theorem:upperbounds-capacities}.
\end{IEEEproof}

These closed-form upper bounds provide important insights on the achievable DL and UL performance under transceiver hardware impairments. In particular, the following two corollaries provide some ultimate capacity limits in the asymptotic regimes of many BS antennas or large transmit powers.

\begin{corollary} \label{cor:upper_bound}
The downlink upper capacity bound in  \eqref{eq:downlink_capacity_upper2} has the following asymptotic properties:
\begin{align}
\lim_{p^{\mathrm{BS}} \rightarrow \infty} {\tt C}^{\mathrm{DL}}_{\mathrm{upper}} &= \frac{  T^{\mathrm{DL}}_{\mathrm{data}}   }{ T_{\mathrm{coher}} } \log_2\left( 1 + \frac{N}{\kappa_t^{\mathrm{BS}} +\kappa_r^{\mathrm{UE}} N} \right) \label{eq:upper-asymptotics-q}\\
\lim_{N \rightarrow \infty } {\tt C}^{\mathrm{DL}}_{\mathrm{upper}} &= \frac{  T^{\mathrm{DL}}_{\mathrm{data}}   }{ T_{\mathrm{coher}} } \log_2\left( 1 + \frac{1}{\kappa_r^{\mathrm{UE}}} \right) \label{eq:upper-asymptotics-N}.
\end{align}
\end{corollary}
\begin{IEEEproof}
The diagonal elements of $\vect{R}$ satisfy $r_{ii}>0 \,\, \forall i$, by definition, thus
$G^{\mathrm{DL}} \rightarrow \sum_{i=1}^N \frac{1}{\kappa_t^{\mathrm{BS}}} = \frac{N}{\kappa_t^{\mathrm{BS}}}$ as $p^{\mathrm{BS}} \rightarrow \infty$ for fixed $N$, giving \eqref{eq:upper-asymptotics-q}. The positive diagonal elements also implies $\frac{1}{N} G^{\mathrm{DL}} > 0$ as $N \rightarrow \infty$, thus $\frac{G^{\mathrm{DL}} }{1 +\kappa_r^{\mathrm{UE}} G^{\mathrm{DL}}}  - \frac{G^{\mathrm{DL}} }{\kappa_r^{\mathrm{UE}} G^{\mathrm{DL}}} \rightarrow 0$ as $N \rightarrow \infty$ which turns \eqref{eq:downlink_capacity_upper2} into \eqref{eq:upper-asymptotics-N}.
\end{IEEEproof}

This corollary shows that the DL capacity has finite ceilings when either the DL transmit power $p^{\mathrm{BS}}$ or the number of BS antennas $N$ grow large. The ceilings depend on the impairment parameters $\kappa_t^{\mathrm{BS}}$ and $\kappa_r^{\mathrm{UE}}$, but the UE impairments are clearly $N$ times more influential. Note that even very small hardware impairments will ultimately limit the capacity. In other words, the ever-increasing capacity observed in the high-SNR and large-$N$ regimes with ideal transceiver hardware (cf.~\cite{Jose2011b,Marzetta2010a,Hoydis2013a,Ngo2013a,Rusek2013a}) is not easily achieved in practice.

The next corollary provides analogous results for the UL.

\begin{corollary} \label{cor:upper_bound-uplink}
The uplink upper capacity bound in  \eqref{eq:uplink_capacity_upper2} has the following asymptotic properties:
\begin{align}
\lim_{p^{\mathrm{UE}} \rightarrow \infty} {\tt C}^{\mathrm{UL}}_{\mathrm{upper}} &= \frac{  T^{\mathrm{UL}}_{\mathrm{data}}   }{ T_{\mathrm{coher}} } \log_2\left( 1 + \frac{N}{\kappa_r^{\mathrm{BS}} +\kappa_t^{\mathrm{UE}} N} \right) \label{eq:upper-asymptotics-q-uplink}\\
\lim_{N \rightarrow \infty } {\tt C}^{\mathrm{UL}}_{\mathrm{upper}} &= \frac{  T^{\mathrm{UL}}_{\mathrm{data}}   }{ T_{\mathrm{coher}} } \log_2\left( 1 + \frac{1}{\kappa_t^{\mathrm{UE}}} \right) \label{eq:upper-asymptotics-N-uplink}.
\end{align}
\end{corollary}
\begin{IEEEproof}
This is proved analogously to Corollary \ref{cor:upper_bound}.
\end{IEEEproof}

As seen from Corollary \ref{cor:upper_bound-uplink}, the UL capacity also has finite ceilings when either the UL transmit power $p^{\mathrm{UE}}$ or the number of antennas $N$ grow large. Analogous to the DL, the UE impairments are $N$ times more influential than the BS impairments and thus dominate as $N \rightarrow \infty$.

The upper bounds in Corollaries \ref{cor:upper_bound} and \ref{cor:upper_bound-uplink} show that the DL and UL capacities are fundamentally limited by the transceiver hardware impairments. To be certain of the cause of these limits, we also need lower bounds on the channel capacities.

\subsection{Lower Bounds on Channel Capacities}
\label{subsec:lower-bounds-capacity}

We obtain lower capacity bounds by making the potentially limiting assumptions of Gaussian codebooks, treating interference as Gaussian noise, using linear single-stream beamforming in the DL, using linear receive combining in the UL, pilot-based channel estimation as in Theorem \ref{theorem:LMMSE-estimator}, and the entropy-maximizing Gaussian distribution on the CSI uncertainty at the receiver of the DL and UL.\footnote{The linear processing assumption is motivated by its asymptotic optimality as $N \rightarrow \infty$ \cite{Rusek2013a}.} The resulting lower bound is given in the following theorem.

\begin{theorem} \label{theorem:lower-bound-on-capacities}
{ Let $\tilde{\mathcal{H}}^{\textrm{UE}}$ and $\tilde{\mathcal{H}}^{\textrm{BS}}$ denote the CSI available in the decoding at the receiver in the downlink and uplink, respectively. These are degraded as compared to $\mathcal{H}^{\textrm{UE}}$ and $\mathcal{H}^{\textrm{BS}}$ or equal.} The downlink and uplink capacities in \eqref{eq:downlink_capacity} and \eqref{eq:uplink_capacity}, respectively, are then bounded as
\begin{align} \label{eq:lower-bound-capacity}
\! {\tt C}^{\mathrm{DL}} &\geq {\tt C}^{\mathrm{DL}}_{\mathrm{lower}} = \frac{  T^{\mathrm{DL}}_{\mathrm{data}}   }{ T_{\mathrm{coher}} } \mathbb{E} \left\{ \log_2 \left( 1 + {\tt SINR}^{\mathrm{DL}}_{\mathrm{lower}} (\vect{v}^{\mathrm{DL}}) \right) \right\} \\
\! {\tt C}^{\mathrm{UL}} &\geq {\tt C}^{\mathrm{UL}}_{\mathrm{lower}} = \frac{  T^{\mathrm{UL}}_{\mathrm{data}}   }{ T_{\mathrm{coher}} } \mathbb{E} \left\{ \log_2 \left( 1 + {\tt SINR}^{\mathrm{UL}}_{\mathrm{lower}} (\vect{v}^{\mathrm{UL}}) \right) \right\}
\label{eq:lower-bound-capacity-uplink}
\end{align}
where the beamforming vector $\vect{v}^{\mathrm{DL}} = [v^{\mathrm{DL}}_1 \, \ldots \, v^{\mathrm{DL}}_{K_r}]^T$ and the receive combining vector $\vect{v}^{\mathrm{UL}} = [v^{\mathrm{UL}}_1 \, \ldots \, v^{\mathrm{UL}}_{K_r}]^T$ are functions of  $\hat{\vect{h}}$ and have unit norms. The expectations are taken over $\tilde{\mathcal{H}}^{\textrm{UE}}$ and $\tilde{\mathcal{H}}^{\textrm{BS}}$, while the SINR expressions for DL and UL are given in \eqref{eq:SINR-approximation-lower-bound} and \eqref{eq:SINR-approximation-lower-bound-uplink}, respectively, at the top of the next page.
\end{theorem}
\begin{IEEEproof}
This theorem is obtained by taking lower bounds on the mutual information in the same way as was previously proposed in \cite{Medard2000a} and \cite{Yoo2006b}. This bounding technique was applied to massive MIMO systems with ideal hardware in \cite{Jose2011b,Hoydis2013a,Ngo2013a} (among others), by making the limiting assumptions listed in the beginning of this subsection. The distortion noises from non-ideal hardware  act as additional noise sources with spatially correlated covariance matrices, thus these can easily be incorporated into the proofs used in previous works.
\end{IEEEproof}

\begin{figure*}[!t]

\normalsize

\setcounter{MYtempeqncnt}{\value{equation}}

\setcounter{equation}{35}

\begin{align} \label{eq:SINR-approximation-lower-bound}
\!\!\! {\tt SINR}^{\mathrm{DL}}_{\mathrm{lower}}(\vect{v}^{\mathrm{DL}}) & \!=\!
\frac{ \left| \mathbb{E}\{ \vect{h}^H \vect{v}^{\mathrm{DL}} \, | \tilde{\mathcal{H}}^{\textrm{UE}}\}  \right|^2 }{  (\! 1\!+\! \kappa_r^{\mathrm{UE}} \!) \mathbb{E} \left\{ | \vect{h}^H \vect{v}^{\mathrm{DL}} |^2 \, | \tilde{\mathcal{H}}^{\textrm{UE}} \right\} \!-\! \left| \mathbb{E}\{ \vect{h}^H \vect{v}^{\mathrm{DL}} \, | \tilde{\mathcal{H}}^{\textrm{UE}} \}  \! \right|^2 \!+\! \kappa_t^{\mathrm{BS}} \! \fracSumtwo{i=1}{N} \mathbb{E}\{ |h_i|^2 |v_i^{\mathrm{DL}}|^2 \, | \tilde{\mathcal{H}}^{\textrm{UE}}\} \!+\! \frac{\mathbb{E}\{ I_{\mathcal{H}}^{\mathrm{UE}}   \, | \tilde{\mathcal{H}}^{\textrm{UE}}\}  }{p^{\mathrm{BS}}} \!+\! \frac{ \sigma_{\mathrm{UE}}^2     }{p^{\mathrm{BS}}} } \\
\label{eq:SINR-approximation-lower-bound-uplink}
\!\!\! {\tt SINR}^{\mathrm{UL}}_{\mathrm{lower}}(\vect{v}^{\mathrm{UL}}) & \!=\!
\frac{ \left| \mathbb{E}\{ \vect{h}^H \vect{v}^{\mathrm{UL}} \, | \tilde{\mathcal{H}}^{\textrm{BS}}\}  \right|^2 }{  (\! 1\!+\! \kappa_t^{\mathrm{UE}} \!) \mathbb{E} \left\{ | \vect{h}^H \vect{v}^{\mathrm{UL}} |^2 \, | \tilde{\mathcal{H}}^{\textrm{BS}} \right\} \!-\! \left| \mathbb{E}\{ \vect{h}^H \vect{v}^{\mathrm{UL}} \, | \tilde{\mathcal{H}}^{\textrm{BS}} \}  \! \right|^2 \!+\! \kappa_r^{\mathrm{BS}} \! \fracSumtwo{i=1}{N} \mathbb{E}\{ |h_i|^2 |v_i^{\mathrm{UL}}|^2 \, | \tilde{\mathcal{H}}^{\textrm{BS}} \} \!+\! \frac{\mathbb{E}\left\{ (\vect{v}^{\mathrm{UL}})^H (\vect{Q}_{\mathcal{H}}+\sigma_{\mathrm{BS}}^2 \vect{I}) \vect{v}^{\mathrm{UL}} \, | \tilde{\mathcal{H}}^{\textrm{BS}} \right\} }{p^{\mathrm{UE}}} }
\end{align}

\setcounter{equation}{\value{MYtempeqncnt}}
\hrulefill
\vskip-3mm
\end{figure*}

\setcounter{equation}{37}

This theorem is the key to the lower capacity bounding in this paper. The lower bounds in \eqref{eq:lower-bound-capacity} and \eqref{eq:lower-bound-capacity-uplink} can be computed numerically for any channel distribution and any way of selecting the beamforming vector (in the DL) and receiver combining vector (in the UL) from the channel estimate $\hat{\vect{h}}$, provided that the conditional distribution of $(\vect{h},\hat{\vect{h}})$ given $\tilde{\mathcal{H}}$ can be characterized.\footnote{Finding such a characterization is a challenging task, except for the case $\tilde{\mathcal{H}}^{\textrm{BS}} = \tilde{\mathcal{H}}^{\textrm{UE}} = \emptyset$ considered in Theorem \ref{theorem:asymptotic-equivalence}.}
To bring explicit insights on the behavior when the number of antennas, $N$, grows large, we have the following result for the cases of (approximate) maximum ratio transmission (MRT) in the DL and (approximate) maximum ratio combining (MRC) in the UL.

\begin{theorem} \label{theorem:asymptotic-equivalence}
Assume that no instantaneous CSI is utilized for decoding (i.e., $\tilde{\mathcal{H}}^{\textrm{BS}} = \tilde{\mathcal{H}}^{\textrm{UE}} = \emptyset$). For $\vect{v}= \frac{\hat{\vect{h}}}{\|\hat{\vect{h}}\|_2}$ the
terms in \eqref{eq:SINR-approximation-lower-bound} and \eqref{eq:SINR-approximation-lower-bound-uplink} behave as
\begin{align}\label{eq:det-equiv-signalpart}
 \left| \mathbb{E}\{ \vect{h}^H \vect{v} \}  \right|^2  &=  \left|  \mathbb{E}\left\{ \varphi \right\} \right|^2 \tr(\vect{R}-\vect{C}) + \mathcal{O}(\sqrt{N}) \\ \label{eq:det-equiv-signalpart2}
 \mathbb{E} \left\{ | \vect{h}^H \vect{v} |^2 \right\}  &=   \mathbb{E}\left\{ |\varphi|^2
\right\} \tr(\vect{R}-\vect{C}) + \mathcal{O}(\sqrt{N}) \\
\sum_{i=1}^{N} \mathbb{E}\{ |h_i|^2 |v_i|^2\} &= \mathcal{O}(1) \label{eq:det-equiv-extra-interference}
\end{align}
where
\begin{equation} \label{eq:definition-varrho}
\varphi = \frac{(1+\p^{-1} \eta_{t}^{\mathrm{UE}}) \sqrt{\tr(\vect{R}-\vect{C})}}{\sqrt{ \tr\big(\vect{A} ( |\p + \eta_{t}^{\mathrm{UE}}|^2 \vect{R} + \boldsymbol{\Psi}) \vect{A}^H \big) } }
\end{equation}
is a function of the stochastic variable $\eta_{t}^{\mathrm{UE}}$ while $\vect{A} = \p^* \vect{R} \bar{\vect{Z}}^{-1}$ and $\boldsymbol{\Psi} =   p^{\mathrm{UE}} \kappa_r^{\mathrm{BS}}  \vect{R}_{\diag} + \vect{S} + \sigma_{\mathrm{BS}}^2 \vect{I}$ are deterministic matrices.
\end{theorem}
\begin{IEEEproof}
The proof is given in Appendix \ref{proof:theorem:asymptotic-equivalence}.
\end{IEEEproof}

Similar asymptotic behaviors were derived in \cite{Jose2011b,Hoydis2013a,Ngo2013a} for the case of ideal hardware.\footnote{We stress that the assumption in Theorem \ref{theorem:asymptotic-equivalence} that decoding is performed without instantaneous CSI is only made to enable closed-form lower bounds. The BS should certainly exploit the channel estimate $\hat{\vect{h}}$ and the UE might receive a downlink pilot signal that enables estimation of the effective channel $\vect{h}^H \vect{v}^{\mathrm{DL}}$.
While this is relatively easy to handle with ideal hardware, where the channel estimate and estimation error are independent (cf.~\cite{Hoydis2013a}), the extension to non-ideal hardware seems intractable due the statistical dependence between the channel estimate and estimation error.}
In the general case with hardware impairments, the expectations of $\varphi$ and $|\varphi|^2$ must be computed numerically, because the randomness of the scalar distortion noise $\eta_{t}^{\mathrm{UE}}$ at the UE remains even when $N$ grows large. In the special case of $\kappa_{t}^{\mathrm{UE}}=0$ (which implies $\eta_{t}^{\mathrm{UE}}=0$), \eqref{eq:det-equiv-signalpart} and  \eqref{eq:det-equiv-signalpart2} both reduce to $\tr(\vect{R}-\vect{C}) + \mathcal{O}(\sqrt{N})$. For $\kappa_{t}^{\mathrm{UE}}>0$, the terms in Theorem \ref{theorem:asymptotic-equivalence} are easy to compute numerically.

Based on this result, we provide now an asymptotic characterization of the downlink capacity.

\begin{corollary} \label{cor:lower_bound}
Consider the DL with beamforming vector $\vect{v} = \frac{\hat{\vect{h}}^*}{\|\hat{\vect{h}}\|_2}$ and $ \tilde{\mathcal{H}}^{\textrm{UE}} = \emptyset$. If $\mathbb{E}\{ I_{\mathcal{H}}^{\mathrm{UE}} \} \leq  \mathcal{O}(N^n)$ for some $n<1$, the lower capacity bound in \eqref{eq:lower-bound-capacity} can be expressed as
\begin{equation} \label{eq:capacity-lower-equivalent}
\begin{split}
& {\tt C}^{\mathrm{DL}} \geq  \frac{  T^{\mathrm{DL}}_{\mathrm{data}}   }{ T_{\mathrm{coher}} } \times \\ & \log_2 \! \left( \!  1 \! + \! \frac{  \left|  \mathbb{E} \left\{ \varphi \right\} \right|^2 \! +\! \mathcal{O} \left(\frac{1}{\sqrt{N}} \right)
 }{ (1 \!+ \! \kappa_r^{\mathrm{UE}})
   \mathbb{E}\left\{ | \varphi |^2 \right\} \!-\!  \left|  \mathbb{E} \left\{ \varphi \right\} \right|^2  \!+\! \mathcal{O} \left(\frac{1}{\sqrt{N}} \!+\! \frac{1}{N^{1-n}} \right) }  \right)
   \end{split}
\end{equation}
where $\varphi$ is given in \eqref{eq:definition-varrho}. The terms $\mathcal{O} \left(\frac{1}{\sqrt{N}} \right)$ and $\mathcal{O} \left( \frac{1}{N^{1-n}} \right)$ vanish when $N \rightarrow \infty$, while the other terms are strictly positive in the limit.
\end{corollary}
\begin{IEEEproof}
The expression \eqref{eq:capacity-lower-equivalent} is obtained from \eqref{eq:lower-bound-capacity} by plugging in the expressions in Theorem \ref{theorem:asymptotic-equivalence} and multiplying each term by $ \frac{1}{\tr(\vect{R}-\vect{C})} = \frac{1}{p^{\mathrm{UE}} \tr(\vect{R} \bar{\vect{Z}}^{-1} \vect{R} )} = \mathcal{O}(N^{-1})$. The interference term becomes $ \frac{\mathbb{E} \{ I_{\mathcal{H}}^{\mathrm{UE}} \} }{ p^{\mathrm{BS}} \tr(\vect{R}-\vect{C})} \!=\!  \mathcal{O} \left( \frac{1}{N^{1-n}}\right)$.
\end{IEEEproof}

Combining the upper bound in Corollary \ref{cor:upper_bound} with the lower bound in Corollary \ref{cor:lower_bound}, we have a clear characterization of the DL capacity behavior when $N \rightarrow \infty$. Both bounds are independent of $\kappa_t^{\mathrm{BS}}$ in the limit, thus the transmitter hardware of the BS plays little role in massive MIMO systems. Contrary to the upper bound, the level of receiver hardware impairments at the BS ($\kappa_r^{\mathrm{BS}}$) is present in the lower bound \eqref{eq:capacity-lower-equivalent}, through $\vect{A}$ and $\boldsymbol{\Psi}$ in $\varphi$. However, the numerical results in Section \ref{subsec:capacity-numerical-illustrations} reveal that the asymptotic impact of BS impairments is negligible also in the lower bound. This can also be seen analytically in certain cases; if $\kappa_{t}^{\mathrm{UE}}=0$ we get $\varphi = 1$ and therefore
\begin{align}
\lim_{N \rightarrow \infty } {\tt C}^{\mathrm{DL}}_{\mathrm{lower}} &= \frac{  T^{\mathrm{DL}}_{\mathrm{data}}   }{ T_{\mathrm{coher}} } \log_2\left( 1 + \frac{1}{\kappa_r^{\mathrm{UE}}} \right) \label{eq:lower-asymptotics-N}.
\end{align}
In this special case, the lower bound actually approaches the upper bound in \eqref{eq:upper-asymptotics-N} asymptotically, and any DL capacity can be achieved by making $\kappa_r^{\mathrm{UE}}$ sufficiently small. The opposite is not true; setting $\kappa_r^{\mathrm{BS}} = 0$ will \emph{not} make the impact of UE impairments vanish. We therefore conclude that the DL capacity limit is mainly determined by the level of impairments at the UE, both in the uplink estimation ($\kappa_{t}^{\mathrm{UE}}$) and the downlink transmission ($\kappa_r^{\mathrm{UE}}$)---although the former connection was not visible in the upper bound since it was based on perfect CSI.

For the uplink, we have the following similar asymptotic capacity characterization.

\begin{corollary} \label{cor:lower_bound-uplink}
Consider the UL with receive combining vector $\vect{v} = \frac{\hat{\vect{h}}}{\|\hat{\vect{h}}\|_2}$ and $\tilde{\mathcal{H}}^{\textrm{BS}} = \emptyset$. If $\mathbb{E}\{ \| \vect{Q}_{\mathcal{H}}  \|_2 \} \leq \mathcal{O}(N^n)$ for some $n<1$, the lower capacity bound in \eqref{eq:lower-bound-capacity-uplink} can be expressed as
\begin{equation} \label{eq:capacity-lower-equivalent-uplink}
\begin{split}
&{\tt C}^{\mathrm{UL}} \geq  \frac{  T^{\mathrm{UL}}_{\mathrm{data}}   }{ T_{\mathrm{coher}} }  \times \\
& \log_2 \! \left( \!  1 \! + \! \frac{  \left|  \mathbb{E} \left\{ \varphi
  \right\} \right|^2 \! + \! \mathcal{O} \left(\frac{1}{\sqrt{N}} \right)
 }{ (1 \!+\!  \kappa_t^{\mathrm{UE}})
   \mathbb{E} \left\{ | \varphi|^2 \right\} \!-\!  \left|  \mathbb{E} \left\{ \varphi
\right\} \right|^2  \!+\! \mathcal{O} \left(\frac{1}{\sqrt{N}} \!+\! \frac{1}{N^{1-n}} \right) } \right)
\end{split}
\end{equation}
where $\varphi$ is given in \eqref{eq:definition-varrho}. The terms $\mathcal{O} \left(\frac{1}{\sqrt{N}} \right)$ and $\mathcal{O} \left( \frac{1}{N^{1-n}} \right)$ vanish when $N \rightarrow \infty$, while the other terms are strictly positive in the limit.
\end{corollary}
\begin{IEEEproof}
The expression \eqref{eq:capacity-lower-equivalent-uplink} is obtained from \eqref{eq:lower-bound-capacity-uplink} by plugging in the expressions from Theorem \ref{theorem:asymptotic-equivalence} and multiplying each term by  $ \frac{1}{\tr(\vect{R}-\vect{C})} = \frac{1}{p^{\mathrm{UE}} \tr(\vect{R} \bar{\vect{Z}}^{-1} \vect{R} )} = \mathcal{O}(N^{-1})$.
The interference term becomes $ \frac{\mathbb{E}\{ \vect{v}^H \vect{Q}_{\mathcal{H}} \vect{v} \} }{ p^{\mathrm{UE}} \tr(\vect{R}-\vect{C})} \!=\!  \mathcal{O} \left( \frac{1}{N^{1-n}} \right)$.
\end{IEEEproof}

The upper bound in Corollary \ref{cor:upper_bound-uplink} and the lower bound in Corollary \ref{cor:lower_bound-uplink} provide a joint characterization of the uplink capacity when $N$ grows large. The UE impairments manifest the behavior in both bounds; the BS impairments are present in \eqref{eq:capacity-lower-equivalent} since $\varphi$ depends on $\vect{A}$ and $\boldsymbol{\Psi}$, but their impact vanish when $\kappa_{t}^{\mathrm{UE}} \rightarrow 0$. By making $\kappa_{t}^{\mathrm{UE}}$ appropriately small, we can thus achieve any UL capacity as $N$ grows large. We therefore conclude that it is of main importance to have high quality hardware at the UE, which is analog to our observations for the DL. These observations are illustrated numerically in the next subsection and are explained by the following remark.

\begin{remark}[BS Impairments Vanish Asymptotically] \label{remark:vanishing-BS}
The lower and upper bounds show that it is the quality of the UE's transceiver hardware that limits the DL and UL capacities as $N \rightarrow \infty$. Thus, the detrimental effect of hardware impairments at the BS vanishes completely, or almost completely, when the number of BS antennas grows large. This is, simply speaking, since the BS's distortion noises are spread in arbitrary directions in the $N$-dimensional vector space while the increased spatial resolution of the array enables very exact transmit beamforming and receive combining for the useful signal. This is a very promising result since large arrays are more prone to impairments, due to implementation limitations and the will to use antenna elements of lower quality (to avoid having deployment costs that increase linearly with $N$). In contrast, the UE's distortion noises are non-vanishing since they behave as interferers with the same effective channels as the useful signals.
\end{remark}

Corollaries \ref{cor:lower_bound} and \ref{cor:lower_bound-uplink} assumed that the inter-user interference satisfy $\mathbb{E}\{ I_{\mathcal{H}}^{\mathrm{UE}} \} \leq \mathcal{O}(N^n)$ and $\mathbb{E}\{ \| \vect{Q}_{\mathcal{H}}  \|_2 \} \leq \mathcal{O}(N^n)$, respectively, for some $n<1$. These conditions imply that the interference terms only vanish asymptotically if the scaling with $N$ is slower than linear. { This is satisfied by \emph{regular} interference which has constant variance (i.e., $n=0$), but there is a special type of non-regular pilot contaminated interference in multi-cell systems that scales linearly with $N$.} This adds an additional non-vanishing term to the denominators of \eqref{eq:capacity-lower-equivalent} and \eqref{eq:capacity-lower-equivalent-uplink}. We detail this scenario in Section \ref{sec:multi-cell-scenario}.

Finally, we stress that the DL and UL capacity bounds in Corollaries \ref{cor:upper_bound} and \ref{cor:upper_bound-uplink}, respectively, have a very similar structure. The main difference is that the UL is only affected by UL hardware impairments (i.e., $\kappa_{t}^{\mathrm{UE}},\kappa_{r}^{\mathrm{BS}}$), while the DL is affected by both DL and UL hardware impairments (i.e., all $\kappa$-parameters) due to the reverse-link channel estimation.

\subsection{Numerical Illustrations}
\label{subsec:capacity-numerical-illustrations}

Next, we illustrate the lower and upper bounds on the capacity that were derived earlier in this section. We consider a scenario without interference, $\vect{Q}_{\mathcal{H}}=\vect{S}=\vect{0}$ and $I_{\mathcal{H}}^{\mathrm{UE}}=0$, and define the average SNRs as $p^{\mathrm{UE}} \frac{\tr (\vect{R})}{N \sigma_{\mathrm{BS}}^2}$  and $p^{\mathrm{BS}} \frac{\tr (\vect{R})}{N \sigma_{\mathrm{UE}}^2}$ in the UL and DL, respectively. We consider different fixed SNR values, while we vary the number of antennas $N$ and the levels of hardware impairments. We assume that the transmitter and receiver hardware of each device are of the same quality: $\kappa^{\mathrm{BS}} \triangleq \kappa_t^{\mathrm{BS}} =\kappa_r^{\mathrm{BS}} $ at the BS and $\kappa^{\mathrm{UE}} \triangleq \kappa_t^{\mathrm{UE}} =\kappa_r^{\mathrm{UE}} $ at the UE.\footnote{The transmitter and receiver hardware both involve converters, mixers, filters, and oscillators; see \cite[Fig.~1]{Cui2005a} for a typical transceiver model. The main difference is the type of amplifiers, thus the assumption of identical levels of impairments makes sense when the non-linearities of the amplifiers at the transmitter are not the dominating source of distortion noise.} Furthermore, we assume $\frac{  T^{\mathrm{DL}}_{\mathrm{data}}   }{ T_{\mathrm{coher}} } = \frac{  T^{\mathrm{UL}}_{\mathrm{data}}   }{ T_{\mathrm{coher}} } = 0.45$, which are the percentage of DL data and UL data. These assumptions make the bounds for the DL and UL capacities become identical, thus we can simulate the DL and UL simultaneously.

\begin{figure}
         \centering
         \begin{subfigure}[b]{\columnwidth}
                 \includegraphics[width=\columnwidth]{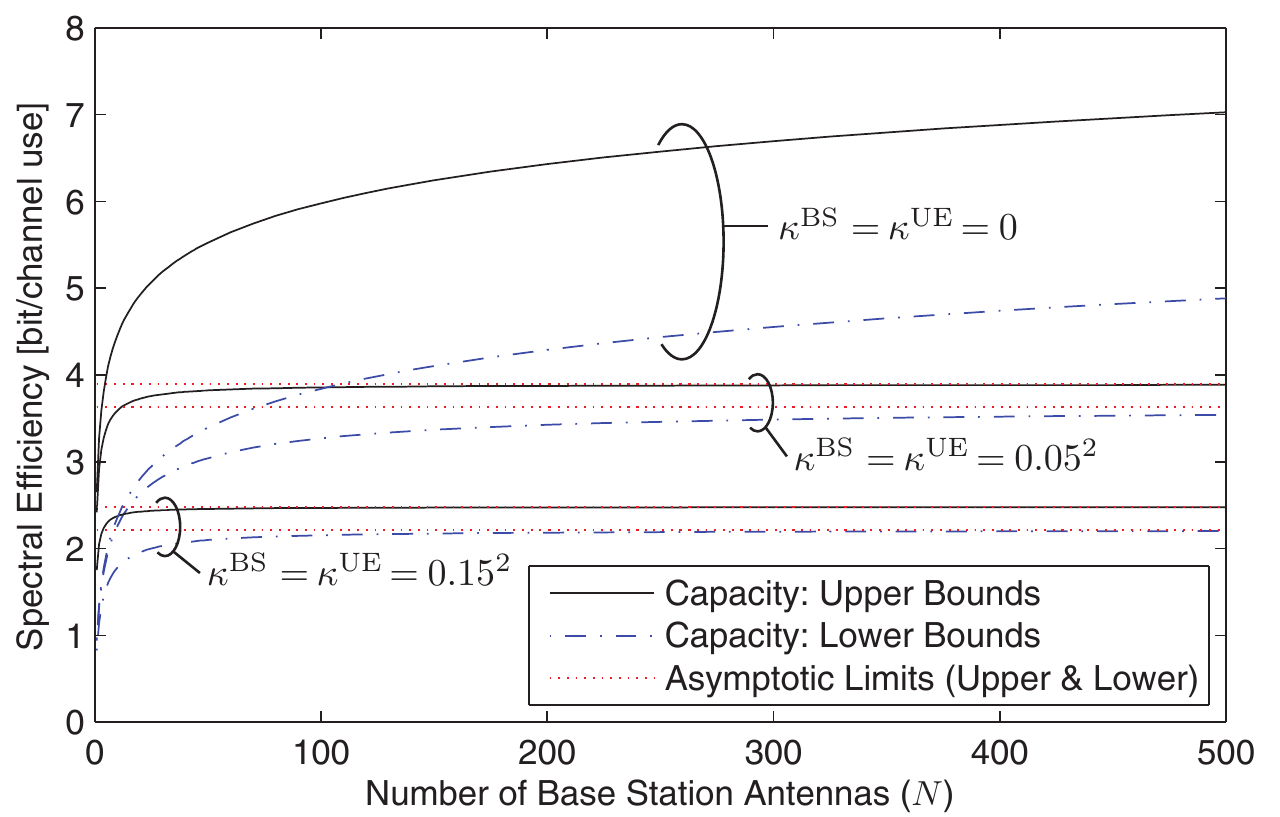}
                 \caption{SNR: 20 dB}
                 \label{figure_capacity_20dB}
         \end{subfigure} \\
         \begin{subfigure}[b]{\columnwidth}
                 \includegraphics[width=\columnwidth]{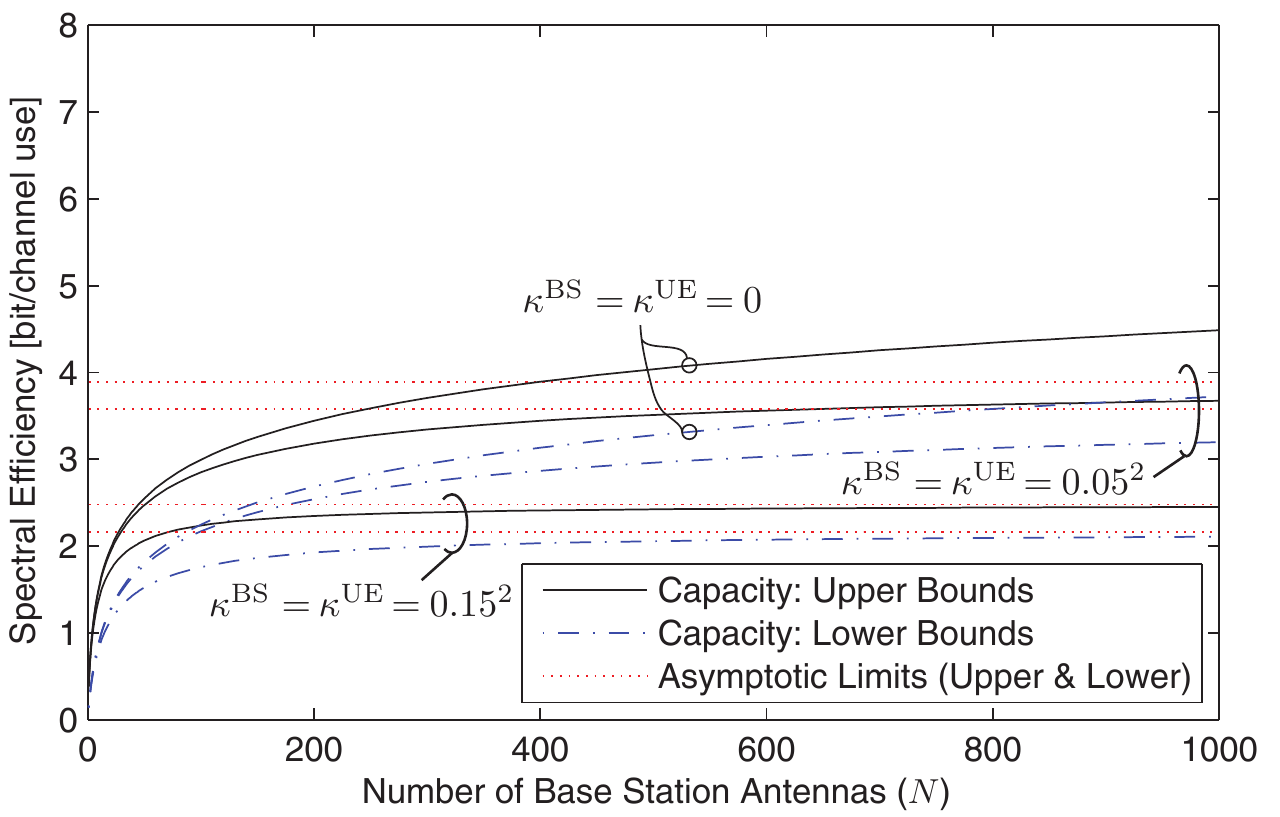}
                 \caption{SNR: 0 dB}
                 \label{figure_capacity_0dB}
         \end{subfigure}
         \caption{Lower and upper bounds on the capacity. Hardware impairments have a fundamental impact on the asymptotic behavior as $N$ grows large.}\label{figure_capacity}
\end{figure}

Fig.~\ref{figure_capacity} considers a spatially uncorrelated scenario with $\vect{R}=\vect{I}$ for different levels of impairments:  $\kappa_t^{\mathrm{UE}} =\kappa_r^{\mathrm{BS}} \in \{ 0, 0.05^2, 0.15^2 \}$. The meaning of these parameter values was discussed in Remark \ref{remark:distortion-noise}. { Simulation results are given for SNRs of 20 dB and 0 dB. The capacity with ideal hardware grows without bound as $N \rightarrow \infty$, while the lower and upper bounds converge to finite limits under transceiver hardware impairments. The main difference between the two SNR values is the convergence speed, while the upper bounds are exactly the same and the lower bounds are approximately the same. Recall that these bounds hold under any CSI $\mathcal{H}^{\mathrm{BS}}$ at the BS and $\mathcal{H}^{\mathrm{UE}}$ at the UE;} the lower bounds represent no instantaneous CSI in the decoding step and the upper bounds represent perfect CSI. Although the gap between these extremes is large for ideal hardware, the difference is remarkably small under non-ideal hardware due to the finite capacity limit (caused by distortion noise) and the channel hardening that makes stochastic inner product such as $\vect{h}^H \vect{v}$ become increasingly deterministic as $N$ grows large. Since a main difference between the lower and upper bounds is the quality of the CSI, the small difference shows that the estimation errors have only a minor impact on the capacity; hence, the estimation error floors described in Section \ref{sec:channel-estimation} has no dominating impact in the large-$N$ regime.

The asymptotic capacity limits in Fig.~\ref{figure_capacity} are characterized by the level of impairments, thus the hardware quality has a fundamental impact on the achievable spectral efficiency. { If the SNRs are sufficiently high (e.g., 20 dB), the majority of the multi-antenna gain is achieved at relatively low $N$}; in particular, only minor improvements can be achieved by having more than $N=100$ antennas. Larger numbers are, however, useful for inter-user interference suppression and multiplexing; see Section \ref{sec:multi-cell-scenario}. We need many more antennas to achieve convergence at 0 dB SNR than at 20 dB, because a 100 times larger array gain is required to compensate for the lower SNR. Hence, we conclude that the massive MIMO gains are much more attractive at higher SNRs (which matches well with the results in Section \ref{sec:channel-estimation} where 20--30 dB SNR was needed to achieve a close-to-perfect channel estimate). Therefore, we only consider an SNR of 20 dB it the remainder of this section.

\begin{figure}
\begin{center}
\includegraphics[width=\columnwidth]{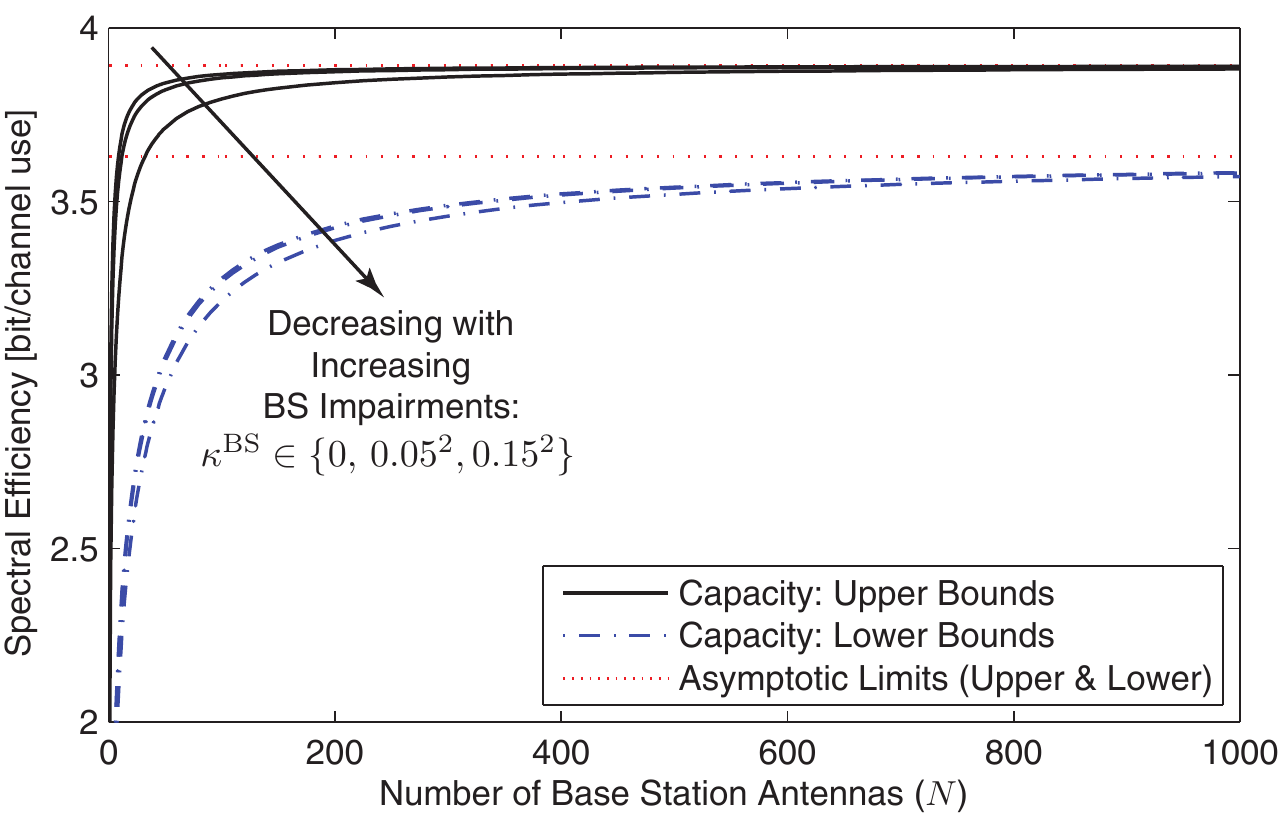} \vskip-2mm
\caption{Lower and upper bounds on the capacity for $\kappa^{\mathrm{UE}} = 0.05^2$. The impact of hardware impairments at the BS vanishes asymptotically.}\label{figure_capacity_diffimpairments}
\end{center} \vskip-4mm
\end{figure}

Fig.~\ref{figure_capacity_diffimpairments} considers the same scenario as in Fig.~\ref{figure_capacity}  but with a fixed level of impairments $\kappa^{\mathrm{UE}} = 0.05^2$ at the UE  and different values at the BS. As expected from the analysis, the lower and upper capacity bounds increase with $\kappa^{\mathrm{BS}}$, but the difference is only visible at small $N$ since the curves converge to virtually the same value as $N \rightarrow \infty$. This validates that the impact of impairments at the BS vanishes as $N$ grows large.

\begin{figure}
\begin{center}
\includegraphics[width=\columnwidth]{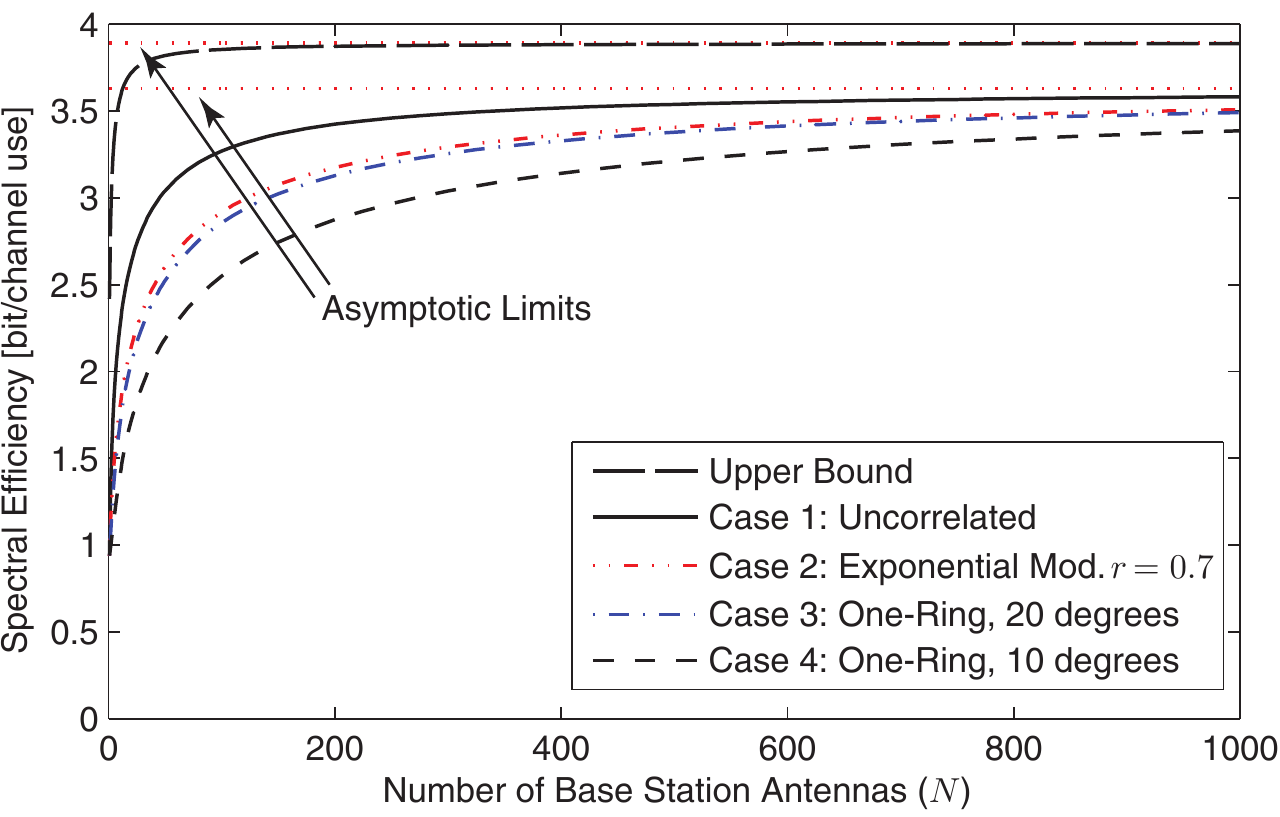} \vskip-2mm
\caption{Lower and upper bounds on the capacity as a function of the number of BS antennas. Four different channel covariance models are considered and the hardware impairments are characterized by $\kappa^{\mathrm{BS}} =\kappa^{\mathrm{UE}} = 0.05$.}\label{figure_capacity_diffchannelmodels}
\end{center} \vskip-4mm
\end{figure}

Finally, we consider the capacity behavior for different channel covariance models, namely the four propagation scenarios described in Section \ref{subsec:estimation-numerical}. The lower and upper capacity bounds are shown in Fig.~\ref{figure_capacity_diffchannelmodels} for $\kappa^{\mathrm{BS}} =\kappa^{\mathrm{UE}} = 0.05$. The upper bound is identical for all the models, since it only utilizes the diagonal elements of $\vect{R}$. However, there are clear differences between the lower bounds. The spatially uncorrelated covariance model provides the highest performance, while the strongly spatially correlated one-ring model from \cite{Shiu2000a} with 10 degrees angular spread gives the lowest performance. This stands in contrast to Section \ref{subsec:estimation-numerical}, where the highly correlated channels gave the lowest estimation errors (i.e., highest estimation accuracy). However, the differences between the channel covariance models vanish asymptotically as $N \rightarrow \infty$.

\section{Improving Energy Efficiency and Reducing Hardware Quality}
\label{sec:energy-efficiency}

Next, we analyze how the energy efficiency (EE) can be optimized in massive MIMO systems. The EE is measured in bit/Joule and a common EE definition is the ratio of the spectral efficiency (in bit/channel use) to the emitted power (in Joule/channel use). It has recently been shown that the array gain in massive MIMO systems can be utilized to reduce the emitted power; see \cite{Hoydis2013a,Ngo2013a} for  systems with ideal hardware and \cite{Pitarokoilis2014a} for systems with phase noise from free-running oscillators. More specifically, these prior works show that one can reduce the transmit powers as $1/N^t$, for $0<t<\frac{1}{2}$, and still achieve  non-zero spectral efficiencies as $N \rightarrow \infty$. By following this power scaling law, we can achieve an \emph{infinitely} high EE as $N \rightarrow \infty$ because the numerator has a non-zero limit and the denominator goes to zero as $1/N^t$ \cite{Bjornson2013f}. Although this property indicates that massive MIMO systems can be very energy efficient, the unboundedness also shows that the conventional EE metric needs to be revised when applied to massive MIMO systems. In this section, we consider a refined metric of overall EE (based on prior work in \cite{Auer2011a,Xu2011b,Ng2012a,Yang2013a,Bjornson2013e}) and use it to analyze the overall EE of massive MIMO systems.

Under the TDD protocol, the energy consumed in the amplifiers of the transmitters (per coherence period) is
\begin{equation}
E_{\mathrm{amp}} = (T^{\mathrm{DL}}_{\mathrm{pilot}} + T^{\mathrm{DL}}_{\mathrm{data}}) \frac{ p^{\mathrm{BS}}}{\omega^{\mathrm{BS}}} + (T^{\mathrm{UL}}_{\mathrm{pilot}} + T^{\mathrm{UL}}_{\mathrm{data}}) \frac{ p^{\mathrm{UE}}}{\omega^{\mathrm{UE}}} \,\,\, \mathrm{[Joule]}
\end{equation}
where the parameters $\omega^{\mathrm{BS}}, \omega^{\mathrm{UE}} \in [0,1]$ are the efficiencies of the power amplifiers at the BS and UE, respectively.\footnote{The efficiency of a specific amplifier depends on the transmit power, but to facilitate analysis we assume that the amplifier is optimized jointly with the transmit power to give a specific efficiency at the particular power level. The efficiency also depends on the PAPR and acceptable distortion noise, which are two properties that we also keep fixed when optimizing the EE.}
The average power (in Joule/channel use) can then be separated as
\begin{equation}
\begin{split}
\frac{E_{\mathrm{amp}}}{T_{\mathrm{coher}}} &= \underbrace{\alpha_{\mathrm{DL}} \left( \frac{T^{\mathrm{DL}}_{\mathrm{pilot}}}{T_{\mathrm{coher}}}  \frac{ p^{\mathrm{BS}}}{\omega^{\mathrm{BS}}} + \frac{T^{\mathrm{UL}}_{\mathrm{pilot}}}{T_{\mathrm{coher}}}  \frac{ p^{\mathrm{UE}}}{\omega^{\mathrm{UE}}} \right) +  \frac{ T^{\mathrm{DL}}_{\mathrm{data}}}{ T_{\mathrm{coher}} } \frac{ p^{\mathrm{BS}}}{\omega^{\mathrm{BS}}}}_{\textrm{Downlink power}} \\
& + \underbrace{\alpha_{\mathrm{UL}} \left( \frac{T^{\mathrm{DL}}_{\mathrm{pilot}}}{T_{\mathrm{coher}}}  \frac{ p^{\mathrm{BS}}}{\omega^{\mathrm{BS}}} + \frac{T^{\mathrm{UL}}_{\mathrm{pilot}}}{T_{\mathrm{coher}}}  \frac{ p^{\mathrm{UE}}}{\omega^{\mathrm{UE}}} \right) +  \frac{ T^{\mathrm{UL}}_{\mathrm{data}}}{ T_{\mathrm{coher}} } \frac{ p^{\mathrm{UE}}}{\omega^{\mathrm{UE}}}}_{\textrm{Uplink power}}
\end{split}
\end{equation}
where the ratios of DL and UL transmission are, respectively,
\begin{align}
\alpha_{\mathrm{DL}} &= \frac{T^{\mathrm{DL}}_{\mathrm{data}}}{T^{\mathrm{DL}}_{\mathrm{data}}+ T^{\mathrm{UL}}_{\mathrm{data}}} \\
 \alpha_{\mathrm{UL}} &= \frac{T^{\mathrm{UL}}_{\mathrm{data}}}{T^{\mathrm{DL}}_{\mathrm{data}}+ T^{\mathrm{UL}}_{\mathrm{data}}}.
\end{align}
In addition to the power consumed by the amplifiers, there is generally a baseband circuit power consumption which we model as $N \rho + \zeta$ \cite{Auer2011a,Xu2011b,Ng2012a,Yang2013a,Bjornson2013e}. The parameter $\rho  \geq 0$ [Joule/channel use] describes the circuit power that scales with the number antennas; for example, hardware components that are needed at each antenna branch (e.g., converters, mixers, and filters) and computational complexity that is proportional to $N$ (e.g., channel estimation and computing MRT/MRC). In contrast, the parameter $\zeta >0$ [Joule/channel use] is a static circuit power term that is independent of $N$ (but might scale with the number UEs); for example, it models baseband processing at the BS and circuit power at the UE.\footnote{This term can also model the overhead power consumption of the network as a whole, which enables comparison of network architectures with different BS density, amounts of backhaul signaling, etc.}

Based on the power consumption model described above, and inspired by the seminal work in \cite{Verdu1990a}, we define the overall energy efficiency (in bit/Joule) as follows.

\begin{definition} \label{def:EE}
The downlink energy efficiency is
\begin{equation} \label{eq:energy-efficiency}
{\tt EE}^{\mathrm{DL}} = \frac{{\tt C}^{\mathrm{DL}} }{  \alpha_{\mathrm{DL}} \! \left( \frac{T^{\mathrm{DL}}_{\mathrm{pilot}}}{T_{\mathrm{coher}}}  \frac{ p^{\mathrm{BS}}}{\omega^{\mathrm{BS}}} \!+\! \frac{T^{\mathrm{UL}}_{\mathrm{pilot}}}{T_{\mathrm{coher}}}  \frac{ p^{\mathrm{UE}}}{\omega^{\mathrm{UE}}} + N \rho + \zeta \right) \!+  \! \frac{ T^{\mathrm{DL}}_{\mathrm{data}}}{ T_{\mathrm{coher}} } \frac{ p^{\mathrm{BS}}}{\omega^{\mathrm{BS}}}  }
\end{equation}
and the uplink energy efficiency is
\begin{equation} \label{eq:energy-efficiency-uplink}
{\tt EE}^{\mathrm{UL}} = \frac{{\tt C}^{\mathrm{UL}} }{  \alpha_{\mathrm{UL}} \! \left( \frac{T^{\mathrm{DL}}_{\mathrm{pilot}}}{T_{\mathrm{coher}}}  \frac{ p^{\mathrm{BS}}}{\omega^{\mathrm{BS}}} \!+\! \frac{T^{\mathrm{UL}}_{\mathrm{pilot}}}{T_{\mathrm{coher}}}  \frac{ p^{\mathrm{UE}}}{\omega^{\mathrm{UE}}} + N \rho + \zeta  \right) \!+ \! \frac{ T^{\mathrm{UL}}_{\mathrm{data}}}{ T_{\mathrm{coher}} } \frac{ p^{\mathrm{UE}}}{\omega^{\mathrm{UE}}}}.
\end{equation}
The EE of any suboptimal transmission scheme is obtained by replacing the capacities ${\tt C}^{\mathrm{DL}}$ and ${\tt C}^{\mathrm{UL}}$ with the corresponding achievable spectral efficiencies.
\end{definition}

This definition considers a single link, which can be any of the links in a massive MIMO system---the parameters $\zeta$ and $\rho$ should then be interpreted as the energy per channel use \emph{per user}. In the process of maximizing the EE metric in Definition~\ref{def:EE}, we first extend the power scaling laws from \cite{Hoydis2013a,Ngo2013a,Pitarokoilis2014a} to our general system model with non-ideal hardware.

\begin{theorem} \label{theorem:energy-efficiency}
Suppose the downlink transmit power $p^{\mathrm{BS}}$ and uplink pilot power $p^{\mathrm{UE}}$ are reduced with $N$ proportionally to $1/N^{t_{\mathrm{BS}}}$ and $1/N^{t_{\mathrm{UE}}}$, respectively. If $t_{\mathrm{BS}}+t_{\mathrm{UE}}<1$, $t_{\mathrm{BS}} \geq 0, 0 < t_{\mathrm{UE}} < \frac{1}{2}$, and $\mathbb{E}\{I_{\mathcal{H}}^{\mathrm{UE}}\} = \mathcal{O}(1)$, we have
\begin{equation} \label{eq:lower-bound-capacity-energy-eff}
\lim_{N \rightarrow \infty} {\tt C}^{\mathrm{DL}} \geq \frac{  T^{\mathrm{DL}}_{\mathrm{data}}   }{ T_{\mathrm{coher}} } \log_2 \left(1 + \frac{1 }{\kappa_r^{\mathrm{UE}} +\kappa_t^{\mathrm{UE}} + \kappa_r^{\mathrm{UE}} \kappa_t^{\mathrm{UE}} } \right).
\end{equation}

Similarly, suppose the uplink transmit/pilot power $p^{\mathrm{UE}}$ is reduced with $N$ proportionally to $1/N^{t_{\mathrm{UE}}}$. If $0 < t_{\mathrm{UE}} < \frac{1}{2}$ and $\mathbb{E}\{\| \vect{Q}_{\mathcal{H}}\|_2 \} = \mathcal{O}(1)$, we have
\begin{equation} \label{eq:lower-bound-capacity-energy-eff-uplink}
\lim_{N \rightarrow \infty} {\tt C}^{\mathrm{UL}} \geq  \frac{  T^{\mathrm{UL}}_{\mathrm{data}}   }{ T_{\mathrm{coher}} } \log_2 \left(1 + \frac{1 }{ 2 \kappa_t^{\mathrm{UE}}  + (\kappa_t^{\mathrm{UE}})^2  } \right).
\end{equation}
\end{theorem}
\begin{IEEEproof}
The proof is given in Appendix \ref{proof:theorem:energy-efficiency}.
\end{IEEEproof}

Theorem \ref{theorem:energy-efficiency} shows that one can reduce the downlink and uplink transmit powers as $N$ grows large (e.g., roughly proportionally to $1/\sqrt{N}$) and converge to non-zero spectral efficiencies. The asymptotic DL capacity is lower bounded by \eqref{eq:lower-bound-capacity-energy-eff} and the UL capacity by \eqref{eq:lower-bound-capacity-energy-eff-uplink}.
As expected from Section \ref{sec:downlink-uplink-capacity}, these lower bounds only depend on the levels of impairments at the UE. The conditions $\mathbb{E}\{ I_{\mathcal{H}}^{\mathrm{UE}} \} = \mathcal{O}(1)$ and $\mathbb{E}\{\| \vect{Q}_{\mathcal{H}}\|_2 \} = \mathcal{O}(1)$ in Theorem \ref{theorem:energy-efficiency} are stronger than the ones in Corollaries \ref{cor:lower_bound} and \ref{cor:lower_bound-uplink}, thus the interfering transmissions might have to reduce their transmit powers as well if their impact should vanish asymptotically. We note that the lower bounds in Theorem \ref{theorem:energy-efficiency} are achieved by using the LMMSE estimator in Theorem \ref{theorem:LMMSE-estimator} for channel estimation and simple linear processing at the BS (approximate MRT in the DL and MRC in the UL).

Based on Theorem \ref{theorem:energy-efficiency} and the upper capacity bounds in Section \ref{sec:downlink-uplink-capacity}, the following corollary describes how to maximize the EE.

\begin{corollary} \label{cor:energy-efficiency}
Suppose we want to maximize the EE metrics with respect to the transmit powers and the number of antennas.
Let $\mathbb{E}\{ I_{\mathcal{H}}^{\mathrm{UE}} \} = \mathcal{O}(1)$ and $\mathbb{E}\{\| \vect{Q}_{\mathcal{H}}\|_2 \} = \mathcal{O}(1)$.
If $\rho=0$, the maximal EEs are bounded as
\begin{equation} \label{eq:bounds-on-EE-DL}
\frac{ \log_2 \! \left(1 \! +\!  \frac{1 }{\kappa_r^{\mathrm{UE}} +\kappa_t^{\mathrm{UE}} + \kappa_r^{\mathrm{UE}} \kappa_t^{\mathrm{UE}} } \right) \! }{ \frac{ T_{\mathrm{coher}} }{  T^{\mathrm{DL}}_{\mathrm{data}}   } \alpha_{\mathrm{DL}}  \zeta} \! \leq  \! \!\! \max_{\begin{subarray}{c}
  p^{\mathrm{BS}},p^{\mathrm{UE}} \geq 0 \\
  N \geq 0
  \end{subarray}} \!\!\! {\tt EE}^{\mathrm{DL}} \!\! \leq \!  \frac{ \log_2 \! \left(1 \! +\!  \frac{1 }{\kappa_r^{\mathrm{UE}} } \right) \! }{ \frac{ T_{\mathrm{coher}} }{  T^{\mathrm{DL}}_{\mathrm{data}}} \alpha_{\mathrm{DL}} \zeta}
\end{equation}
\begin{equation} \label{eq:bounds-on-EE-UL}
\frac{ \log_2 \! \left(1 + \frac{1 }{ 2 \kappa_t^{\mathrm{UE}}  + (\kappa_t^{\mathrm{UE}})^2  } \right) }{ \frac{ T_{\mathrm{coher}} }{  T^{\mathrm{UL}}_{\mathrm{data}} } \alpha_{\mathrm{UL}} \zeta} \!  \leq \! \!  \max_{p^{\mathrm{UE}},N \geq 0}  {\tt EE}^{\mathrm{UL}} \! \leq \!  \frac{ \log_2 \! \left(1 + \frac{1 }{ \kappa_t^{\mathrm{UE}} } \right) }{ \frac{ T_{\mathrm{coher}} }{  T^{\mathrm{UL}}_{\mathrm{data}} }\alpha_{\mathrm{UL}} \zeta}
\end{equation}
where the lower bounds are achieved as $N \rightarrow \infty$ using the power scaling law in Theorem \ref{theorem:energy-efficiency}.

If $\rho>0$, the upper bounds in \eqref{eq:bounds-on-EE-DL} and \eqref{eq:bounds-on-EE-UL} are still valid, but the asymptotic EEs are
\begin{equation}
\lim_{N \rightarrow \infty} \max_{p^{\mathrm{BS}},p^{\mathrm{UE}} \geq 0}  {\tt EE}^{\mathrm{DL}} = \lim_{N \rightarrow \infty} \max_{p^{\mathrm{UE}} \geq 0}  {\tt EE}^{\mathrm{UL}} = 0
\end{equation}
and, consequently, the EEs are maximized at some finite $N$.
\end{corollary}
\begin{IEEEproof}
The lower bounds for $\rho=0$ are achieved as described in corollary, while the upper bounds follow from neglecting the transmit power term in the denominator and applying the capacity upper bounds from Corollaries \ref{cor:upper_bound} and \ref{cor:upper_bound-uplink}. In the case of $\rho>0$, we note that the EE is non-zero for $N=1$ for any non-zero transmit power, while the EE goes to zero as $N \rightarrow \infty$ since the denominators of the EE metrics grow to infinity and the numerators are bounded.
\end{IEEEproof}

This corollary reveals that the maximal overall EE is finite, also in massive MIMO systems. If the circuit power consumption does not scale with $N$, such that $\rho=0$, we can achieve an EE very close to the upper bounds in \eqref{eq:bounds-on-EE-DL} and \eqref{eq:bounds-on-EE-UL} by having very many antennas. This changes completely when there is a non-zero circuit power per antenna: $\rho>0$. The maximal EE is then achieved at some finite $N$, which naturally depends on the parameters $\rho$, $\zeta$, $\omega^{\mathrm{BS}}$, and $ \omega^{\mathrm{UE}}$. We illustrate this dependence numerically in the next subsection.

Since $\rho$ has a dominating impact on the maximal EE in massive MIMO systems, one would like to find a way to reduce $\rho$.
Generally speaking, the hardware power consumption depends on the circuit architecture and the hardware resolution \cite{Mezghani2010a,Zhang2012a}; by tolerating larger hardware impairments we can also reduce the power dissipation in the corresponding circuits. Now recall from Section \ref{sec:downlink-uplink-capacity} that the impact of hardware impairments at the BS vanishes as $N \rightarrow \infty$. This fact raises the important question: Can we increase the levels of impairments at the BS as $N$ grows and still obtain non-zero capacities? The answer is given by the following corollary.

\begin{corollary} \label{cor:degrading-hardware-quality}
Suppose the levels of impairments $\kappa_t^{\mathrm{BS}},\kappa_r^{\mathrm{BS}}$ are increased with $N$ proportionally to $N^{\tau_t}$ and $N^{\tau_r}$, respectively.
The lower capacity bounds in Corollaries \ref{cor:lower_bound} and \ref{cor:lower_bound-uplink} (for $n\leq \frac{1}{2}$)
converge to non-zero quantities as $N \rightarrow \infty$ if $\tau_r < \frac{1}{2}$ in the UL and $\tau_t+\tau_r < 1$ and $\tau_r < \frac{1}{2}$ in the DL.
\end{corollary}
\begin{IEEEproof}
The proof is given in Appendix \ref{proof:cor:degrading-hardware-quality}.
\end{IEEEproof}

This corollary shows that we can indeed increase the levels of impairments, $\kappa_t^{\mathrm{BS}}$ and $\kappa_r^{\mathrm{BS}}$, at the BS roughly proportionally to $\sqrt{N}$ and still have a non-zero asymptotic capacity. The numerical results in the next subsection shows that only minor degradations of the lower capacity bounds appear when the impairment scaling law in Corollary \ref{cor:degrading-hardware-quality} is followed.

Recall from \eqref{eq:EVM-definition} that the conventional EVM measure of transceiver quality equals the square root of the $\kappa$-parameters, thus Corollary \ref{cor:degrading-hardware-quality} shows that the EVMs can be increased proportionally to $N^{1/4}$. A high-quality BS antenna element with an EVM of $0.03$ can thus be replaced by 256 low-quality antenna elements with an EVM of $0.12$, while the loss in capacity is negligible. This is a very encouraging result, since it indicates that massive MIMO can be deployed with BS hardware components that are inexpensive, have lower quality and thus low power consumption than conventional ones (i.e., $\rho$ is smaller). If the hardware components are treated as optimization variables, the maximal EE is achieved by jointly reducing the transmit power and the circuit power consumption with $N$. This optimization is, however, strongly dependent on the practical hardware setup (e.g., how an increased EVM maps to a smaller circuit power dissipation) and is outside the scope of this paper.  Finally, we note that the ability to degrade the hardware quality with $N$ comes in addition to all other benefits of massive MIMO, such as the array gain and the decorrelation of user channels; see the multi-cell results in Section \ref{sec:multi-cell-scenario}.

\subsection{Numerical Illustrations}

Next, we illustrate how the overall EE depends on the number of antennas, transmit powers, and circuit power parameters $\zeta$ and $\rho$. Based on the power consumption numbers reported in \cite[Table 7]{EARTH_D23}, we consider two setups: $\zeta + \rho = 2 \,\mu \mathrm{J}/\textrm{channel use}$ and $\zeta + \rho = 0.02 \,\mu \mathrm{J}/\textrm{channel use}$.\footnote{As a reference, these numbers correspond to 18 W and 0.18 W, respectively, for a system with an effective bandwidth of 9 MHz, since then there are $9 \cdot 10^6$ channel uses per second.} These represent the total circuit power consumptions in a system with $N=1$ antenna. Since the total circuit power for arbitrary $N$ is $\zeta + N \rho$, we consider three different splittings between $\rho$ and $\zeta$: $\frac{\rho}{\zeta + \rho} \in \{0, \, 0.01, \, 0.1\}$. From an EE optimization perspective, any scaling of the power amplifier efficiencies is equivalent to an inverse scaling of $\zeta$ and $\rho$. Hence, we can set the efficiencies to $\omega^{\mathrm{BS}}= \omega^{\mathrm{UE}} = 0.3$, which corresponds to $30\%$, without limiting the generality of our numerical results.

The transmit powers in the UL and DL are assumed to be equal and upper bounded by $p_{\max} = 0.0222 \, \mu \mathrm{J}/\textrm{channel use}$.\footnote{As a reference, this number corresponds to 200 mW, or 23 dBm, for a system with an effective bandwidth of 9 MHz.} We consider a scenario without interference (i.e., $\vect{Q}_{\mathcal{H}} = \vect{S} = \vect{0}$ and $I_{\mathcal{H}}^{\mathrm{UE}}=0$), the channel covariance matrix $\vect{R}$ is generated by the exponential correlation model in \eqref{eq:exponential-model} with correlation coefficient $r=0.7$. We let $\frac{N \sigma_{\mathrm{UE}}^2}{\tr(\vect{R})} = \frac{N \sigma_{\mathrm{BS}}^2}{\tr(\vect{R})} =  \frac{p_{\max}}{100} \, \mu \mathrm{J}/\textrm{channel use}$ which gives an SNR of 20 dB if the maximal transmit power $p_{\max}$ is used; { recall from Section \ref{subsec:capacity-numerical-illustrations} that this SNR is desirable if one should operate close to the asymptotic capacity limits.} To make the EE of the DL and UL equal, we consider a symmetric scenario with $\alpha_{\mathrm{DL}} = \alpha_{\mathrm{UL}} = 0.5$ and $\frac{  T^{\mathrm{DL}}_{\mathrm{pilot}} }{ T_{\mathrm{coher}} } = \frac{  T^{\mathrm{UL}}_{\mathrm{pilot}} }{ T_{\mathrm{coher}} } = 0.05$.

\begin{figure}
\begin{center}
\includegraphics[width=\columnwidth]{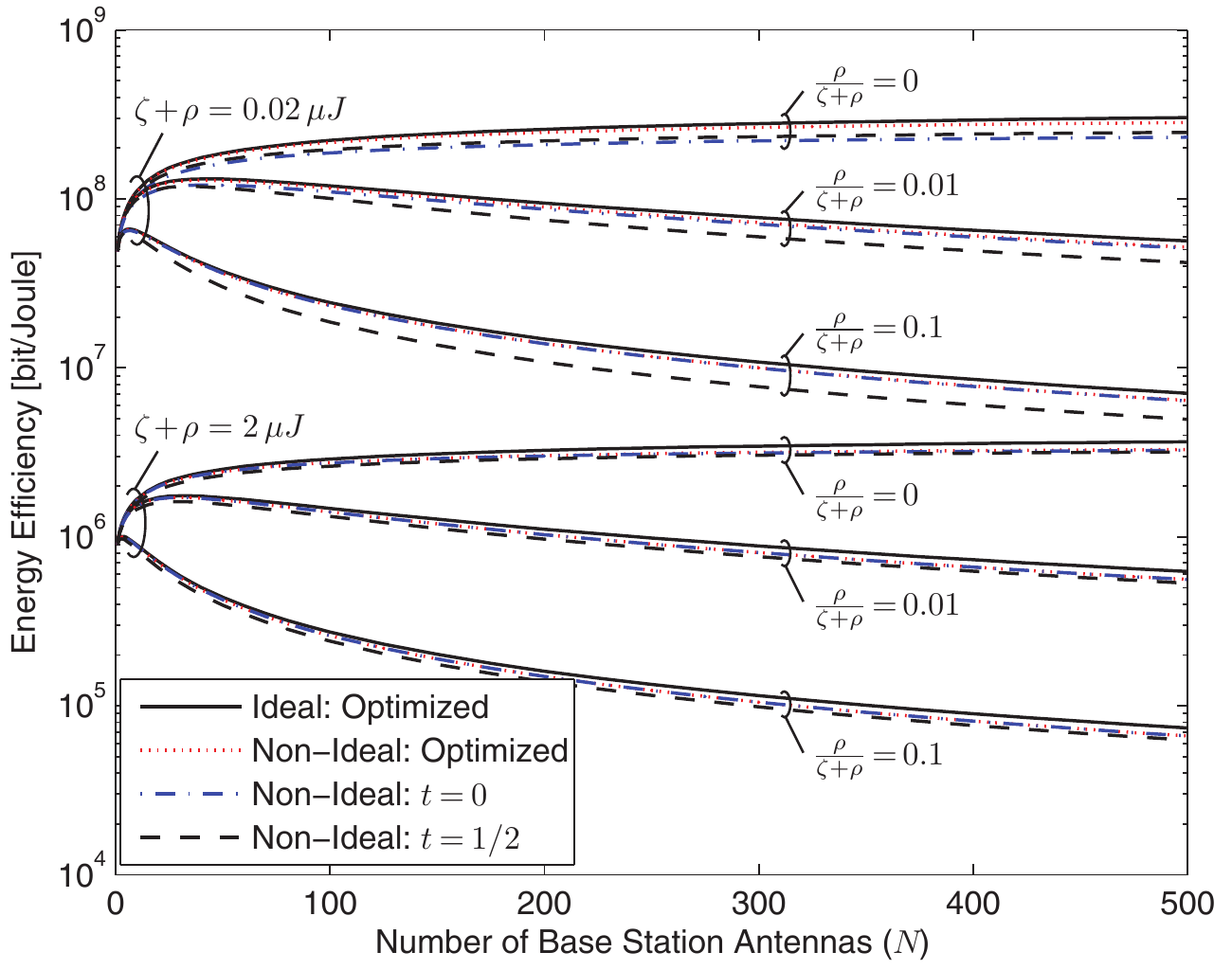} \vskip-2mm
\caption{Achievable energy efficiency with ideal and non-ideal hardware for fixed transmit power ($t=0$), transmit power that decreases as $1/N^t$ for $t= \frac{1}{2}$, and the transmit power that maximizes the EE. The EE is computed using the lower bounds in Theorem \ref{theorem:lower-bound-on-capacities} and are valid for both DL and UL transmissions.}\label{figure_EE_new1}
\end{center} \vskip-3mm
\end{figure}

Fig.~\ref{figure_EE_new1} shows the achievable DL/UL energy efficiencies using the lower capacity bounds in Theorem \ref{theorem:lower-bound-on-capacities}. The levels of impairments are set to $\kappa_t^{\mathrm{BS}} =\kappa_r^{\mathrm{BS}} = \kappa_t^{\mathrm{UE}} =\kappa_r^{\mathrm{UE}} = 0.05^2$; see Remark \ref{remark:distortion-noise} for the interpretation of these parameter values. The transmit powers are either optimized numerically for maximal EE at each $N$, fixed at the value that is optimal for $N=1$, or reduced from this value according to the power scaling law in Theorem \ref{theorem:energy-efficiency} with $t=\frac{1}{2}$. Fig.~\ref{figure_EE_new1} shows that the EE is almost the same in all three cases, but varies a lot with the circuit power parameters $\zeta$ and $\rho$. If $\rho=0$, the EE increases monotonically with $N$ and eventually converge according to \eqref{eq:bounds-on-EE-DL} and \eqref{eq:bounds-on-EE-UL} in Corollary \ref{cor:energy-efficiency}. On the contrary, the EE has a unique maximum when $\rho>0$ and then decreases towards zero. The maximum is in the range of $5 \leq N \leq 50$ in the figure, but the exact position depends on the circuit power parameters. If $\frac{\rho}{\zeta + \rho}$ is sufficiently small we can use a larger $N$ without losing much in EE. Hence, it is important to make the power that scales with $N$ (e.g., the number of extra hardware components and the computational complexity) as low as possible if massive MIMO systems should excel in terms of energy efficiency. The EE with ideal hardware is also shown in Fig.~\ref{figure_EE_new1}, which reveals that the difference in EE between ideal and non-ideal hardware is small. This is because the performance loss from hardware impairments is relatively small at reasonable number of antennas.

\begin{figure}
\begin{center}
\includegraphics[width=\columnwidth]{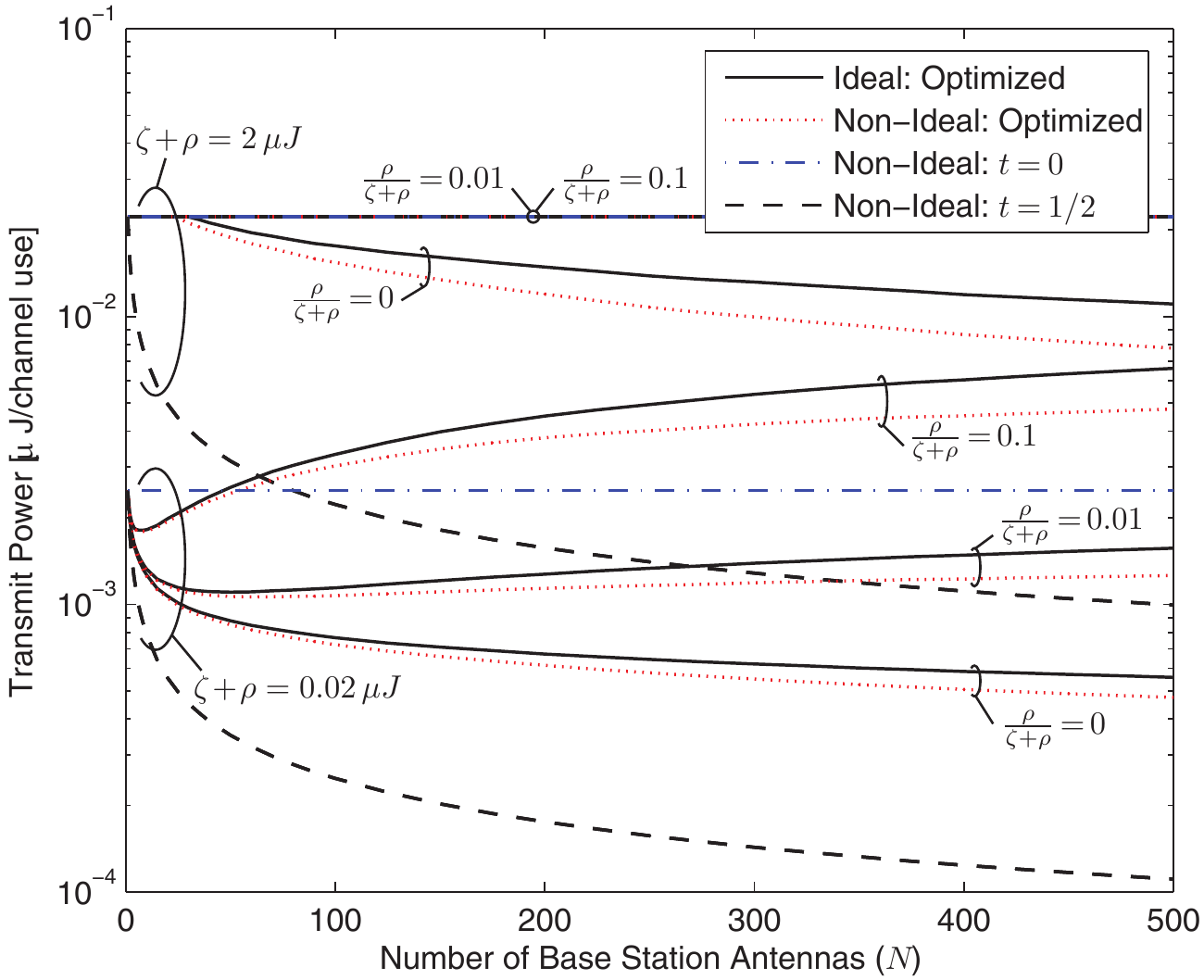} \vskip-2mm
\caption{The transmit powers that correspond to the curves in Fig.~\ref{figure_EE_new1}.}\label{figure_EE_new_power1}
\end{center} \vskip-3mm
\end{figure}

In order to compare the three different power allocations, we show the corresponding transmit powers in Fig.~\ref{figure_EE_new_power1}. Recall that the power is either fixed, reduced with $N$ according to the scaling law in Theorem \ref{theorem:energy-efficiency}, or optimized for maximal EE at each $N$. Despite the very similar EEs in Fig.~\ref{figure_EE_new1}, these three power allocations behave very differently. If $\rho=0$, the optimal transmit power decreases with $N$ but at a clearly slower pace than $1/\sqrt{N}$ (which is the fastest power scaling that gives a non-zero asymptotic rate according to Theorem \ref{theorem:energy-efficiency}). However, the optimal transmit for $\rho>0$ only decreases until the maximal EE is achieved (which is in the range of $5 \leq N \leq 50$) and then \emph{increases} with $N$. This makes much sense, because when the circuit power increases we can also afford using more transmit power to get a higher spectral efficiency. In the case when the circuit power is large (i.e., $\zeta + \rho = 2 \, \mu J$), we see that it is often optimal to use full transmit power, as represented by the upper straight line. To summarize, the transmit power in massive MIMO systems can be decreased monotonically with $N$, but this is generally not the way to maximize the EE since we have $\rho>0$ in most practical systems. The loss in EE by decreasing the power appears to be small, but the loss in spectral efficiency is naturally larger due to the definition of the EE. If we want a simple design rule, it is better to keep the total power fixed for all $N$ than to decrease it with $N$.

\begin{figure}
\begin{center}
\includegraphics[width=\columnwidth]{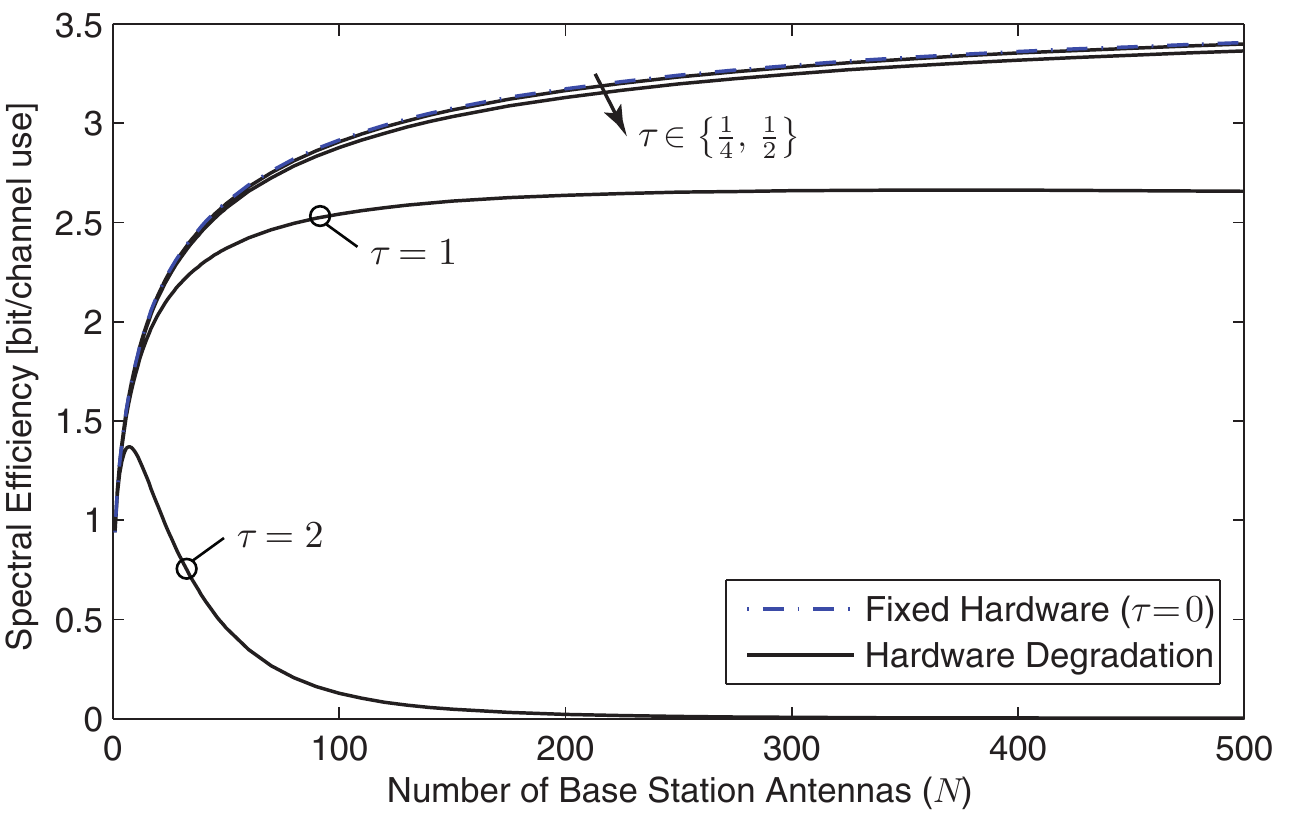} \vskip-2mm
\caption{Lower bounds on the capacity when the levels of impairments at the BS are increased with $N$ as $N^{\tau}$ for $\tau \in \{ 0, \frac{1}{4}, \frac{1}{2}, 1, 2\}$. The results are valid for both DL and UL transmissions.}\label{figure_hardwaredegradation}
\end{center} \vskip-3mm
\end{figure}

Finally, the ability to increase the levels of impairments at the BS with $N$ is illustrated in Fig.~\ref{figure_hardwaredegradation}.
We consider the same symmetric scenario as in the previous two figures, but the average SNR is set to 20 dB in the DL and UL.
We have $\kappa_t^{\mathrm{UE}} =\kappa_r^{\mathrm{UE}} = 0.05^2$ at the UE, while the levels of impairments at the BS are scaled as $\kappa_t^{\mathrm{BS}} =\kappa_r^{\mathrm{BS}} = 0.05^2 N^{\tau}$ for different $\tau$-values: $\tau \in \{ 0, \frac{1}{4}, \frac{1}{2}, 1, 2\}$. The lower capacity bounds are shown in Fig.~\ref{figure_hardwaredegradation} as a function of $N$. The simulation confirms that the performance degradation is small when the impairment scaling law in Corollary \ref{cor:degrading-hardware-quality} is followed (e.g., for $\tau=\frac{1}{4}$ and $\tau=\frac{1}{2}$). A larger performance loss is observed for $\tau=1$ and the curve begins to bend downwards at $N \approx 350$. In the extreme case of $\tau=2$, the lower bound goes quickly to zero. While these asymptotic results are based on the lower capacity bounds, we note that whenever $\tau>1$ the upper capacity bounds in Corollaries \ref{cor:upper_bound} and \ref{cor:upper_bound-uplink} converge to zero as well.

\section{Extensions to Multi-Cell Scenarios}
\label{sec:multi-cell-scenario}

The previous sections focused on a single link of a massive MIMO system, which experiences interference from other concurrent transmissions. Recall that $I_{\mathcal{H}}^{\mathrm{UE}} \in \mathbb{C}$ and $\vect{Q}_{\mathcal{H}} \in \mathbb{C}^{N \times N}$ are the conditional covariances of these transmissions in the DL and UL, respectively, for a given set of channel realizations $\mathcal{H}$. The asymptotic capacity analysis in Section \ref{sec:downlink-uplink-capacity}, particularly the lower capacity bounds in Corollaries \ref{cor:lower_bound} and \ref{cor:lower_bound-uplink}, are based on the assumptions that $\mathbb{E}\{ I_{\mathcal{H}}^{\mathrm{UE}}\} \leq \mathcal{O}(N^n)$ and $\mathbb{E}\{ \| \vect{Q}_{\mathcal{H}}\|_2 \} \leq \mathcal{O}(N^n)$ for some $n<1$. However, the interference variance can actually increase faster than this, particularly under the so-called pilot contamination which gives rise to terms that { scale linearly with $N$} \cite{Marzetta2010a,Jose2011b,Hoydis2013a,Ngo2013a,Rusek2013a}. This section investigates the impact of inter-user interference on massive MIMO systems with non-ideal hardware. The BS and UE from the previous sections are referred to as the ones \emph{under study}.

\subsection{Inter-User Interference in the Uplink}

To exemplify the impact of inter-user interference, we assume that there is a set $\mathcal{U}$ of co-users that are scheduled for UL transmission in the current coherence period. Each co-user is served by the BS under study or any of the neighboring BSs, thus the total number of co-users $| \mathcal{U}|$ is generally large. The association of UEs to BSs is arbitrary since the association has no impact on the UE under study in the UL. The block-fading channel from UE $l \in \mathcal{U}$ to the BS under study is modeled as $\vect{h}_l \sim \mathcal{CN}(\vect{0},\vect{R}_l)$, where $\vect{R}_l$ has bounded spectral norm and the channel is ergodic and block fading.  { Recall that $\mathcal{H}$ is the set of channel realizations for all channels in the system, thus we have $\vect{h}_l \in \mathcal{H}$ for all $l \in \mathcal{U}$.}
The co-user channels are assumed to be independent, which in practice means that users are selected to have no common scatterers \cite{Gao2011a,Hoydis2012a}---this is a basic criterion of spatial user separability in the scheduler. A more refined scheduling criteria would be the one in \cite{Huh2012a}, where the coverage area is divided into location bins. The users in a bin are roughly equivalent in terms of channel statistics and should not be served simultaneously. Users in different bins have independent channels and sufficiently different spatial properties, thus selecting one user per location bin for parallel transmission is a reasonable scheduling decision.

The UL pilot signaling is limited to $T^{\mathrm{UL}}_{\mathrm{pilot}}$ channel uses in the TDD protocol depicted in Fig.~\ref{figure:tdd_operation}. Since the number of active co-users generally satisfies $| \mathcal{U} | > T^{\mathrm{UL}}_{\mathrm{pilot}}$, each pilot channel use must be allocated to multiple users. We divide $\mathcal{U}$ into two disjoint sets: $\mathcal{U}_{\parallel}$ are the users that transmit in parallel with the pilot of the UE under study, while $\mathcal{U}_{\perp}$ are the remaining users.\footnote{Only one pilot channel use is allocated per UE in this section. For other pilot lengths $B>1$, one can construct up to $B$ parallel pilot signals that are orthogonal in space. This increases the pilot power per UE, but does not increase the total number of orthogonal pilots.} The co-users in the same cell as the UE under study are usually in $\mathcal{U}_{\perp}$, but this is not necessary. The interference vector during UL pilot signaling is
\begin{equation} \label{eq:uplink-interference-pilot}
 \boldsymbol{\nu}_{\mathrm{interf}}^{\mathrm{pilot}} = \sum_{l \in  \mathcal{U}_{\parallel} } \p_l \vect{h}_l
\end{equation}
where $\p_l$ is the signal transmitted by UE $l \in \mathcal{U}_{\parallel}$. These signals can be either deterministic or stochastic, thus some of the interfering transmissions can in principle carry data instead of pilot signals (cf.~Remark 5 in \cite{Ngo2013a}). Assuming $\mathbb{E}\{|\p_l|^2\} = p^{\mathrm{UE}}$, the interference covariance matrix during pilot signaling is
\begin{equation} \label{eq:uplink-interference-covariance-pilot}
\vect{S} = \mathbb{E} \left\{ \boldsymbol{\nu}_{\mathrm{interf}}^{\mathrm{pilot}} ( \boldsymbol{\nu}_{\mathrm{interf}}^{\mathrm{pilot}})^H \right\} = p^{\mathrm{UE}} \sum_{l \in  \mathcal{U}_{\parallel} } \vect{R}_l.
\end{equation}

The LMMSE estimator in Theorem \ref{theorem:LMMSE-estimator} and the corresponding analysis in Section \ref{sec:channel-estimation} holds for any covariance matrix $\vect{S}$, thus the explicit ways of computing $\boldsymbol{\nu}_{\mathrm{interf}}^{\mathrm{pilot}}$ and $\vect{S}$ in \eqref{eq:uplink-interference-pilot}--\eqref{eq:uplink-interference-covariance-pilot} can be plugged in directly. The channel estimate $\hat{\vect{h}}$ is now correlated with the co-user channels $\vect{h}_l$ for $l \in \mathcal{U}_{\parallel}$, which has an important impact on the spectral efficiency. More specifically, the interference vector during UL data transmission becomes
\begin{equation} \label{eq:uplink-interference-data}
\boldsymbol{\nu}_{\mathrm{interf}}^{\mathrm{data}} = \sum_{l \in  \mathcal{U} } \p_l \vect{h}_l
\end{equation}
where $\p_l$ is the independent zero-mean stochastic data signal sent by UE $l \in  \mathcal{U}$ and has power $\mathbb{E}\{|\p_l|^2\} = p^{\mathrm{UE}}$. The conditional interference covariance matrix during data transmission is
\begin{equation} \label{eq:uplink-interference-covariance-data}
\vect{Q}_{\mathcal{H}} = \mathbb{E} \left\{ \boldsymbol{\nu}_{\mathrm{interf}}^{\mathrm{data}}  (\boldsymbol{\nu}_{\mathrm{interf}}^{\mathrm{data}} )^H | \mathcal{H} \right\} = p^{\mathrm{UE}}  \sum_{l \in  \mathcal{U} } \vect{h}_l  \vect{h}_l^H.
\end{equation}
Note that $\vect{Q}_{\mathcal{H}}$ depends on the channel realizations in $\mathcal{H}$ and has \emph{not} bounded spectral norm; in fact, there are $|\mathcal{U}|$ eigenvalues of $\vect{Q}_{\mathcal{H}}$ that grow without bound as $N \rightarrow \infty$. This property affects the lower capacity bound in \eqref{eq:lower-bound-capacity-uplink} of Theorem \ref{theorem:lower-bound-on-capacities} where the conditional interference term now becomes
\begin{equation} \label{eq:uplink-explicit-impact-interference}
\begin{split}
\mathbb{E}\left\{ (\vect{v}^{\mathrm{UL}})^H \vect{Q}_{\mathcal{H}} \vect{v}^{\mathrm{UL}} \, | \tilde{\mathcal{H}}^{\textrm{BS}} \right\} =
p^{\mathrm{UE}} \sum_{l \in  \mathcal{U} } \mathbb{E}\left\{ | \vect{h}_l^H \vect{v}^{\mathrm{UL}} |^2 \, | \tilde{\mathcal{H}}^{\textrm{BS}} \right\}.
\end{split}
\end{equation}
The following theorem shows how interference terms of the type $\mathbb{E}\left\{ | \vect{h}_l^H \vect{v}^{\mathrm{UL}} |^2 \, | \tilde{\mathcal{H}}^{\textrm{BS}} \right\}$ in \eqref{eq:uplink-explicit-impact-interference} behave as $N$ grows large.

\begin{theorem} \label{theorem:asymptotic-equivalence-interference}
Assume that no instantaneous CSI is utilized for decoding (i.e., $\tilde{\mathcal{H}}^{\textrm{BS}} = \tilde{\mathcal{H}}^{\textrm{UE}} = \emptyset$), interference is treated as noise, and the receive combining vector $\vect{v}= \frac{\hat{\vect{h}}}{\|\hat{\vect{h}}\|_2}$. Under the interference model in \eqref{eq:uplink-interference-pilot}--\eqref{eq:uplink-interference-covariance-data}, the terms in \eqref{eq:uplink-explicit-impact-interference} are
\begin{align} \label{eq:det-equiv-inter-user-interference}
\!\! &\mathbb{E}\left\{ | \vect{h}_l^H \vect{v} |^2 \right\} \\ \!\! & =\! \begin{cases}
 \mathbb{E}\left\{ \frac{ p^{\mathrm{UE}}  \left( \tr ( \vect{A} \vect{R}_l ) \right)^2 }{\tr\big(\vect{A} ( |\p + \eta_{t}^{\mathrm{UE}}|^2 \vect{R} + \boldsymbol{\Psi}) \vect{A}^H \big) }   \right\} \!+\! \mathcal{O}(\sqrt{N}), \! & l \in \mathcal{U}_{\parallel}, \! \\
\mathcal{O}(1), & l \in \mathcal{U}_{\perp}, \!
\end{cases} \!\! \notag
\end{align}
where $\eta_{t}^{\mathrm{UE}}$ is stochastic and $\vect{A},\boldsymbol{\Psi}$ are given in Theorem \ref{theorem:asymptotic-equivalence}.

The lower capacity bound in \eqref{eq:capacity-lower-equivalent-uplink} is generalized by replacing the term $\mathcal{O}(\frac{1}{N^{1-n}})$ in the denominator by
\begin{align} \notag
&\mathbb{E}\left\{ \frac{ \tr(\vect{R}-\vect{C}) }{\tr\big(\vect{A} ( |\p + \eta_{t}^{\mathrm{UE}}|^2 \vect{R} + \boldsymbol{\Psi}) \vect{A}^H \big) }   \right\}
 \sum_{l \in  \mathcal{U}_{\parallel} }  \left( \frac{ \tr ( \vect{A} \vect{R}_l ) }{  \tr ( \vect{A} \vect{R} )  } \right)^2 \\ & \quad + \mathcal{O} \left( \frac{1}{N} \right) . \label{eq:contamination-term}
\end{align}
\end{theorem}
\begin{IEEEproof}
The proof is given in Appendix \ref{proof:theorem:asymptotic-equivalence-interference}.
\end{IEEEproof}

This theorem shows that the effective interference from a co-user depends strongly on whether it interfered with the pilot transmission of the UE under study or not.
The interference from co-users in $\mathcal{U}_{\perp}$, which were silent when the UE under study sent its pilot, vanishes asymptotically since the user channels decorrelate with $N$. { This is the classical type of interference and is called \emph{regular} interference in this section.} In contrast, the interference from co-users in $\mathcal{U}_{\parallel}$, which were active during the pilot transmission remains and even scales with $N$. This is the very essence of \emph{pilot contaminated interference} and Theorem \ref{theorem:asymptotic-equivalence-interference} generalizes previous results from \cite{Marzetta2010a,Jose2011b,Hoydis2013a,Ngo2013a,Rusek2013a} (among others) to non-ideal hardware. The explanation to the diverse behavior in Theorem \ref{theorem:asymptotic-equivalence-interference} is that the channel estimate $\hat{\vect{h}}$ used in the receive combining is independent of the co-user channels $\vect{h}_l$ for $l \in \mathcal{U}_{\perp}$, but correlated with $\vect{h}_l$ for $l \in \mathcal{U}_{\parallel}$ since these vectors appeared in the interference term \eqref{eq:uplink-interference-pilot} during pilot transmission. Note that Theorem \ref{theorem:asymptotic-equivalence-interference} was derived using MRC, while minimum mean squared error (MMSE) receive combining is generally a better choice in multi-cell multi-user scenarios since it actively suppresses interference \cite{Hoydis2013a}. Nevertheless, the theorem establishes the baseline behavior: only the pilot contaminated interference may have a substantial impact when $N$ is large (if a judicious receive combining is used). The severity of the pilot contamination depends on how the sets $\mathcal{U}_{\perp}$ and $\mathcal{U}_{\parallel}$ are chosen \cite{Yin2013a}.

\subsection{Inter-User Interference in the Downlink}

The downlink transmission can also suffer from pilot contamination, especially if the numbers of antennas at neighboring BSs also grow linearly with $N$. { The conditional interference variance in the DL } takes a similar form as in \eqref{eq:uplink-interference-data}--\eqref{eq:uplink-explicit-impact-interference}:
\begin{equation}
I_{\mathcal{H}}^{\mathrm{UE}} =  p^{\mathrm{BS}} \sum_{l \in  \mathcal{U} } \mathbb{E}\left\{ | \tilde{\vect{h}}_l^H \vect{v}_l^{\mathrm{DL}} |^2 \, | \tilde{\mathcal{H}}^{\textrm{UE}} \right\}
\end{equation}
where $\vect{v}_l^{\mathrm{DL}}$ is the beamforming vector for DL transmission to UE $l \in \mathcal{U}$ from its (arbitrary) serving BS and $\tilde{\vect{h}}_l$ is the channel from that BS to the UE under study. For brevity, we will not dive into the details since these require assumptions on the decision making at other BSs. The general behavior is however the same: UEs with parallel UL pilots cause non-vanishing interference to each other in the DL, while the impact of all other interfering DL transmissions vanish as $N$ grows large.

\subsection{Numerical Illustrations}
\label{subsec:contamination-simulations}

The impact of inter-user interference and pilot contamination on multi-cell systems with non-ideal hardware is now studied numerically. We consider UL scenarios with spatially uncorrelated channels, define the average SNR as $p^{\mathrm{UE}} \frac{\tr (\vect{R})}{N \sigma_{\mathrm{BS}}^2}$, and let $\frac{  T^{\mathrm{UL}}_{\mathrm{data}}   }{ T_{\mathrm{coher}} } = 0.45$ be the fraction of UL data transmission.

\begin{figure}
\begin{center}
\includegraphics[width=\columnwidth]{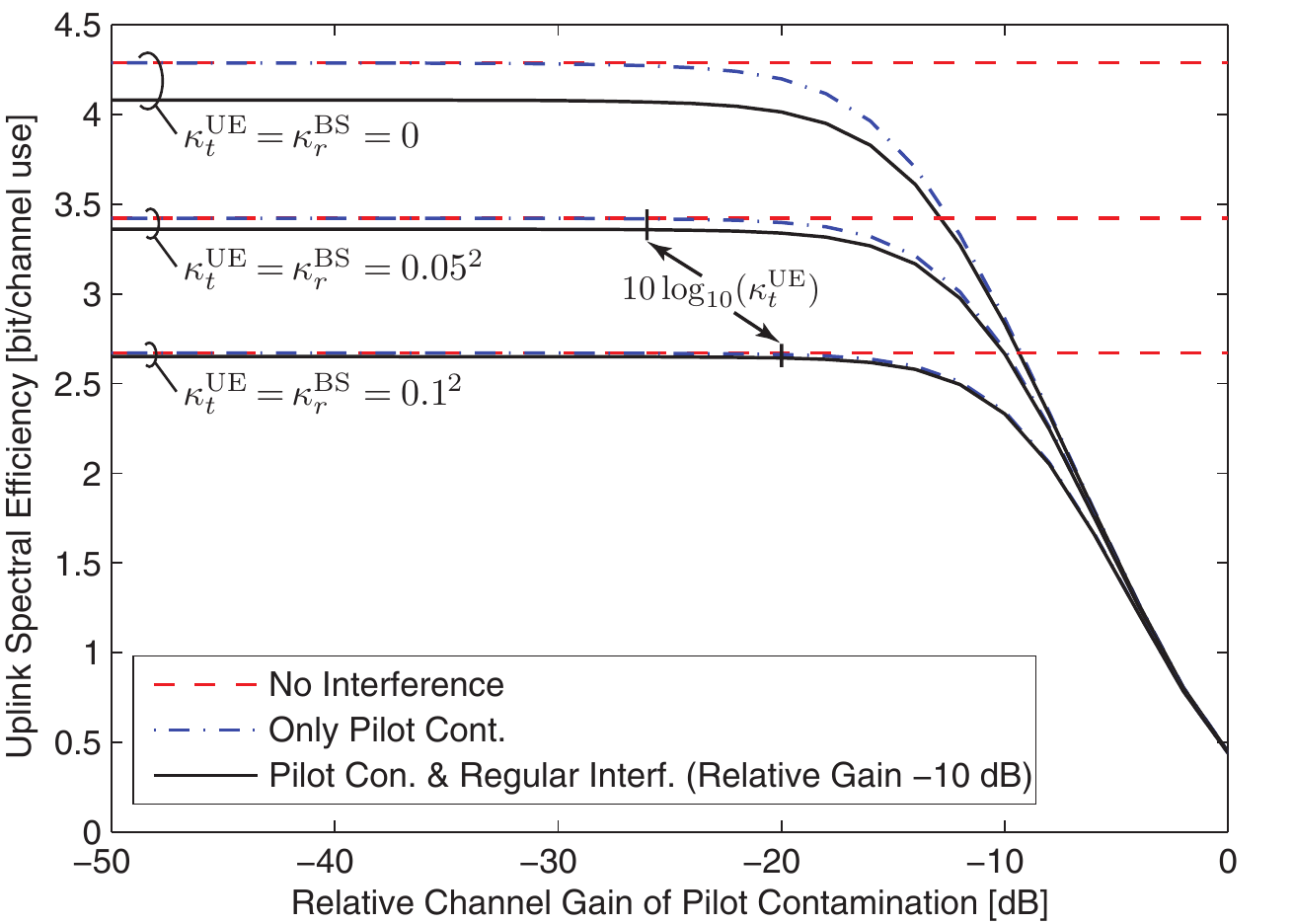} \vskip-1mm
\caption{Lower capacity bounds of a user that experiences pilot contaminated interference of varying strength and possibly regular inter-user interference that is $-10$ dB weaker than the useful channel. The interference drowns in the distortion noise if it is weaker than the level of impairments at the UE.}\label{figure_multicell_gains}
\end{center} \vskip-3mm
\end{figure}

In Fig.~\ref{figure_multicell_gains} we consider the two types of inter-user interference from Theorem \ref{theorem:asymptotic-equivalence-interference}: regular interference from a UE whose pilot is orthogonal to the UE under study and pilot contaminated interference from a UE with an overlapping pilot. { We want to investigate how the achievable per-user spectral efficiency in massive MIMO systems depends on the strength of the pilot contaminated interference, thus we consider a scenario where we operate close to the asymptotic limits: the SNR is 20 dB and the number of antennas is set to $N = 200$ (see Fig.~\ref{figure_capacity}).} We consider three levels of impairments: $\kappa_t^{\mathrm{UE}}=\kappa_r^{\mathrm{BS}} \in \{ 0, 0.05^2, 0.1^2 \}$. The lower capacity bounds are shown without interference, with only pilot contaminated interference, and with both types of interference. The horizontal axis in Fig.~\ref{figure_multicell_gains} shows the performance as a function of the relative channel gain of the pilot contaminated interference (with respect to the useful channel).

{ We make several observations.} Firstly, the ideal hardware case is more sensitive to interference than the non-ideal hardware case. This is particularly evident when it comes to regular inter-user interference, which gives a much larger performance gap in the ideal case.
With non-ideal hardware, the regular interference (from a channel that is only $-10$ dB weaker than the useful channel) has little impact. This is due to the large number of antennas, which decorrelate the user channels. Secondly, the figure shows that pilot contaminated interference has a negligible impact when it arrives over a channel that is much weaker than the useful channel, but there are breaking points where the degradation effect suddenly becomes immense. Interestingly, the breaking points are close to $10 \log_{10}( \kappa_{t}^{\mathrm{UE}})$; that is, how much weaker the distortion noise caused by the UE is compared to the useful signal. This is very intuitive if we compare the size of the distortion term $\kappa_t^{\mathrm{UE}} \mathbb{E}\left\{ | \varphi |^2 \right\}$ in lower capacity bound in \eqref{eq:capacity-lower-equivalent-uplink} with the interference term in \eqref{eq:contamination-term}. { This is formalized as follows.}

\begin{corollary} \label{cor:pilot-contamination}
The pilot contaminated interference is negligible, when $N$ grows large, if
\begin{equation} \label{eq:negligible-contamination}
\kappa_{t}^{\mathrm{UE}} \gg \sum_{l \in  \mathcal{U}_{\parallel} }  \left( \frac{  \tr ( \vect{A} \vect{R}_l ) }{  \tr ( \vect{A} \vect{R} )  } \right)^2.
\end{equation}
\end{corollary}

This corollary shows that pilot contaminated interference drowns in the distortion noise under certain conditions, which are independent of the absolute SNRs but depend on relative SNR differences of the type $\tr ( \vect{A} \vect{R}_l ) /  \tr ( \vect{A} \vect{R} )$. Since the distortion noise typically is $20$--$30$ dB weaker than the useful signal, the same is needed for the pilot contaminated interference to make its impact negligible. This is not a big deal in cellular deployments; the scheduler should simply allocate different pilots within each cell and to cell-edge users of neighboring cells.\footnote{As an example, suppose $\vect{R}=\delta^{-3.7}\vect{I}$ and $\vect{R}_l = \delta_l^{-3.7} \vect{I}$ where $3.7$ is the path loss exponent and $\delta,\delta_l$ are the distances between the BS under study and the two users. The right-hand side of \eqref{eq:negligible-contamination} becomes $\left( \tr ( \vect{A} \vect{R}_l ) /  \tr ( \vect{A} \vect{R} )  \right)^2 = (\delta_l/\delta)^{-7.4}$ which is in the range $-20$ to $-30$ dB if UE $l$ is $1.9$--$2.5$ times further away from the BS than the UE under study. This is the case for most UEs in neighboring cells, but to be sure one can apply a fractional reuse pattern such that adjacent cells use different pilots. All interfering UEs will then be, at least, $2$ times further away from the BS than the UE under study. \label{footnote:pilot-example}}
This can be achieved by the pilot allocation algorithm in \cite{Yin2013a}, but also by simple predefined cell sectorization as illustrated next.

\begin{figure}
\begin{center}
\includegraphics[width=\columnwidth]{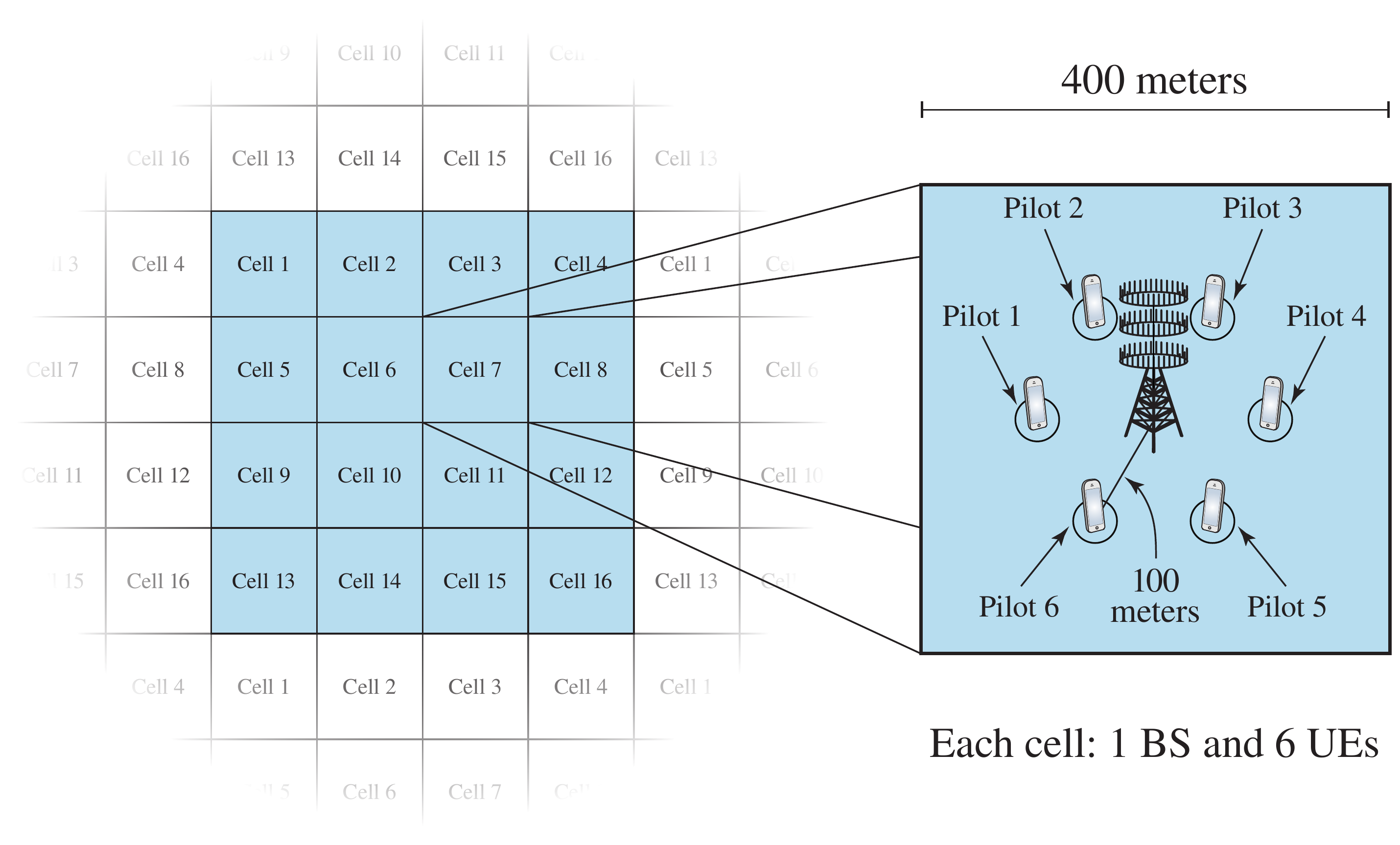} \vskip-1mm
\caption{Illustration of a multi-cell scenario consisting of 16 square cells with wrap-around
to avoid edge effects. Each cell is $400 \, \mathrm{m} \times 400 \, \mathrm{m}$ and contains
of 6 UEs equally spaced on a circle of radius $100 \, \mathrm{m}$.}\label{figure_multi-cell-setup}
\end{center} \vskip-3mm
\end{figure}

Fig.~\ref{figure_multi-cell-setup} shows an illustration of the realistic multi-cell scenario that we use validate Corollary \ref{cor:pilot-contamination}. The setup consists of 16 square cells, each of size $400 \, \mathrm{m} \times 400 \, \mathrm{m}$. To avoid edge effects, we use wrap-around as illustrated in Fig.~\ref{figure_multi-cell-setup}. For simplicity, six UEs are scheduled per cell using a simple angular sectorization technique; the UEs are equally spaced on a circle of radius $100 \, \mathrm{m}$. We assume that orthogonal pilots are allocated to the UEs in each cell, while the same pilots are reused across cells with the same pattern. The channel covariance matrices are identity matrices that are scaled by the channel attenuations, which are based on the 3GPP propagation model in \cite{LTE2010b}: the path loss is $10^{-1.53}/D^{3.76}$ where $D$ is the distance in meters. The transmit powers are $p^{\mathrm{UE}} = 0.0222 \, \mu \mathrm{J}/\textrm{channel use}$ and the noise variance is $\sigma_{\mathrm{BS}}^2 = 10^{-7.9} \, \mu \mathrm{J}/\textrm{channel use}$. { This gives an SNR of 32 dB to the serving BS and 0--13 dB to the surrounding BSs.}

\begin{figure}
\begin{center}
\includegraphics[width=\columnwidth]{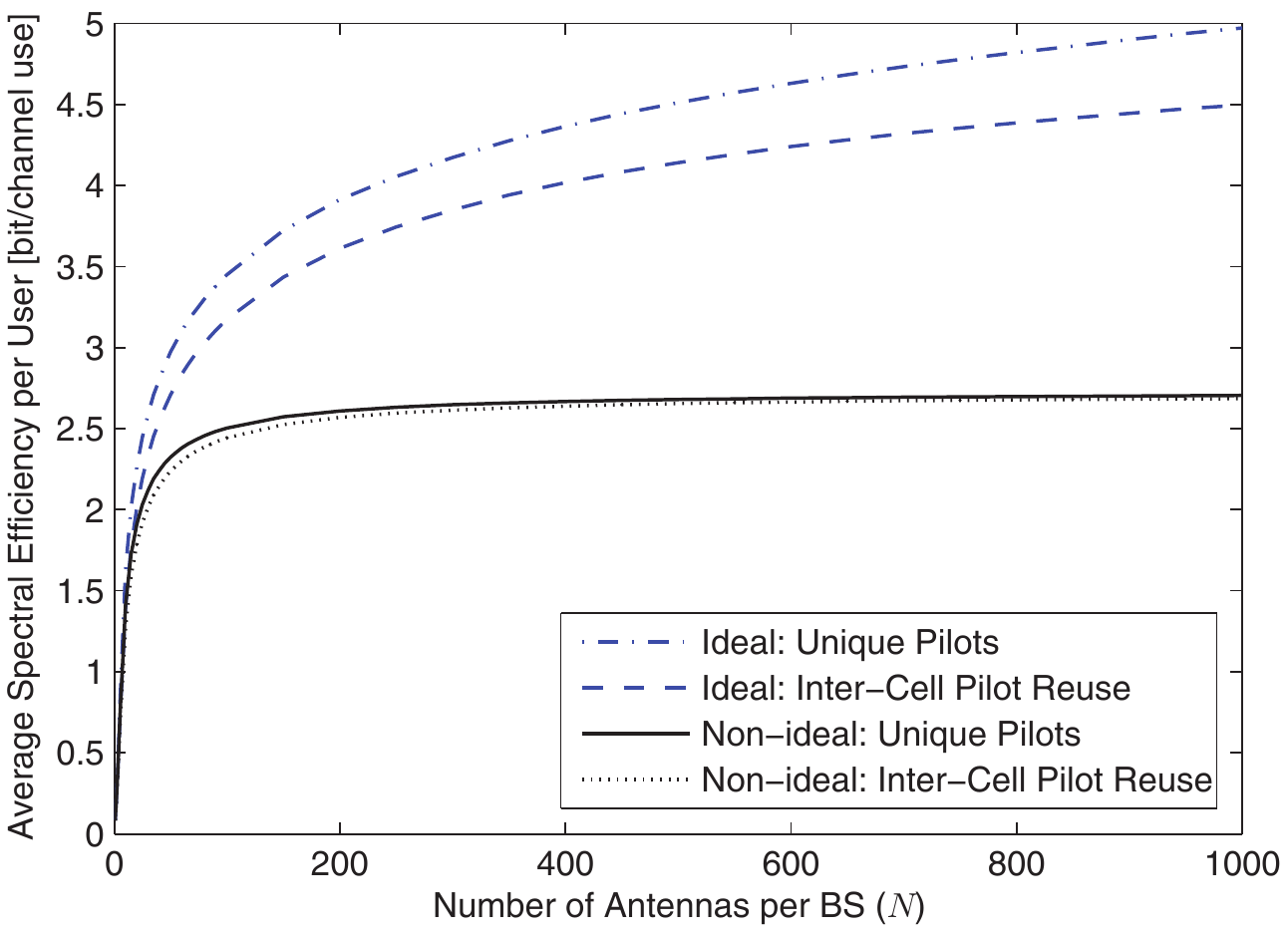} \vskip-2mm
\caption{Achievable UL spectral efficiency for an average user in the multi-cell scenario depicted in Fig.~\ref{figure_multi-cell-setup}. Each UE has either a unique pilot signal or the same pilots are reused in every cell.
Pilot contamination degrades the spectral efficiency under ideal hardware, while the impact on a system with hardware impairments ($\kappa_t^{\mathrm{UE}} =\kappa_r^{\mathrm{BS}} =0.1^2$) is negligible.}\label{figure_multicell_simulation}
\end{center} \vskip-3mm
\end{figure}

{ Fig.~\ref{figure_multicell_simulation} shows the average achievable rates (based on the lower capacity bounds) with MMSE receive combining, which exploits the estimated intra-cell channels to suppress intra-cell interference.} We consider ideal hardware and hardware impairments with $\kappa_t^{\mathrm{UE}}=\kappa_r^{\mathrm{BS}} = 0.1^2$. To illustrate the impact of pilot contamination, we compare the inter-cell pilot reuse pattern described above with the ideal case when all  UEs are allocated unique pilots. We observe that pilot contamination has a substantial impact on the ideal hardware case, and the relative loss will continue to increase with $N$ since only the curve for unique pilots grows towards infinity.
In contrast, there is almost no difference between the unique and reused pilots cases in the system with non-ideal hardware---particularly not when $N$ is large. This implies that pilot contamination might have a negligible impact on massive MIMO systems with hardware impairments, as also shown in Corollary \ref{cor:pilot-contamination}. The explanation is that the distortion noise at the UE is the main limiting factor in the considered scenario, thus regular inter-user interference and the pilot contamination drowns in the distortions.

Informally speaking, the distortion noise acts as a fog that prevents the BS from seeing distant interferers in the pilot transmission phase. The number of orthogonal pilot signals is limited by $ T^{\mathrm{UL}}_{\mathrm{pilot}}$ within the sight of each BS, but can otherwise be reused freely. A simple location-based pilot allocation was sufficient for the multi-cell scenario depicted in Fig.~\ref{figure_multi-cell-setup}, but in general we believe that cell center and cell edge UEs need to be treated differently. If one can afford a fractional reuse pattern, where adjacent cells never use the same pilots, then Corollary \ref{cor:pilot-contamination} will be satisfied in most cases; see Footnote \ref{footnote:pilot-example}.

Finally, we recall from Section \ref{subsec:capacity-numerical-illustrations} that the gain of increasing the number of antennas beyond $N=100$ was small in the single-user case with non-ideal hardware. More antennas can be used in multi-cell scenarios to suppress the regular interference. The convergence to the asymptotic limit is, however, much faster with non-ideal hardware, because it is sufficient to suppress the regular interference to a level below the distortion noise.

\section{Refinements of the System Model and the Possible Implications}
\label{sec:advanced-impairment-models}

Using the system model defined in Section \ref{sec:system-model}, we have shown how the additive distortion noises from hardware impairments limit the estimation accuracy and channel capacities. The practical relevance of the system model has been verified experimentally in \cite{Studer2010a,Wenk2010a,Zetterberg2011a}. It can also be motivated theoretically when the impairment characteristics are static within each coherence period (e.g., due to the use of strong compensation algorithms). In this case, one can apply the Bussgang theorem which shows that any nonlinear distortion function of a Gaussian signal can be reduced to an affine function where the signal is multiplied with an effective channel and corrupted by uncorrelated Gaussian noise \cite{Zhang2012a}. This results in additive distortion noise similar to the one in our system model, but not identical. For analytic tractability, we assumed in the system model of Section \ref{sec:system-model} that the distortion noises are \emph{independent} of the data signals (not only uncorrelated) and Gaussian distributed (even if the data signals are not); the same assumptions were made in \cite{Schenk2008a,Goransson2008a,Studer2010a,Wenk2010a,Zetterberg2011a,Bjornson2012b}. If one would consider an alternative model where these two assumptions are not made, then the lower capacity bounds in this paper will still hold (because the mutual information is always reduced by adding the two assumptions \cite{Hassibi2003a}). The upper bounds in Section \ref{sec:downlink-uplink-capacity} would not hold without the two assumptions, and new upper bounds can only be derived if we impose alternative assumptions on the exact dependence between signal and distortion. In other words, the model in Section \ref{sec:system-model} is a tractable canonical approximation of communication systems with hardware imperfections, but it is not a perfect model of reality.

In general, the time-varying nature of hardware impairments cannot be completely mitigated, which also give rise to multiplicative distortions that vary within each coherence period. Furthermore, the covariance matrices of the additive distortion noises, given in \eqref{eq:distortion-statistics-DL-BS}, \eqref{eq:distortion-statistics-DL-UE}, \eqref{eq:distortion-statistics-UL-UE}, and \eqref{eq:distortion-statistics-UL-BS}, can be refined in several ways. This section outlines some possible refinements of the system model in Section \ref{sec:system-model} and how each one is expected to affect the main results---the exact analysis is not straightforward and is left for future work. Most of these model refinements will further degrade the performance, thus the upper capacity bounds in Theorem \ref{theorem:upperbounds-capacities} is typically valid, while the lower capacity bounds need to be reduced.

\subsection{Power Loss}

It is difficult to model the total emitted power under non-ideal hardware, because some distortions are created independently in the hardware, other distortions take their power from the useful signals, and some impairment sources (e.g., non-linearities) can even reduce the emitted power. In this paper, we have implicitly assumed that the compensation algorithms scale the total emitted power such that it equals $p^{\mathrm{BS}} (1+\kappa_t^{\mathrm{BS}})$ in the DL and $p^{\mathrm{UE}} (1+\kappa_t^{\mathrm{UE}})$ in the UL. This simplification creates a small bias when comparing systems with different levels of impairments, but the simulations in \cite[Section 4.3]{Bjornson2013d} showed that this has a negligible impact on the spectral efficiencies. Nevertheless, it is important to note that although the distortion noise caused by the BS vanishes as $N \rightarrow \infty$, there remains a \emph{power loss} of $\frac{\kappa_t^{\mathrm{BS}}}{1+\kappa_t^{\mathrm{BS}}}$ that should be taken into account when designing massive MIMO systems.

\subsection{High-Power Scalings}
\label{subsec:power-increases-EVM}

The levels of impairments in the transmitter hardware, $\kappa_t^{\mathrm{BS}}$ and $\kappa_t^{\mathrm{UE}}$, were taken as constants in Sections \ref{sec:system-model}--\ref{sec:downlink-uplink-capacity}. This is reasonable when operating within the dynamic/linear range of the respective power amplifiers. Outside these ranges, the proportionality coefficients increase rapidly with the transmit powers $p^{\mathrm{BS}}$ and $p^{\mathrm{UE}}$ due  non-linearities. This behavior was accurately modeled by polynomials in \cite{Bjornson2012b} and \cite{Studer2011a}; for example, $\kappa_t^{\mathrm{BS}}$ could have two terms: a constant term describing the low-power EVM and a term $ (p^{\mathrm{BS}} / c)^q $, for some exponent $q$, describing the severity/order of the dominating non-linearity and a constant $c>0$ that marks the end of the dynamic range \cite{Bjornson2012b}. Note that the distortion noise added by the low-noise amplifiers in radio receivers will typically not become worse with the received power, thus it is reasonable to let $\kappa_r^{\mathrm{BS}}$ and $\kappa_r^{\mathrm{UE}}$ be constants.

The consequence of having proportionality coefficients that scale with the transmit powers is that the distortion noise power increases faster than the signal power. Hence, the capacity and estimation accuracy are no longer monotonically increasing in $p^{\mathrm{BS}}$ and $p^{\mathrm{UE}}$ when using non-ideal hardware---these metrics are instead maximized at some finite transmit powers \cite{Studer2011a}. The reason that we took $\kappa_t^{\mathrm{BS}}$ and $\kappa_t^{\mathrm{UE}}$ as constants herein is that the high-power regime is not our main focus. Consequently, the high-power limits that we derived are optimistic and might not be achievable in practice---alternatively, they are the result of decreasing the propagation distance instead of increasing the actual emitted power. The results when $N \rightarrow \infty$ are however accurate since the total power (or at least the power per antenna) decreases with $N$; see the  discussion in Section \ref{sec:energy-efficiency}.

\subsection{Alternative Distortion Noise Distributions}

The distortion covariance matrices in \eqref{eq:distortion-statistics-DL-BS} and \eqref{eq:distortion-statistics-UL-BS} are based on the assumption of independent distortion at the different BS antennas. This implies that the distortion noise has a different spatial signature than the useful signal, which is the reason why the detrimental impact of the distortion noise caused by the BS vanishes as $N \rightarrow \infty$. The underlying assumption is that the hardware chains of different antennas are decoupled. Nevertheless, there can exist cross-correlation since the same useful signal is transmitter/received over the array, thus making the hardware react similarly. Such correlation was predicted and characterized in \cite{Moghadam2012a} but is typically small. Thus, we believe that also in practice  the distortion noise and useful signal have different spatial signatures as $N$ grows large.

The distortion noises were assumed to be Gaussian distributed (for any fixed channel realization), but this can also be relaxed. As can be seen in the appendices, the proofs rely on that the cross-moments between the signal and the distortion are weak. The independence can, probably, be replaced with uncorrelation and that the higher-order moments are sufficiently weak, but the corresponding generalized proofs will be rather tedious and the convergence as $N \rightarrow \infty$ might be slower.

\subsection{Multiplicative Distortions}

The additive distortion model in this paper has been verified experimentally for systems that apply compensation algorithms to mitigate the main hardware impairments. It is also an accurate model for uncompensated inter-carrier interference caused by phase noise and I/Q imbalance, amplitude-amplitude nonlinearities in power amplifiers, and quantization errors \cite{Mezghani2007a,Schenk2008a,Holma2011a}.
As described in the beginning of Section \ref{sec:advanced-impairment-models}, hardware impairments also cause channel attenuations and phase shifts that are multiplied with the channel vector $\vect{h}$. If these multiplicative distortions are sufficiently static (after compensation), they can be included in the channel vector $\vect{h}$ by an appropriate scaling of the covariance matrix $\vect{R}$ or by exploiting that the channel distribution is circularly symmetric. However, phase noise is a prime example of an impairment that causes multiplicative distortions that
drift and accumulate within the channel coherence period \cite{Demir2000a,Petrovic2007a,Durisi2013a,Pitarokoilis2014a}. We now take a look closer at this type of distortions, to investigate in which ways it behaves differently from additive distortion noise. The actual channel under phase noise can be described as $\diag( e^{\jmath  \phi_{1,t}}, \ldots, e^{\jmath  \phi_{N,t}}) \vect{h}$, where $\jmath=\sqrt{-1}$ is the imaginary unit and $\{ \phi_{i,t} \}$ is the stochastic process at the $i$th channel element. The phase drift of free-running oscillators is commonly modeled as a Wiener process
\begin{equation}
\phi_{i,t} = \phi_{i,t-1}  + \theta_{i,t}^{\mathrm{BS}} + \theta_{t}^{\mathrm{UE}} \quad \forall i
\end{equation}
where the initial value is $\phi_{i,0}=0$ since $t=0$ denotes the time of the channel estimation. The innovations that occur $t$ channel uses after the channel estimation are $\theta_{i,t}^{\mathrm{BS}} \sim  \mathcal{N}(0,\Delta^{\mathrm{BS}})$ and $\theta_{t}^{\mathrm{UE}} \sim  \mathcal{N}(0,\Delta^{\mathrm{UE}})$  at the BS and UE, respectively. Note that the single-antenna UE's hardware causes identical drifts on all channel elements, while the BS can cause identical or independent drifts depending on the use of a common oscillator (CO) or separate oscillators (SOs) at each antenna element. The phase drifts are temporally white, thus $\phi_{i,t} \sim \mathcal{N}(0, t \Delta^{\mathrm{BS}}+ t \Delta^{\mathrm{UE}})$ which shows that the variance increases with time.

To comprehend the impact of phase noise, we note that the signal part of the received UL signal under MRC, $\vect{v}= \frac{\hat{\vect{h}}}{\|\hat{\vect{h}}\|_2}$, is
\begin{equation} \label{eq:phase-noise-signals}
\begin{split}
\vect{v}^H &\diag( e^{\jmath  \phi_{1,t}}, \ldots, e^{\jmath  \phi_{N,t}})  \vect{h} d \\ &\approx
\underbrace{\vect{v}^H \vect{h} d}_{\textrm{Ideal signal}} + \underbrace{\jmath \vect{v}^H \diag( \phi_{1,t}, \ldots, \phi_{N,t} ) \vect{h} d}_{\textrm{Distortion from phase noise}}
\end{split}
\end{equation}
using the Taylor approximation $e^{\jmath  \phi_{i,t}} \approx 1 + \jmath  \phi_{i,t}$ because the drifts are small\footnote{Since the variance of $\phi_{i,t}$ increases linearly with $t$, the Taylor approximation is only valid for a certain time. The time dependence can however be mitigated by tracking the phase noise within each coherence period \cite{Mehrpouyan2012a}.} \cite{Nasser2010a,Mehrpouyan2012a}. The first term in \eqref{eq:phase-noise-signals} is the same as without phase noise, while the second term characterizes the mismatch from the phase drift. Since $\phi_{i,t}$ has zero mean, the two terms are uncorrelated (irrespective of if $\vect{h}$ and $d$ are deterministic or stochastic). We can therefore obtain a lower bound on the mutual information by treating the uncorrelated second term of \eqref{eq:phase-noise-signals} as independent Gaussian noise \cite[Theorem 1]{Hassibi2003a}.
By taking the average over channel realizations, data signals, and phase drifts, the variance of this distortion is
\begin{equation} \label{eq:phase-noise-interference-variance}
\begin{split}
\mathbb{E} & \left\{ \left| \jmath \vect{v}^H \diag( \phi_{1,t}, \ldots, \phi_{N,t} ) \vect{h} d \right|^2 \right\} \\ &=
\sum_{i_1=1}^N \sum_{i_2=1}^N \mathbb{E} \{ v^*_{i_1} v_{i_2}  h_{i_1} h^*_{i_2} \}
\mathbb{E} \{ \phi_{i_1,t} \phi_{i_2,t}^* \}
\mathbb{E} \{ |d|^2\} \\
&= p^{\mathrm{UE}} \mathbb{E} \{ | \vect{v}^H \vect{h}|^2 \} t \Delta^{\mathrm{UE}} \\
 & \quad  + \begin{cases}
p^{\mathrm{UE}} t \Delta^{\mathrm{BS}} \mathbb{E} \{ | \vect{v}^H \vect{h}|^2 \} , & \textrm{if CO}, \\
p^{\mathrm{UE}} t \Delta^{\mathrm{BS}} \sum_{i=1}^{N} \mathbb{E}\{ |h_i|^2 |v_i|^2\}, & \textrm{if SO},
\end{cases}
\end{split}
\end{equation}
where the first term originates from the UE and the second term is due to the BS having either a CO or SOs at each antenna. Recall from Theorem \ref{theorem:asymptotic-equivalence} that $\mathbb{E} \{ | \vect{v}^H \vect{h}|^2 \} = \mathcal{O}(N)$ and $\mathbb{E}\{ |h_i|^2 |v_i|^2\} = \mathcal{O}(1)$. This means that a BS with a CO causes distortion that scales as $\mathcal{O}(t N)$, while it only scales as $\mathcal{O}(t)$ when having SOs.\footnote{Since the useful signal power $p^{\mathrm{UE}} \mathbb{E} \{ | \vect{v}^H \vect{h}|^2 \}$ also scales as $\mathcal{O}(N)$, the relative distortion power behaves as $\mathcal{O}(t)$ with CO and $\mathcal{O}(t/N)$ with SOs.} In other words, { it appears to be} preferable to have independent oscillators at each antenna element in massive MIMO systems, which was also noted in \cite{Pitarokoilis2014a}. This is also consistent with our results for additive distortion noise: impairments at the UE are $N$ times more influential on the capacity, thus we can degrade the quality of the BS's oscillators with $N$ and only get a minor loss in performance. This property is also positive for distributed massive MIMO deployments where the antenna separation is large and prevents the use of a CO. A major difference from the additive distortion noise in Section \ref{sec:system-model} is that the distortion in \eqref{eq:phase-noise-interference-variance} increases linearly with the time $t$, thus it eventually grows large and it becomes necessary to send pilot/calibration signals more often to mitigate it \cite{Pitarokoilis2014a}.

The narrowband phase-noise analysis above only considered a lower capacity bound, thus it is possible to achieve higher rates. In particular, the analysis assumed uncompensated free-running oscillators, while it might be better to track the phase noise process at the BS; for example, by using previous received signals, extra calibration signals (see \cite{Mehrpouyan2012a} and references therein), and utilizing correlation between subcarriers in multi-carrier systems. The tracking might be more accurate for a CO since there are $O(N)$ observations of a single phase drift parameter, instead of  $O(N)$ observations for $N$ parameters as with SOs. Another important aspect of phase noise is that the standard deviations $\sqrt{\Delta^{\mathrm{BS}}}$ and $\sqrt{\Delta^{\mathrm{UE}}}$ are typically  proportional to the carrier frequency \cite{Petrovic2007a}, thus phase noise might be a major challenge in higher frequency bands (e.g., mmWave) \cite{Baldemair2013a}---unless the symbol time is also sufficiently reduced by increasing the bandwidth.

\subsection{Imperfect Channel Reciprocity}
\label{subsec:imperfect-reciprocity}

The downlink beamforming in  massive MIMO TDD systems relies on channel reciprocity; that is, if $\vect{h}$ is the uplink channel then $\vect{h}^T$ is the downlink channel. This property holds for the radio-frequency propagation channels, but the end-to-end channels are also affected by the hardware since different transceiver chains are used for transmission/reception at the BS and the UE. The actual downlink channel is $\vect{h}^T \vect{D}_{\vect{b}}$ where the diagonal matrix $\vect{D}_{\vect{b}} = \diag( b_1,\ldots,b_{N})$ contains $N$ calibration parameters. These are $b_i = 1 \,\, \forall i$ for ideal hardware but we generally have $b_i \neq 1 \,\, \forall i$ due to non-ideal hardware. The mismatch is fully specified by $b_1,\ldots,b_{N}$ and fortunately these parameters change slowly with time, thus one can compute estimates $\hat{b}_1,\ldots,\hat{b}_{N}$ using a negligible amount of overhead signaling \cite{Zetterberg2011a} (even in massive MIMO systems \cite{Shepard2012a}). Since the transmit beamforming mainly depends on the channel direction, it is often sufficient for the BS to compute the downlink channel up to an unknown scaling factor; see \cite{Zetterberg2011a,Nishimori2001a,Shepard2012a,Rogalin2013a} for different techniques that exploit uplink pilot transmissions. The estimates are naturally imperfect, thus $b_i = c( \hat{b}_i  + e_i)$ where $e_i$ is the estimation error and $c$ is the unknown common scaling factor.

Imperfect channel reciprocity has no impact on the UL and is not expected to change anything fundamentally in the DL. There is a loss in received signal power since the beamforming direction is perturbed, but there is no extra self-interference since all the CSI available at the receiving UE is estimated in the downlink and thus reflects the actual downlink channel $\vect{h}^T \vect{D}_{\vect{b}}$. In other words, the lower capacity bound in \eqref{eq:lower-bound-capacity} is still valid if we replace $\vect{h}$ by $\vect{D}_{\vect{b}} \vect{h}$ everywhere and compute the expectations with respect to the actual distributions. The beamforming vector $\vect{v}^{\mathrm{DL}}$ is now a function of $\diag( \hat{b}_1,\ldots,\hat{b}_{N}) \hat{\vect{h}}$. This perturbation of $\vect{v}^{\mathrm{DL}}$, as compared to having perfect reciprocity, behaves like a channel estimation error and its impact is expected to vanish as $N$ grows large. Moreover, it should only have a minor impact on the inter-user interference in multi-cell scenarios, since the reciprocity calibration errors are independent of the co-user channels.

\section{Conclusion}
\label{sec:conclusion}

This paper analyzed the capacity and estimation accuracy of massive MIMO systems with non-ideal transceiver hardware. The analysis was based on a new system model that models the hardware impairment at each  antenna by an additive distortion noise that is proportional to the signal power at this antenna. This model has several attractive features: it is mathematically tractable, it has been verified experimentally in previous works, and it can be motivated theoretically in systems that apply compensation algorithms to mitigate the hardware impairments.

We proved analytically that hardware impairments create non-zero estimation error floors and finite capacity ceilings in the uplink and downlink---irrespective of the SNR and the number of base station antennas $N$. This stands in contrast to the very optimistic asymptotic results previously reported for ideal hardware. Despite these discouraging results, we showed that massive MIMO systems can still achieve a huge array gain, in the sense that relatively high spectral efficiency and energy efficiency can be obtained. Furthermore, we proved that only the hardware impairments at the UEs limit the capacities as $N$ grows large. This implies that the hardware quality at the BS can be decreased as $N$ grows, which is an important insight and might become a key enabler for future network deployments.

In multi-cell scenarios, we proved that the detrimental effect of inter-user interference and pilot contamination drowns in the distortion noise if a simple pilot allocation algorithm is used to avoid the strongest forms of pilot contaminated interference. Many quantitative conclusions can be drawn from the numerical results in Sections \ref{sec:channel-estimation}--\ref{sec:multi-cell-scenario}; for example, that there is little gain in having more than 100 antennas for a single-user link, but additional antennas are useful to suppress inter-user interference in multi-cell scenarios. The asymptotic limits under non-ideal hardware are generally reached at much fewer antennas than the asymptotic limits for ideal hardware, which implies that we can expect practical systems to benefit from the asymptotic results. We also gave a brief description of how the system model considered in this paper can be refined to model hardware impairments in even greater detail and how such refinements would affect our results.

\appendices

\section{New and Old Results on Random Vectors} \label{app:general-lemmas}

\begin{lemma} \cite[Eq.~(2.2)]{Silverstein1995a} \label{lemma:woodbury-identity}
For invertible matrices $\vect{B}$ and $\tau \geq 0$, it holds that
\begin{equation}
(\vect{B}+\tau \vect{x}\vect{x}^H)^{-1} \vect{x} = \frac{\vect{B}^{-1} \vect{x}}{1+\tau \vect{x}^H \vect{B}^{-1} \vect{x}}.
\end{equation}
\end{lemma}

\vskip 3mm

\begin{lemma} \label{lemma:integration_h-ah-b}
Suppose $h \sim \mathcal{CN}(0,r)$ and $a,b>0$, then
\begin{equation} \label{eq:integrationresult}
\mathbb{E}  \left\{ \frac{|h|^2}{ a |h|^2 + b} \right\} = \frac{1}{a} \left( 1 - \frac{b}{a r} E_1 \bigg( \frac{b}{a r} \bigg) e^{\frac{b}{a r}} \right)
\end{equation}
where $E_1(x) = \int_{1}^{\infty} \frac{e^{-tx} }{t} dt$ denotes the exponential integral.
\end{lemma}
\begin{IEEEproof}
Since $\varrho=|h|^2$ has the exponential distribution with mean value $r$, the expectation in \eqref{eq:integrationresult} equals
\begin{equation}
\int_{0}^{\infty} \frac{\varrho}{ a \varrho + b} \frac{e^{-\varrho/r}}{r} d\varrho = \frac{b }{a^2 r} e^{\frac{b}{ar}} \int_{1}^{\infty} \bigg(1- \frac{1}{x} \bigg) e^{-\frac{\varrho b}{ar}} dx
\end{equation}
where the equality follows from a change of variable $x= \frac{a}{b} \varrho +1$. Straightforward integration and identification of the exponential integral yield the right-hand side of \eqref{eq:integrationresult}.
\end{IEEEproof}

\vskip 3mm

\begin{lemma} \label{lemma:split-absolute-values}
For any $a,b \in \mathbb{C}$ and non-zero $c,d \in \mathbb{C}$, we have
\begin{equation}
\left|  \frac{a}{c} - \frac{b}{d} \right| \leq \frac{|b| \, |c-d|  }{|c| \, |d| } + \frac{|a-b|}{|c|}.
\end{equation}
\end{lemma}
\begin{IEEEproof}
This is straightforward to prove by using that $|ad-bc| = | ad-bc+bd-bd | \leq |b| \, |c-d| + |d| \, |a-b|$.
\end{IEEEproof}

\vskip 3mm

\begin{lemma} \label{lemma:trace-bound-O-N}
Consider $M$ arbitrary matrices  $\vect{M}_1,\ldots,\vect{M}_{M} \in \mathbb{C}^{N \times N}$ and an Hermitian positive semi-definite matrix $\vect{B} \in \mathbb{C}^{N \times N}$. It follows that
\begin{equation} \label{eq:trace-inequality-multiplications}
| \tr( \vect{M}_1 \cdots \vect{M}_M \vect{B}) | \leq \tr (  \vect{B}  ) \, \prod_{i =1}^{M} \| \vect{M}_i \|_2 .
\end{equation}
If $\vect{M}_1,\ldots,\vect{M}_{M},\vect{B}$ have uniformly bounded spectral norms, then
\begin{equation}
| \tr( \vect{M}_1 \cdots \vect{M}_M \vect{B}) | = \mathcal{O}(N).
\end{equation}
\end{lemma}
\begin{IEEEproof}
The bound in \eqref{eq:trace-inequality-multiplications} follows from that $\vect{B}$ has non-negative eigenvalues and each matrix $\vect{M}_i$ cannot amplify these by more than $\| \vect{M}_i \|_2 $. The $\mathcal{O}(N)$-scaling follows from \eqref{eq:trace-inequality-multiplications} by using the assumptions and $\tr (  \vect{B}  ) \leq N \| \vect{B} \|_2$.
\end{IEEEproof}

\vskip 3mm

\begin{lemma} \cite[Lemma B.26]{Bai2009a} \label{lemma:classic-trace-result}
Let $\vect{B} \in \mathbb{C}^{N \times N}$ be deterministic and $\vect{x} = [x_1 \, \ldots \, x_N ]^T  \in \mathbb{C}^{N}$ be a stochastic vector of independent entries. Assume that $\mathbb{E}\{ x_i\} = 0$, $\mathbb{E}\{ |x_i|^2\} = 1$, and $\mathbb{E}\{ |x_i|^{\ell}\} = \chi_{\ell}< \infty$ for $\ell \leq 2 q$. Then, for any $q \geq 1$,
\begin{equation}
\mathbb{E} \left\{ \left| \vect{x}^H \vect{B} \vect{x} - \tr( \vect{B} ) \right|^q \right\} \leq C_q \left( \tr(\vect{B} \vect{B}^H) \right)^{\frac{q}{2}} \big( \chi_4^{\frac{q}{2}} + \chi_{2q}\big)
\end{equation}
where $C_q$ is a constant depending on $q$ only.
\end{lemma}

\section{Application-Related Random Vector Results} \label{app:specific-lemmas}

\begin{lemma} \label{lemma:hhat-characterization}
The channel estimate $\hat{\vect{h}}$ can be decomposed as
\begin{equation}
\hat{\vect{h}} = \vect{A} \left(  \left( (\p + \eta_{t}^{\mathrm{UE}}) \vect{I} + \vect{D}_r \right) \vect{h}   + \boldsymbol{\nu} \right)
\end{equation}
where $\vect{A}$ is defined in \eqref{eq:LMMSE-estimator} and the diagonal matrix $\vect{D}_r$ has independent $\mathcal{CN}(0,\kappa_r^{\mathrm{BS}} p^{\mathrm{UE}})$-entries such that $\boldsymbol{\eta}_{r}^{\mathrm{BS}} = \vect{D}_r \vect{h}$.

For any realizations of $\eta_{t}^{\mathrm{UE}}$ and $\vect{D}_r$, the conditional distribution is
\begin{equation} \label{eq:conditional-channel-estimate}
\hat{\vect{h}} | \eta_{t}^{\mathrm{UE}},\vect{D}_r \sim \mathcal{CN} \left( \vect{0}, \vect{A} (\boldsymbol{\Phi} + \vect{S} + \sigma_{\mathrm{BS}}^2 \vect{I}) \vect{A}^H  \right)
\end{equation}
where $\boldsymbol{\Phi} = ((d+\eta_{t}^{\mathrm{UE}})\vect{I} + \vect{D}_r) \vect{R} ((d+\eta_{t}^{\mathrm{UE}})\vect{I} + \vect{D}_r)^H$.
\end{lemma}
\begin{IEEEproof}
This characterization follows directly from Theorem \ref{theorem:LMMSE-estimator} and the system model defined in Section \ref{sec:system-model}.
\end{IEEEproof}

\vskip 3mm

\begin{lemma} \label{lemma:asymptotic-differences-innerproduct}
For the channel $\vect{h}$ and its estimate $\hat{\vect{h}}$ it holds that
\begin{align} \label{eq:innerproduct-hhbar}
\mathbb{E} \left\{ \left| \vect{h}^H \hat{\vect{h}} - (1+\p^{-1} \eta_{t}^{\mathrm{UE}}) \tr(\vect{R}-\vect{C}) \right|^2 \right\} &= \mathcal{O}(N) \\
\mathbb{E} \left\{ \left| |\vect{h}^H \hat{\vect{h}}| - |1+\p^{-1} \eta_{t}^{\mathrm{UE}}| \tr(\vect{R}-\vect{C}) \right|^2 \right\} &= \mathcal{O}(N). \label{eq:innerproduct-hhbar-variation}
\end{align}
\end{lemma}
\begin{IEEEproof}
Recall that $\hat{\vect{h}} = \vect{A} \big( \vect{h} (\p + \eta_{t}^{\mathrm{UE}}) + \boldsymbol{\nu} + \boldsymbol{\eta}_{r}^{\mathrm{BS}} \big)$ for $\vect{A} =
\p^* \vect{R} \bar{\vect{Z}}^{-1}$. To prove \eqref{eq:innerproduct-hhbar}, we expand the argument as
\begin{align} \label{eq:bound-inner-products_initialbound1}
&\left| \vect{h}^H \hat{\vect{h}} - (1+\p^{-1} \eta_{t}^{\mathrm{UE}}) \tr(\vect{R}-\vect{C}) \right|^2 \leq 4 | \vect{h}^H \vect{A} \boldsymbol{\nu} |^2  \\ & +4 | \vect{h}^H \vect{A} \boldsymbol{\eta}_{r}^{\mathrm{BS}} |^2+  2 |\p + \eta_{t}^{\mathrm{UE}}|^2  \left| \vect{h}^H \vect{A} \vect{h} - \p^{-1} \tr(\vect{R}-\vect{C}) \right|^2 \notag
\end{align}
by using the rule $|a+b|^q \leq 2^{q-1} ( |a|^q + |b|^q)$ (from H\"older's inequality) twice.
Next, we observe that
\begin{align} \label{eq:bound-inner-products1}
\mathbb{E} \{ | \vect{h}^H \vect{A} \boldsymbol{\nu} |^2 \} & \!=\!
 \tr \left( \vect{A} (\vect{S} + \sigma_{\mathrm{BS}}^2 \vect{I}) \vect{A}^H \vect{R} \right) = \mathcal{O}(N) \\
\mathbb{E} \{ | \vect{h}^H \vect{A} \boldsymbol{\eta}_{r}^{\mathrm{BS}} |^2 \} & \!=\!
\kappa_{r}^{\mathrm{BS}} p^{\mathrm{UE}}  \tr \left( \vect{A} \vect{R}_{\diag} \vect{A}^H \vect{R} \right) \notag \\
 & +  \kappa_{r}^{\mathrm{BS}} p^{\mathrm{UE}} \sum_{i=1}^{N} |\vect{e}_i^H \vect{R} \vect{A} \vect{e}_i|^2 = \mathcal{O}(N) \label{eq:bound-inner-products1b}
\end{align}
where $\vect{e}_i$ is the $i$th column of an $N \times N$ identity matrix. The expression \eqref{eq:bound-inner-products1}
follows from the independence of $\vect{h}, \boldsymbol{\nu}$ and \eqref{eq:bound-inner-products1b} follows by straightforward computation using the characterization $\boldsymbol{\eta}_{r}^{\mathrm{BS}} = \vect{D}_r \vect{h}$ in Lemma \ref{lemma:hhat-characterization}. The $\mathcal{O}(N)$-properties follows from Lemma \ref{lemma:trace-bound-O-N} since $\vect{R},\vect{S},\vect{A}$ have uniformly bounded spectral norms (by assumption).

Since $\vect{h} \sim \vect{R}^{1/2} \tilde{\vect{h}}$ for $\tilde{\vect{h}} \sim \mathcal{CN}(\vect{0},\vect{I})$ and $\p^{-1} \tr(\vect{R}-\vect{C}) = \tr(\vect{R}^{1/2} \vect{A}  \vect{R}^{1/2})$ we can apply Lemma \ref{lemma:classic-trace-result} to obtain
\begin{equation} \label{eq:bound-inner-products2}
\begin{split}
&\mathbb{E} \left\{ |\p + \eta_{t}^{\mathrm{UE}}|^2  \left| \vect{h}^H \vect{A} \vect{h} - \p^{-1} \tr(\vect{R}-\vect{C}) \right|^2 \right\} \\
&\leq (1+\kappa_{t}^{\mathrm{UE}}) 2\chi_4 C_2 \tr \left( (\vect{R}-\vect{C})^2  \right) = \mathcal{O}(N).
\end{split}
\end{equation}
We obtain \eqref{eq:innerproduct-hhbar} by combining \eqref{eq:bound-inner-products_initialbound1}--\eqref{eq:bound-inner-products2}. Expression \eqref{eq:innerproduct-hhbar-variation} follows directly, since it is upper bounded similarly to \eqref{eq:bound-inner-products_initialbound1}.
\end{IEEEproof}

\vskip 3mm

\begin{lemma} \label{lemma:asymptotic-differences-norm}
For the channel $\vect{h}$ and its estimate $\hat{\vect{h}}$ it holds that
\begin{align} \label{eq:norm-hbarhbar}
\mathbb{E}\left\{  \Big| \|\hat{\vect{h}}\|_2^2 - \tr \Big( \vect{A} ( |\p \!+\! \eta_{t}^{\mathrm{UE}}|^2 \vect{R} \!+\! \boldsymbol{\Psi}) \vect{A}^H \Big)  \Big|^2  \right\} \!=\! \mathcal{O}(N) \\
\mathbb{E}\left\{  \Big| \|\hat{\vect{h}}\|_2 - \sqrt{\tr \Big( \vect{A} ( |\p \!+\! \eta_{t}^{\mathrm{UE}}|^2 \vect{R} \!+\! \boldsymbol{\Psi}) \vect{A}^H \Big)}  \Big|^2  \right\} \!=\! \mathcal{O}(1) \label{eq:norm-hbarhbar-sqrt}
\end{align}
where $\vect{A}$ is defined in \eqref{eq:LMMSE-estimator} and $\boldsymbol{\Psi} =   p^{\mathrm{UE}} \kappa_r^{\mathrm{BS}}  \vect{R}_{\diag} + \vect{S} + \sigma_{\mathrm{BS}}^2 \vect{I}$.
\end{lemma}
\begin{IEEEproof}
By injecting the term $\tr(\vect{A} (\boldsymbol{\Phi} + \vect{S} + \sigma_{\mathrm{BS}}^2 \vect{I}) \vect{A}^H)$ that appeared in Lemma \ref{lemma:hhat-characterization} and using the rule $|a+b|^q \leq 2^{q-1} (|a|^q + |b|^q )$ (from H\"older's inequality), we bound the left-hand side of \eqref{eq:norm-hbarhbar} as
\begin{equation} \label{eq:norm-hbarhbar-firstbound}
\begin{split}
&\mathbb{E}\left\{  \Big| \|\hat{\vect{h}}\|_2^2 - \tr \Big( \vect{A} ( |\p \!+\! \eta_{t}^{\mathrm{UE}}|^2 \vect{R} \!+\! \boldsymbol{\Psi}) \vect{A}^H \Big)  \Big|^2  \right\} \\
&\leq 2 \mathbb{E} \left\{ \left| \| \hat{\vect{h}}\|_2^2 - \tr \Big(\vect{A} (\boldsymbol{\Phi} + \vect{S} + \sigma_{\mathrm{BS}}^2 \vect{I}) \vect{A}^H \Big) \right|^2 \right\} \\
&+ 2 \mathbb{E} \bigg\{ \Big|  \tr \Big(\vect{A} (\boldsymbol{\Phi} + \vect{S} + \sigma_{\mathrm{BS}}^2 \vect{I}) \vect{A}^H \Big) \\ & \quad \quad - \tr \Big( \vect{A} ( |\p \!+\! \eta_{t}^{\mathrm{UE}}|^2 \vect{R} \!+\! \boldsymbol{\Psi}) \vect{A}^H \Big) \Big|^2 \bigg\}.
\end{split}
\end{equation}
The first term in \eqref{eq:norm-hbarhbar-firstbound} satisfies
\begin{equation}
\begin{split}
& \mathbb{E} \left\{ \left| \| \hat{\vect{h}}\|_2^2 - \tr(\vect{A} (\boldsymbol{\Phi} + \vect{S} + \sigma_{\mathrm{BS}}^2 \vect{I}) \vect{A}^H) \right|^2 \right\} \\
&\leq \!
4 C_2 \mathbb{E} \left\{  \tr(\vect{A} (\boldsymbol{\Phi} + \vect{S} + \sigma_{\mathrm{BS}}^2 \vect{I}) \vect{A}^H \vect{A} (\boldsymbol{\Phi} + \vect{S} + \sigma_{\mathrm{BS}}^2 \vect{I})^H \vect{A}^H)  \right\} \\
& \leq \! 4 C_2 \| \vect{A} \|_2^4 \, \mathbb{E} \left\{ \| \boldsymbol{\Phi} + \vect{S} + \sigma_{\mathrm{BS}}^2 \vect{I}\|_F^2 \right\} = \mathcal{O}(N)
\end{split}
\end{equation}
where the first inequality follows from applying Lemma \ref{lemma:classic-trace-result} on \eqref{eq:norm-hbarhbar} for fixed $\eta_{t}^{\mathrm{UE}},\vect{D}_r$ (note that the fourth-order moment is
$\chi_4 =2$ for complex Gaussian variables), while the second inequality follows from applying Lemma \ref{lemma:trace-bound-O-N} twice. The scaling $\mathcal{O}(N)$ follows since $\sigma_{\mathrm{BS}}^2$ is constant,
$\|\vect{A} \|_2 = \mathcal{O}(1)$, $\|\vect{S} \|_F^2 = \mathcal{O}(N)$, $\mathbb{E}\{ \tr(\boldsymbol{\Phi} \vect{S})  \} \leq \| \vect{R} \|_2 \| \vect{S} \|_2 \mathbb{E}\{ \|  (d+\eta_{t}^{\mathrm{UE}})\vect{I} + \vect{D}_r \|_F^2 \} =  \mathcal{O}(N)$, and $\mathbb{E}\{ \tr(\boldsymbol{\Phi} \boldsymbol{\Phi}^H)  \} \leq  \| \vect{R} \|_2  \mathbb{E}\{ \|  ((d+\eta_{t}^{\mathrm{UE}})\vect{I} + \vect{D}_r) ((d+\eta_{t}^{\mathrm{UE}})\vect{I} + \vect{D}_r)^H \|_F^2 \} =  \mathcal{O}(N)$ using Lemma \ref{lemma:trace-bound-O-N}.

Next, we characterize the second term in \eqref{eq:norm-hbarhbar-firstbound} as
\begin{align} \notag
&\mathbb{E} \bigg\{ \Big|  \tr \Big(\vect{A} (\boldsymbol{\Phi} + \vect{S} + \sigma_{\mathrm{BS}}^2 \vect{I}) \vect{A}^H \Big) \\ &\quad \quad \notag - \tr \Big( \vect{A} ( |\p \!+\! \eta_{t}^{\mathrm{UE}}|^2 \vect{R} \!+\! \boldsymbol{\Psi}) \vect{A}^H \Big) \Big|^2 \bigg\} \\ \notag
& = \mathbb{E} \Bigg\{ \bigg|  \tr \Big(\vect{A} \big( \vect{D}_r \vect{R} \vect{D}_r^H + \vect{D}_r \vect{R} (d+\eta_{t}^{\mathrm{UE}})^*  \\ \notag
& \quad + (d+\eta_{t}^{\mathrm{UE}})  \vect{R} \vect{D}_r^H
 \big) \vect{A}^H \Big) \!-\! p^{\mathrm{UE}} \kappa_r^{\mathrm{BS}} \tr \Big( \vect{A}  \vect{R}_{\diag} \vect{A}^H \Big) \bigg|^2 \Bigg\}
\\ \notag
&\leq 2 \mathbb{E} \left\{ \left|  \tr \Big(\vect{A} ( \vect{D}_r \vect{R} \vect{D}_r^H   - p^{\mathrm{UE}} \kappa_r^{\mathrm{BS}}  \vect{R}_{\diag} ) \vect{A}^H \Big) \right|^2 \right\} \\
&+4 \mathbb{E} \left\{ | d+\eta_{t}^{\mathrm{UE}} |^2  \right\}   \mathbb{E} \left\{  \left| \tr \Big(\vect{A}  \vect{D}_r \vect{R} \vect{A}^H \Big) \right|^2 \right\} \!=\! \mathcal{O}(N). \label{eq:norm-hbarhbar-secondbound}
\end{align}
where the equality follows from plugging in the expressions for $\boldsymbol{\Phi}$ and $\boldsymbol{\Psi}$ and noting that the terms  $|\p \!+\! \eta_{t}^{\mathrm{UE}}|^2 \tr ( \vect{A}  \vect{R} \vect{A}^H )$,
$\tr(\vect{A} \vect{S} \vect{A}^H )$, and $\sigma_{\mathrm{BS}}^2 \tr(\vect{A} \vect{A}^H )$ cancel out. The inequality follows again from the rule $|a+b|^q \leq 2^{q-1} (|a|^q + |b|^q )$.
The $\mathcal{O}(N)$-scaling follows since the first term in \eqref{eq:norm-hbarhbar-secondbound} is upper bounded by $(p^{\mathrm{UE}} \kappa_r^{\mathrm{BS}})^2 \tr( \vect{A}  \vect{R}_{\diag}  \vect{A}^H \vect{A}  \vect{R}_{\diag}  \vect{A}^H ) = \mathcal{O}(N)$ using Lemma \ref{lemma:classic-trace-result} and some algebra, while $\mathbb{E} \{  | \tr (\vect{A}  \vect{D}_r \vect{R} \vect{A}^H ) |^2 \} \leq  \| \vect{R} \vect{A}^H \vect{A}  \|_2^2 \mathbb{E} \{  | \tr (  \vect{D}_r ) |^2 \} = \mathcal{O}(N)$ using  Lemma \ref{lemma:trace-bound-O-N}.
The expression \eqref{eq:norm-hbarhbar} now follows from combining \eqref{eq:norm-hbarhbar-firstbound}--\eqref{eq:norm-hbarhbar-secondbound}.

Finally, the expression \eqref{eq:norm-hbarhbar-sqrt} is proved as
\begin{align}
&\mathbb{E}\left\{  \Big| \|\hat{\vect{h}}\|_2 - \sqrt{\tr \Big( \vect{A} ( |\p \!+\! \eta_{t}^{\mathrm{UE}}|^2 \vect{R} \!+\! \boldsymbol{\Psi}) \vect{A}^H \Big)}  \Big|^2 \right\} \\ \notag
&\leq \mathbb{E}\left\{ \frac{ \Big| \|\hat{\vect{h}}\|_2^2 - \tr \Big( \vect{A} ( |\p \!+\! \eta_{t}^{\mathrm{UE}}|^2 \vect{R} \!+\! \boldsymbol{\Psi}) \vect{A}^H \Big)  \Big|^2 }{\Big| \|\hat{\vect{h}}\|_2 + \sqrt{\tr \Big( \vect{A} ( |\p \!+\! \eta_{t}^{\mathrm{UE}}|^2 \vect{R} \!+\! \boldsymbol{\Psi}) \vect{A}^H \Big)}  \Big|^2} \right\} \\
&\leq \frac{ \mathbb{E}\left\{  \Big| \|\hat{\vect{h}}\|_2^2 - \tr \Big( \vect{A} ( |\p \!+\! \eta_{t}^{\mathrm{UE}}|^2 \vect{R} \!+\! \boldsymbol{\Psi}) \vect{A}^H \Big)  \Big|^2 \right\} }{  \tr \Big( \vect{A} \boldsymbol{\Psi} \vect{A}^H \Big) } = \mathcal{O}(1) \notag
\end{align}
where the first inequality follows from the rule $(a-b)(a+b) = a^2 -b^2$ and the second inequality is due to removal of non-zero terms from the denominator. The numerator scales as $\mathcal{O}(N)$ and the denominator scales at least linearly with $N$ because $\tr \Big( \vect{A} \boldsymbol{\Psi} \vect{A}^H \Big) \geq \lambda_{\min} (\boldsymbol{\Psi} ) \tr \Big( \vect{A}  \vect{A}^H \Big) \geq p^{\mathrm{UE}} \lambda_{\min} (\boldsymbol{\Psi} ) \lambda_{\max}( \bar{\vect{Z}}) \tr \Big( \vect{R}  \vect{R}^H \Big)$, where $\tr \Big( \vect{R} \Big)$ grows linearly with $N$ (by assumption). Here, $\lambda_{\max}(\cdot)$ and $\lambda_{\min}(\cdot)$ denotes the largest and smallest eigenvalues of a matrix, respectively.
This shows that \eqref{eq:norm-hbarhbar-sqrt} is bounded and finalizes the proof.
\end{IEEEproof}

\vskip 3mm

\begin{lemma} \label{lemma:1overh2}
For the estimated channel $\hat{\vect{h}}$ in \eqref{eq:LMMSE-estimator} it holds that
\begin{align} \label{eq:oneoverhhat-p1}
\mathbb{E} \left\{  \frac{|1 \!+\! d^{-1} \eta_{t}^{\mathrm{UE}}|^k}{\|\hat{\vect{h}}\|_2^2} \right\} & \!\leq\! \frac{2^{k}+2^{k} \left(\frac{k}{2}\right)! (\kappa_{t}^{\mathrm{UE}})^{\frac{k}{2}} }{\lambda^+_{\min}(\vect{B}) (N_B \!-\! 1)} \! = \! \mathcal{O}(N^{-1}) \\
\mathbb{E} \left\{  \frac{|1 \!+\! d^{-1} \eta_{t}^{\mathrm{UE}}|^k}{\|\hat{\vect{h}}\|_2^4} \right\} &\!\leq\! \frac{2^{k}+2^{k} \left(\frac{k}{2}\right)!  (\kappa_{t}^{\mathrm{UE}})^{\frac{k}{2}} }{\lambda^+_{\min}(\vect{B})^2 (N_B \!-\! 1)(N_B \!-\!  2)} \! = \! \mathcal{O}(N^{-2}) \label{eq:oneoverhhat-p2}
\end{align}
for any even integer $k$, where $\lambda^+_{\min}(\vect{B})>0$ denotes the smallest non-zero eigenvalue of $\vect{B} = \sigma_{\mathrm{BS}}^2 \vect{A}  \vect{A}^H$ and $N_B =\rank(\vect{B})$.
\end{lemma}
\begin{IEEEproof}
Using the conditional distribution of the channel estimate in \eqref{eq:conditional-channel-estimate}, it holds for any integer $q>0$ that
\begin{align} \notag
&\mathbb{E} \left\{  \frac{|1+d^{-1} \eta_{t}^{\mathrm{UE}}|^{k}}{\|\hat{\vect{h}}\|_2^{2q}} \right\} = \mathbb{E} \left\{ \mathbb{E} \left\{ \frac{|1+d^{-1}  \eta_{t}^{\mathrm{UE}}|^{k}}{\|\hat{\vect{h}}\|_2^{2q}} \bigg| \eta_{t}^{\mathrm{UE}},\vect{D}_r \right\} \right\} \\
&\quad \quad  \leq \mathbb{E} \left\{ \frac{2^{k} (1+ |d^{-1} \eta_{t}^{\mathrm{UE}}|^{k})}{\lambda^+_{\min}( \sigma_{\mathrm{BS}}^2 \vect{A}   \vect{A}^H )^q} \right\} \mathbb{E} \left\{ \frac{1}{\| \vect{U}_B^H  \vect{v} \|_2^{2q}} \right\}
\end{align}
where the inequality follows from $\|\hat{\vect{h}}\|_2^2 = \vect{v}^H \vect{A} (\boldsymbol{\Phi} + \vect{S} + \sigma_{\mathrm{BS}}^2 \vect{I}) \vect{A}^H \vect{v} \geq \sigma_{\mathrm{BS}}^2
\vect{v}^H \vect{A} \vect{A}^H \vect{v} \geq \lambda^+_{\min}( \sigma_{\mathrm{BS}}^2 \vect{A} \vect{A}^H ) \| \vect{U}_B^H \vect{v}  \|_2^2$ where $\vect{v} \sim \mathcal{CN}(\vect{0},\vect{I})$ and $\vect{U}_B \in \mathbb{C}^{N \times N_B}$ is an orthogonal basis of the span of $\vect{A} \vect{A}^H$ (note that this matrix is generally rank-deficient). The expectations were separated since $\| \vect{U}_B^H \vect{v}  \|_2^2$ is independent of the smallest non-zero eigenvalue.
Furthermore, the rule $|a+b|^q \leq 2^{q} (|a|^q + |b|^q )$ from H\"older's inequality was applied on $|1+d^{-1} \eta_{t}^{\mathrm{UE}}|^{k}$.
Note that $ \vect{U}_B^H  \vect{v} \sim \mathcal{CN}(\vect{0},\vect{I})$ with a dimension reduced from $N$ to $N_B$.

The final result in \eqref{eq:oneoverhhat-p1}--\eqref{eq:oneoverhhat-p2} follows from $\mathbb{E}\{| d^{-1}  \eta_{t}^{\mathrm{UE}}|^{k}\} = \left(\frac{k}{2}\right)! (\kappa_{t}^{\mathrm{UE}})^{\frac{k}{2}}$ and that \cite[Lemma 2.10]{Tulino2004a} with $m=1$ and $n=N_B$ gives
\begin{equation}
\mathbb{E} \left\{ \frac{1}{\| \vect{U}_B^H  \vect{v} \|_2^{2q}} \right\} = \begin{cases} \frac{1}{N_B-1}, & \text{if } q=1,\\ \frac{1}{(N_B-1)(N_B-2)}, & \text{if } q=2, \end{cases}
\end{equation}
and that $N_B$ scales linearly with $N$ (see Section \ref{sec:system-model}).
\end{IEEEproof}

\section{Collection of Proofs}
\label{appendix:full-proofs}

\subsection{Proof of Lemma \ref{lemma:initial-upper-capacity-bounds}}
\label{proof:lemma:initial-upper-capacity-bounds}

The DL capacity in \eqref{eq:downlink_capacity} is upper bounded as
\begin{equation} \label{eq:downlink_capacity_upper1}
{\tt C}^{\mathrm{DL}} \leq  \frac{  T^{\mathrm{DL}}_{\mathrm{data}}   }{ T_{\mathrm{coher}} } \mathbb{E} \left\{ \max_{ \vect{w}(\vect{h}) \, : \,  \| \vect{w}\|_2=1} \log_2( 1+ {\tt SINR}(\vect{w}) ) \right\}
\end{equation}
where
\begin{equation}
\label{eq:downlink_SINR_upper1}
{\tt SINR}(\vect{w}) = \frac{ |\vect{h}^T \vect{w}|^2}{ \kappa_t^{\mathrm{BS}}  \fracSumtwo{i=1}{N} |h_i w_i|^2 + \kappa_r^{\mathrm{UE}} |\vect{h}^T \vect{w}|^2   + \frac{\sigma_{\mathrm{UE}}^2}{p^{\mathrm{BS}}} }
\end{equation}
by assuming that the interference part of $n$ is somehow canceled, perfect CSI is available, and exploiting the corresponding optimality of single-stream Gaussian signaling \cite{Telatar1999a,Bjornson2013c}.
We can write \eqref{eq:downlink_SINR_upper1} as
\begin{equation} \label{eq:downlink_SINR_upper2}
{\tt SINR}(\vect{w}) = \frac{\vect{w}^H \vect{h}^* \vect{h}^T \vect{w} }{ \vect{w}^H \big(\kappa_t^{\mathrm{BS}} \vect{D}_{|\vect{h}|^2} + \kappa_r^{\mathrm{UE}} \vect{h}^* \vect{h}^T   + \frac{\sigma_{\mathrm{UE}}^2}{p^{\mathrm{BS}}} \vect{I} \big)\vect{w}}
\end{equation}
by utilizing $\vect{w}^H \vect{w}=1$. Since the logarithm is a monotonically increasing function, the maximization in \eqref{eq:downlink_capacity_upper1}
can be applied onto ${\tt SINR}(\vect{w})$. Using \eqref{eq:downlink_SINR_upper2}, this optimization is a generalized Rayleigh quotient problem and thus solved by
\begin{equation} \label{eq:solution-to-rayleigh-quotient}
\vect{w} = \frac{ ( \kappa_t^{\mathrm{BS}} \vect{D}_{|\vect{h}|^2} + \kappa_r^{\mathrm{UE}} \vect{h}^* \vect{h}^T + \frac{\sigma_{\mathrm{UE}}^2}{p^{\mathrm{BS}}} \vect{I} )^{-1} \vect{h}^* }{\big\| \big( \kappa_t^{\mathrm{BS}} \vect{D}_{|\vect{h}|^2} + \kappa_r^{\mathrm{UE}} \vect{h}^* \vect{h}^T + \frac{\sigma_{\mathrm{UE}}^2}{p^{\mathrm{BS}}} \vect{I} \big)^{-1} \vect{h}^* \big\|_2}
\end{equation}
which is equivalent to \eqref{eq:optimal_beamforming_downlink_perfectCSI} by using Lemma \ref{lemma:woodbury-identity}. The DL capacity bound in \eqref{eq:downlink_capacity_upper_first} follows from plugging \eqref{eq:solution-to-rayleigh-quotient} into \eqref{eq:downlink_SINR_upper2} (we also took the complex conjugate of the real-valued SINR expression to make it more consistent with the UL).

The UL capacity bound in \eqref{eq:uplink_capacity_upper_first} follows from \cite{Bjornson2013c} and by assuming that the interference part of $\boldsymbol{\nu}$ is somehow canceled.
We note that the uplink SINR with a receive combining vector $\vect{w}$ is
\begin{equation} \label{eq:rayleigh-quotient-uplink}
\frac{\vect{w}^H \vect{h} \vect{h}^H \vect{w} }{ \vect{w}^H \big(
\kappa_t^{\mathrm{UE}}  \vect{h} \vect{h}^H + \kappa_r^{\mathrm{BS}} \vect{D}_{|\vect{h}|^2} + \frac{\sigma_{\mathrm{BS}}^2}{p^{\mathrm{UE}}} \vect{I} \big)\vect{w}}.
\end{equation}
The receiver combining vector in \eqref{eq:optimal_beamforming_uplink_perfectCSI} maximizes  \eqref{eq:rayleigh-quotient-uplink} and achieves the upper bound in \eqref{eq:uplink_capacity_upper_first}.

\subsection{Proof of Theorem \ref{theorem:upperbounds-capacities}}
\label{proof:theorem:upperbounds-capacities}

The DL capacity bound in \eqref{eq:downlink_capacity_upper_first} can be rewritten as
\begin{equation}
\frac{  T^{\mathrm{DL}}_{\mathrm{data}}   }{ T_{\mathrm{coher}} } \mathbb{E} \left\{ \! \log_2 \! \left( 1+ \frac{ \vect{h}^H \big( \kappa_t^{\mathrm{BS}} \vect{D}_{|\vect{h}|^2} + \frac{\sigma_{\mathrm{UE}}^2}{p^{\mathrm{BS}}} \vect{I} \big)^{-1} \vect{h} }{1+ \kappa_r^{\mathrm{UE}} \vect{h}^H \big( \kappa_t^{\mathrm{BS}} \vect{D}_{|\vect{h}|^2} + \frac{\sigma_{\mathrm{UE}}^2}{p^{\mathrm{BS}}} \vect{I} \big)^{-1} \vect{h} } \right) \! \right\}
\end{equation}
using Lemma \ref{lemma:woodbury-identity}. This expression has the structure $m(\psi) = \log_2(1+\frac{\psi}{1+\kappa_r^{\mathrm{UE}} \psi} )$ where $\psi= \vect{h}^H ( \kappa_t^{\mathrm{BS}} \vect{D}_{|\vect{h}|^2} + \frac{\sigma_{\mathrm{UE}}^2}{p^{\mathrm{BS}}} \vect{I} )^{-1} \vect{h}$. Since $m(\psi)$ is a concave function of $\psi$, we apply Jensen's inequality to achieve a new upper bound
\begin{equation}
{\tt C}^{\mathrm{DL}} \leq  \frac{  T^{\mathrm{DL}}_{\mathrm{data}}   }{ T_{\mathrm{coher}} } \mathbb{E} \left\{ m(\psi) \right\} \leq  \frac{  T^{\mathrm{DL}}_{\mathrm{data}}   }{ T_{\mathrm{coher}} } m(\mathbb{E} \left\{ \psi \right\} ).
\end{equation}
The upper bound in \eqref{eq:downlink_capacity_upper2} follows from evaluating $\mathbb{E} \left\{ \psi \right\}$ as
\begin{equation}
\begin{split}
\mathbb{E} \left\{ \psi \right\} &= \mathbb{E} \left\{\vect{h}^H ( \kappa_t^{\mathrm{BS}} \vect{D}_{|\vect{h}|^2} + \frac{\sigma_{\mathrm{UE}}^2}{p^{\mathrm{BS}}} \vect{I} )^{-1} \vect{h} \right\} \\
&= \sum_{i=1}^{N} \mathbb{E}  \left\{ \frac{|h_i|^2}{ \kappa_t^{\mathrm{BS}} |h_i|^2 + \frac{\sigma_{\mathrm{UE}}^2}{p^{\mathrm{BS}}}} \right\} = G^{\mathrm{DL}}
\end{split}
\end{equation}
where the expression for $G^{\mathrm{DL}}$ is obtained from Lemma \ref{lemma:integration_h-ah-b} using $a= \kappa_t^{\mathrm{BS}}$ and $b=\frac{\sigma_{\mathrm{UE}}^2}{p^{\mathrm{BS}}}$.

The closed-form upper bound on the UL capacity in \eqref{eq:uplink_capacity_upper2} is derived analogously to the DL capacity bound.

\subsection{Proof of Theorem \ref{theorem:asymptotic-equivalence}}
\label{proof:theorem:asymptotic-equivalence}

We introduce the notation $\sqrt{\tr(\vect{R}-\vect{C})} \varphi = \frac{\vartheta}{\sqrt{\gamma}}$ where
\begin{align}
\vartheta &= (1+\p^{-1} \eta_{t}^{\mathrm{UE}}) \tr(\vect{R}-\vect{C}) \\
\gamma &= \tr\big(\vect{A} ( |\p + \eta_{t}^{\mathrm{UE}}|^2 \vect{R} + \boldsymbol{\Psi}) \vect{A}^H \big).
\end{align}

Starting with the equivalence in \eqref{eq:det-equiv-signalpart}, we use the rule $a^2-b^2 =(a+b)(a-b)$ to obtain
\begin{equation} \label{eq:initial-upper-bound}
\begin{split}
&\left| \left| \mathbb{E} \left\{ \frac{\vect{h}^H \hat{\vect{h}}}{\|\hat{\vect{h}}\|_2} \right\}  \right|^2 - \left| \mathbb{E} \left\{  \frac{\vartheta}{ \sqrt{ \gamma } } \right\} \right|^2 \right| \\
& = \left| \mathbb{E} \left\{ \frac{\vect{h}^H \hat{\vect{h}}}{\|\hat{\vect{h}}\|_2 } -  \frac{\vartheta}{ \sqrt{ \gamma } } \right\}  \right| \,  \left| \mathbb{E} \left\{ \frac{\vect{h}^H \hat{\vect{h}}}{\|\hat{\vect{h}}\|_2} +  \frac{\vartheta}{ \sqrt{ \gamma } } \right\} \right| \\
& \leq  \mathbb{E} \left\{ \left|\frac{\vect{h}^H \hat{\vect{h}}}{\|\hat{\vect{h}}\|_2} -  \frac{\vartheta}{ \sqrt{ \gamma } } \right|  \right\}   \left( \left| \mathbb{E} \left\{ \frac{\vect{h}^H \hat{\vect{h}}}{\|\hat{\vect{h}}\|_2} \right\} \right| + \left| \mathbb{E} \left\{ \frac{\vartheta}{ \sqrt{ \gamma } } \right\} \right| \right).
\end{split}
\end{equation}
In order to prove the first part of the theorem, we must show that right-hand side of \eqref{eq:initial-upper-bound} behaves as $\mathcal{O}(\sqrt{N})$.
Using Cauchy-Schwartz inequality and that $\gamma \geq \tr (\vect{A}  \boldsymbol{\Psi} \vect{A}^H )$, we have
\begin{equation}
\begin{split}
&\left| \mathbb{E} \left\{ \frac{\vect{h}^H \hat{\vect{h}}}{\|\hat{\vect{h}}\|_2} \right\} \right| + \left| \mathbb{E} \left\{ \frac{\vartheta}{ \sqrt{ \gamma } } \right\} \right| \\
 &\leq  \mathbb{E}\{ \| \vect{h}\|_2 \} + \left| \mathbb{E} \left\{ \frac{\vartheta}{ \sqrt{ \tr\big(\vect{A}  \boldsymbol{\Psi} \vect{A}^H \big) } } \right\} \right| =  \mathcal{O}(\sqrt{N})
\end{split}
\end{equation}
where $\mathbb{E}\{ \| \vect{h}\|_2 \} = \mathcal{O}(\sqrt{N})$ and the second term is bounded in the same way since $\mathbb{E}\{\vartheta\} = \tr(\vect{R}-\vect{C}) = \mathcal{O}(N)$ and $\tr\big(\vect{A}  \boldsymbol{\Psi} \vect{A}^H \big) $ grows at least linearly with $N$ (see the proof of Lemma \ref{lemma:asymptotic-differences-norm}). Hence, it remains to prove that the $  \mathbb{E}\left\{ \left| \vect{h}^H \hat{\vect{h}}/ \|\hat{\vect{h}}\|_2 - \vartheta/\sqrt{ \gamma } \right| \right\} = \mathcal{O}(1).$
To this end, we expand the expression using Lemma \ref{lemma:split-absolute-values}:
\begin{equation} \label{eq:bound_for_signalpart}
\begin{split}
&  \mathbb{E}\left\{ \left| \frac{\vect{h}^H \hat{\vect{h}}}{\|\hat{\vect{h}}\|_2} - \frac{\vartheta }{\sqrt{ \gamma }} \right| \right\}
\leq  \mathbb{E}\left\{  \frac{ \Big| \vect{h}^H \hat{\vect{h}} - \vartheta \Big|}{\|\hat{\vect{h}}\|_2}
 \right\} \\ & \quad + \tr(\vect{R}-\vect{C})  \mathbb{E}\left\{ \!
 \frac{|1+\p^{-1} \eta_{t}^{\mathrm{UE}}| \, \Big| \|\hat{\vect{h}}\|_2 \!-\! \sqrt{ \gamma } \Big|}{\|\hat{\vect{h}}\|_2 \sqrt{ \gamma } } \! \right\}.
\end{split}
\end{equation}
The first term in \eqref{eq:bound_for_signalpart} is asymptotically bounded since
\begin{equation}
\begin{split}
  \mathbb{E}\left\{  \frac{| \vect{h}^H \hat{\vect{h}} - \vartheta|}{\|\hat{\vect{h}}\|_2} \right\} \leq \sqrt{ \underbrace{\mathbb{E}\left\{ | \vect{h}^H \hat{\vect{h}} \! -\! \vartheta|^2 \right\}}_{\stackrel{(a)}{=}\mathcal{O}(N)} \underbrace{\mathbb{E}\left\{ \! \frac{1}{\|\hat{\vect{h}}\|_2^2} \! \right\}}_{\stackrel{(b)}{=}\mathcal{O}(N^{-1})} } =  \mathcal{O}\left( 1 \right)
\end{split}
\end{equation}
where the expectation of the numerator and denominator are separated using H\"older's inequality, $(a)$ follows from Lemma \ref{lemma:asymptotic-differences-innerproduct}, and $(b)$ from Lemma \ref{lemma:1overh2} with $k=0$. The second term of \eqref{eq:bound_for_signalpart} is upper bounded by
\begin{align} \notag
&\underbrace{\tr(\vect{R}-\vect{C})}_{=\mathcal{O}(N)} \underbrace{\sqrt{ \mathbb{E}\left\{
 \frac{|1+\p^{-1} \eta_{t}^{\mathrm{UE}}|^2}{\|\hat{\vect{h}}\|_2^2  \gamma  } \right\} }}_{= \mathcal{O}(N^{-1})}
 \underbrace{\sqrt{
 \mathbb{E}\left\{ \Big| \|\hat{\vect{h}}\|_2 - \sqrt{ \gamma } \Big|^2 \right\} } }_{=\mathcal{O}(1)} \\
 &= \mathcal{O}(1)
\end{align}
where H\"older's inequality was used to separate the expectations. The scaling of the first square root follows from
Lemma \ref{lemma:1overh2} and that $\frac{1}{\gamma} \leq \frac{1}{\tr ( \vect{A} \boldsymbol{\Psi} \vect{A}^H )} = \mathcal{O}(N^{-1})$ for any realization of $\eta_{t}^{\mathrm{UE}}$.
The scaling of the second square root follows from Lemma \ref{lemma:asymptotic-differences-norm}.
By plugging these scaling expressions into \eqref{eq:initial-upper-bound}, we have proved \eqref{eq:det-equiv-signalpart}.

Similarly, the equivalence in \eqref{eq:det-equiv-signalpart2} follows if
\begin{equation} \label{eq:det-equiv-signalpart2-upperbound}
\begin{split}
& \mathbb{E}\left\{ \left| \frac{|\vect{h}^H \hat{\vect{h}}|^2}{\|\hat{\vect{h}}\|_2^2} \!-\! \frac{|1+\p^{-1} \eta_{t}^{\mathrm{UE}}|^2 (\tr(\vect{R}-\vect{C}))^2}{ \gamma } \right| \right\}  \\
& \leq \big( \tr(\vect{R}-\vect{C}) \big)^2 \mathbb{E}\left\{
 \frac{|1+\p^{-1} \eta_{t}^{\mathrm{UE}}|^2  \Big| \|\hat{\vect{h}}\|_2^2 - \gamma \Big|}{ \|\hat{\vect{h}}\|_2^2  \gamma  } \right\}\\
 & \quad +  \mathbb{E}\left\{  \frac{\Big| |\vect{h}^H \hat{\vect{h}}|^2 \!-\! |1+\p^{-1} \eta_{t}^{\mathrm{UE}}|^2 (\tr(\vect{R}-\vect{C}))^2 \Big|}{\|\hat{\vect{h}}\|_2^2}
 \right\}
\end{split}
\end{equation}
scales as $\mathcal{O} ( \sqrt{N} )$, where the inequality follows from Lemma \ref{lemma:split-absolute-values}.
By applying H\"older's inequality on the first term, we obtain
\begin{align} \notag
&\underbrace{\frac{(\tr(\vect{R}-\vect{C}))^2}{\tr(\vect{A}  \boldsymbol{\Psi} \vect{A}^H)}}_{\stackrel{(a)}{=}\mathcal{O}(N)} \sqrt{
\underbrace{\mathbb{E}\left\{
 \frac{|1+\p^{-1} \eta_{t}^{\mathrm{UE}}|^4}{\|\hat{\vect{h}}\|_2^4    } \right\}}_{\stackrel{(b)}{=}\mathcal{O}(N^{-2})} }  \sqrt{
\underbrace{\mathbb{E}\left\{  \Big| \|\hat{\vect{h}}\|_2^2 - \gamma  \Big|^2  \right\}}_{\stackrel{(c)}{=}\mathcal{O}(N)}
} \\ &= \mathcal{O} \left(\sqrt{N} \right)
\end{align}
where $(a)$ follows from $\frac{1}{\gamma} \leq \frac{1}{\tr ( \vect{A} \boldsymbol{\Psi} \vect{A}^H )} = \mathcal{O}(N^{-1})$ which is a deterministic upper bound, $(b)$ is characterized by Lemma \ref{lemma:1overh2}, and $(c)$ follows from Lemma \ref{lemma:asymptotic-differences-norm}. The second term behaves as
\begin{align} \notag
& \sqrt{\underbrace{\mathbb{E}\left\{ \Big| |\vect{h}^H \hat{\vect{h}}| - |1+\p^{-1} \eta_{t}^{\mathrm{UE}}| \tr(\vect{R}-\vect{C}) \Big|^2 \right\}}_{\stackrel{(d)}{=}\mathcal{O}(N)}   } \\  \notag &\times \sqrt{ \underbrace{\mathbb{E}\left\{ \frac{2 |\vect{h}^H \hat{\vect{h}}|^2}{\|\hat{\vect{h}}\|_2^4}\right\}}_{\stackrel{(e)}{=}\mathcal{O}(1)} + \underbrace{\mathbb{E}\left\{ \frac{2 |1+\p^{-1} \eta_{t}^{\mathrm{UE}}|^2 (\tr(\vect{R}-\vect{C}))^2}{\|\hat{\vect{h}}\|_2^4} \right\}}_{\stackrel{(f)}{=}\mathcal{O}(1)} } \\
&= \mathcal{O}(\sqrt{N})
\end{align}
by using the rule $a^2-b^2 =(a+b)(a-b)$ and H\"older's inequality.
$(d)$ is characterized by Lemma \ref{lemma:asymptotic-differences-innerproduct} and $(f)$ by Lemma \ref{lemma:1overh2}. Moreover, $(e)$ follows since $\frac{ |\vect{h}^H \hat{\vect{h}}|^2}{\|\hat{\vect{h}}\|_2^4} \leq \frac{ \|\vect{h} \|_2^2}{\|\hat{\vect{h}}\|_2^2} = \mathcal{O}(1)$ by Cauchy-Schwartz inequality and
that $\|\vect{h} \|_2^2, \|\hat{\vect{h}}\|_2^2$ have same asymptotic scaling.
By plugging these scaling expressions into \eqref{eq:det-equiv-signalpart2-upperbound}, we have proved \eqref{eq:det-equiv-signalpart2}.

Finally, the equivalence in \eqref{eq:det-equiv-extra-interference} follows since
\begin{align}
& \left| \sum_{i=1}^{N} \mathbb{E} \{ |h_i|^2 |v_i|^2 \} - 0 \right| \stackrel{(a)}{=}
\left| \mathbb{E} \left\{ \frac{\|\hat{\vect{h}}\|_4^2}{\|\hat{\vect{h}} \|_2^2} \sum_{i=1}^{N} |h_i|^2 \frac{|\hat{h}_i|^2}{\|\hat{\vect{h}}\|_4^2}   \right\} \right| \notag \\
& \stackrel{(b)}{\leq} \left| \mathbb{E} \left\{ \frac{\|\hat{\vect{h}}\|_4^2}{\|\hat{\vect{h}}\|_2^2} \sqrt{\sum_{i=1}^{N} |h_i|^4} \sqrt{\sum_{i=1}^{N} \frac{|\hat{h}_i|^4}{\|\hat{\vect{h}}\|_4^4} }   \right\} \right| \notag \\ &= \left| \mathbb{E} \left\{ \frac{\|\hat{\vect{h}}\|_4^2}{\|\hat{\vect{h}}\|_2^2} \| \vect{h} \|_4^2   \right\} \right| \notag \\
& \stackrel{(c)}{\leq} \sqrt{ \sqrt{ \mathbb{E} \left\{  \| \vect{h} \|_4^8 \right\} \mathbb{E} \left\{ \|\hat{\vect{h}}\|_4^8 \right\} } \mathbb{E} \left\{  \frac{1}{\|\hat{\vect{h}} \|_2^4}  \right\}   } \stackrel{(d)}{=} \mathcal{O}(1)
\end{align}
where $(a)$ follows from $|v_i|^2 = \frac{|\hat{h}_i|^2 }{\|\hat{\vect{h}}\|_2^2}$ and by inserting the $L_4$-norms $\|\hat{\vect{h}}\|_4^2$. The reason for this is that the vector $[|\hat{h}_1|^2 \, \ldots \, |\hat{h}_N|^2]^T /\|\hat{\vect{h}}\|_4^2$ now has unit $L_2$-norm, thus we apply Cauchy-Schwarz inequality in $(b)$ to bound the sum by $\| \vect{h} \|_4^2$. Next, $(c)$ is obtained by applying H\"older's inequality twice and $(d)$ follows from that $\mathbb{E} \{  \| \vect{h} \|_4^8 \} = \mathcal{O}(N^2)$ and $\mathbb{E} \{ \| \hat{\vect{h}} \|_4^8 \} = \mathcal{O}(N^2)$
and $\mathbb{E} \left\{  \frac{1}{\|\hat{\vect{h}} \|_2^4}  \right\} = \mathcal{O}(N^{-2})$ from Lemma \ref{lemma:1overh2}.

\subsection{Proof of Theorem \ref{theorem:energy-efficiency}}
\label{proof:theorem:energy-efficiency}

The bounds in this theorem are derived using the capacity lower bounds in Corollaries \ref{cor:lower_bound} and \ref{cor:lower_bound-uplink}. We begin with the DL and note that the arguments of the expectations in \eqref{eq:capacity-lower-equivalent} have deterministic upper bounds since
\begin{align}
\left| \varphi \right| &\leq \sqrt{    \frac{\tr(\vect{R}-\vect{C})}{p^{\mathrm{UE}} \tr(\vect{A} \vect{R} \vect{A}^H)}}
\end{align}
for any realization of $\eta_{t}^{\mathrm{UE}}$. The dominated convergence theorem implies that we can take the limit $N \rightarrow \infty$ inside the expectations.\footnote{To be strict, we first should multiply all terms in \eqref{eq:capacity-lower-equivalent} by $p^{\mathrm{UE}} $.}
Next, we observe that scaling the pilot power $p^{\mathrm{UE}}$ proportionally to $1/N^{t_{\mathrm{UE}}}$ for some $t_{\mathrm{UE}}>0$ means that $N^{t_{\mathrm{UE}}} p^{\mathrm{UE}} \rightarrow B $ as $N \rightarrow \infty$ for some $0<B <\infty$. Therefore, we have
\begin{align} \notag
& N^{t_{\mathrm{UE}}} \tr( \vect{R}-\vect{C}) = N^{t_{\mathrm{UE}}} p^{\mathrm{UE}} \tr \left( \vect{R} \bar{\vect{Z}}^{-1} \vect{R} \right) \\ \label{eq:ee-limits1} &\quad \quad \rightarrow
B \tr( \vect{R} (\vect{S}+\sigma_{\mathrm{BS}}^2 \vect{I})^{-1} \vect{R}), \\
& N^{t_{\mathrm{UE}}} \tr( \vect{A} ( |\p + \eta_{t}^{\mathrm{UE}}|^2 \vect{R} + \boldsymbol{\Psi}) \vect{A}^H ) \notag \\ & \quad \quad \rightarrow B \tr( \vect{R} ( \vect{S}+\sigma_{\mathrm{BS}}^2 \vect{I})^{-1} \vect{R} ),
\label{eq:ee-limits2}
\end{align}
as $N \rightarrow \infty$. Using \eqref{eq:ee-limits1} and \eqref{eq:ee-limits2} and the dominated convergence theorem we obtain
\begin{align} \label{eq:ee-limits_exp1}
&\lim_{N \rightarrow \infty} |  \mathbb{E} \{ \varphi \} |^2 \\ \notag&= \bigg| \mathbb{E}\bigg\{ \!
\frac{(1 \!+\!\p^{-1} \eta_{t}^{\mathrm{UE}}) \sqrt{B \tr \left( \vect{R} (\vect{S}+\sigma_{\mathrm{BS}}^2 \vect{I})^{-1} \vect{R} \right) } }{\sqrt{ B \tr \left( \vect{R} (\vect{S}+\sigma_{\mathrm{BS}}^2 \vect{I})^{-1} \vect{R} \right)  }}
\! \bigg\} \bigg|^2 = 1 \\
\label{eq:ee-limits_exp2}
& \lim_{N \rightarrow \infty}  \mathbb{E}\{  | \varphi|^2 \}  \\ \notag&=
 \mathbb{E}\bigg\{
\frac{ |1\!+\!\p^{-1} \eta_{t}^{\mathrm{UE}} |^2 B \tr \left( \vect{R} (\vect{S}+\sigma_{\mathrm{BS}}^2 \vect{I})^{-1} \vect{R} \right) }{ B \tr \left( \vect{R} (\vect{S} + \sigma_{\mathrm{BS}}^2 \vect{I})^{-1} \vect{R} \right) }
\bigg\} = 1+\kappa_{t}^{\mathrm{UE}}
\end{align}
which holds for any $t_{\mathrm{UE}}>0$. The goal is to make the interference term
$\frac{\mathbb{E}\{ I_{\mathcal{H}}^{\mathrm{UE}} \}}{ p^{\mathrm{BS}} \tr(\vect{R}-\vect{C})} = \frac{\mathbb{E}\{ I_{\mathcal{H}}^{\mathrm{UE}} \}}{p^{\mathrm{UE}} p^{\mathrm{BS}}  \tr(\vect{R} \bar{\vect{Z}}^{-1} \vect{R} )}$ vanish asymptotically
under the assumption that $\mathbb{E}\{ I_{\mathcal{H}}^{\mathrm{UE}} \} = \mathcal{O}(1)$, which is achieved if the denominator grows to infinity with $N$.
We note that $\tr(\vect{R} \bar{\vect{Z}}^{-1} \vect{R} ) \geq \frac{\| \vect{R}\|^2}{\| \bar{\vect{Z}} \|_2}$ scales at least linearly with $N$.
Hence, the product $p^{\mathrm{UE}} p^{\mathrm{BS}}$ must reduce at a slower pace than linear with $N$, which implies  $t_{\mathrm{BS}}+t_{\mathrm{UE}}=t_{\mathrm{sum}}<1$.

Finally, we need the $\mathcal{O}(1/\sqrt{N})$ terms in \eqref{eq:lower-bound-capacity-energy-eff} to still vanish as $N \rightarrow \infty$. Some careful but lengthy algebra reveals that the $\mathcal{O}(N)$ properties in Lemmas \ref{lemma:asymptotic-differences-norm}--\ref{lemma:1overh2} become $\mathcal{O}(N^{1-t_{\mathrm{UE}}})$. The term $\mathcal{O}(1/\sqrt{N})$ in the numerator of \eqref{eq:lower-bound-capacity-energy-eff} becomes $\mathcal{O}(1/N^{\frac{1}{2}-\frac{t_{\mathrm{UE}}}{2}})$ while the $\mathcal{O}(1/\sqrt{N})$ in the denominator becomes $\mathcal{O}(1/N^{\frac{1}{2}-t_{\mathrm{UE}}})$. These terms vanish if $t_{\mathrm{UE}} < \frac{1}{2}$, which finishes the proof for the DL.

The proof for the UL is analogous since the uplink capacity bound in \eqref{eq:capacity-lower-equivalent-uplink} has the same structure and contains the same expectations as the downlink capacity.

\subsection{Proof of Corollary \ref{cor:degrading-hardware-quality}}
\label{proof:cor:degrading-hardware-quality}

Recall from the proof of Theorem \ref{theorem:energy-efficiency} that the dominated convergence theorem can be applied, which means that we can take the limit $N \rightarrow \infty$ inside the expectations in the DL capacity bound of \eqref{eq:capacity-lower-equivalent} and UL capacity bound of \eqref{eq:capacity-lower-equivalent-uplink}. If $\kappa_t^{\mathrm{BS}}$ and $\kappa_r^{\mathrm{BS}}$ grow with $N$, we  obtain
\begin{align} \notag
\lim_{N \rightarrow \infty} |  \mathbb{E} \{ \varphi \} |^2 &= \Bigg| \mathbb{E}\Bigg\{ \!
\frac{(1+\p^{-1} \eta_{t}^{\mathrm{UE}}) \sqrt{ \tr( \vect{R} \vect{R}_{\diag}^{-1} \vect{R}) } }{\sqrt{  \tr( \vect{R} \vect{R}_{\diag}^{-1} \vect{R})  }}
\! \Bigg\} \Bigg|^2 \\ &= 1 \label{eq:degrading-hardware-limits_exp1}\\
\notag
\lim_{N \rightarrow \infty}  \mathbb{E}\{  | \varphi|^2 \}  &=
\mathbb{E}\bigg\{
\frac{ |1+\p^{-1} \eta_{t}^{\mathrm{UE}} |^2 \tr( \vect{R} \vect{R}_{\diag}^{-1} \vect{R}) }{  \tr( \vect{R} \vect{R}_{\diag}^{-1} \vect{R}) }
\bigg\} \\ &= 1+\kappa_{t}^{\mathrm{UE}}.\label{eq:degrading-hardware-limits_exp2}
\end{align}

In the DL, we further note that
\begin{align}
\frac{\kappa_t^{\mathrm{BS}} \! \fracSumtwo{i=1}{N} \mathbb{E}\{ |h_i|^2 |v_i|^2\}}{ \tr( \vect{R}-\vect{C}) } = \mathcal{O}( \frac{\kappa_t^{\mathrm{BS}} \kappa_r^{\mathrm{BS}} }{N}  )
\end{align}
since $\kappa_r^{\mathrm{BS}} \tr( \vect{R}-\vect{C}) \rightarrow \tr( \vect{R} \vect{R}_{\diag}^{-1} \vect{R}) = \mathcal{O}(N)$ as $N \rightarrow \infty$. If this term should vanish asymptotically, it is sufficient that $\frac{\kappa_t^{\mathrm{BS}} \kappa_r^{\mathrm{BS}} }{N} \rightarrow 0$ which corresponds to the condition in the corollary.
The corresponding condition for the UL is obtained analogously and gives $\frac{(\kappa_r^{\mathrm{BS}})^2 }{N} \rightarrow 0$.

Finally, we note that the noise terms (for $n \leq \frac{1}{2}$) and the $\mathcal{O}(\frac{1}{\sqrt{N}})$ terms in \eqref{eq:capacity-lower-equivalent} and \eqref{eq:capacity-lower-equivalent-uplink} all behave as $\mathcal{O}(\frac{\kappa_r^{\mathrm{BS}}}{\sqrt{N}})$ or smaller, after some straightforward but lengthy algebra. These terms thus vanish under the condition $\tau_r < \frac{1}{2}$ stated in the corollary.

\subsection{Proof of Theorem \ref{theorem:asymptotic-equivalence-interference}}
\label{proof:theorem:asymptotic-equivalence-interference}

The interference expressions in \eqref{eq:det-equiv-inter-user-interference} are proved similar to Theorem \ref{theorem:asymptotic-equivalence}. For the case $l\in \mathcal{U}_{\parallel}$ we have
\begin{equation} \label{eq:bound-difference-same-pilot}
\begin{split}
& \mathbb{E}\left\{ \left| \frac{|\vect{h}_l^H \hat{\vect{h}} |^2}{\| \hat{\vect{h}}\|^2_2} - \frac{ p^{\mathrm{UE}}  a_l^2 }{\gamma} \right| \right\} \leq
\mathbb{E}\left\{ \frac{\left| |\vect{h}_l^H \hat{\vect{h}} |^2 - p^{\mathrm{UE}} a_l^2 \right|}{\| \hat{\vect{h}}\|^2_2 } \right\}\\
& \quad \quad \quad \quad +
\mathbb{E}\left\{ \frac{ p^{\mathrm{UE}}  a_l^2 \left|  \| \hat{\vect{h}}\|^2_2 - \gamma \right| }{\| \hat{\vect{h}}\|^2_2 \gamma}
\right\} = \mathcal{O}(\sqrt{N})
\end{split}
\end{equation}
where $\gamma = \tr\big(\vect{A} ( |\p + \eta_{t}^{\mathrm{UE}}|^2 \vect{R} + \boldsymbol{\Psi}) \vect{A}^H \big)$ and $a_l =  \tr ( \vect{A} \vect{R}_l )$. This follows since the first term in \eqref{eq:bound-difference-same-pilot} equals
\begin{align} \notag
&\mathbb{E}\left\{ \frac{\left| |\vect{h}_l^H \hat{\vect{h}} | - \sqrt{p^{\mathrm{UE}}} a_l \right| \left| |\vect{h}_l^H \hat{\vect{h}} | + \sqrt{p^{\mathrm{UE}} a_l} \right|  }{\| \hat{\vect{h}}\|^2_2 } \right\} \\ \notag &\leq
\underbrace{\sqrt{  \mathbb{E}\left\{ \left| |\vect{h}_l^H \hat{\vect{h}} | - \sqrt{p^{\mathrm{UE}}} a_l \right|^2 \right\}}}_{=\mathcal{O}(\sqrt{N})} \underbrace{\sqrt{
\mathbb{E}\left\{\frac{ \|\vect{h}_l \|_2^2 }{\| \hat{\vect{h}}\|^2_2 } + \frac{p^{\mathrm{UE}} a_l^2 \  }{\| \hat{\vect{h}}\|^4_2 } \right\} }}_{=\mathcal{O}(1)} \\
& = \mathcal{O}(\sqrt{N})
\end{align}
by using H\"older's inequality, Lemma \ref{lemma:classic-trace-result}, Cauchy-Schwartz inequality, and Lemma \ref{lemma:1overh2}.
The second term in \eqref{eq:bound-difference-same-pilot} is also upper bounded by $\mathcal{O}(\sqrt{N})$ by using H\"older's inequality and that $a_l = \mathcal{O}(N)$,
$\mathbb{E}\big\{ \big|  \| \hat{\vect{h}}\|^2_2 - \gamma \big|^2 \big\} = \mathcal{O}(N)$ from Lemma \ref{lemma:asymptotic-differences-norm},
$\mathbb{E}\{ \| \hat{\vect{h}}\|^{-4}_2 \} = \mathcal{O}(N^{-2})$ from Lemma \ref{lemma:1overh2}, and
$\frac{1}{\gamma} \leq \frac{1}{\tr ( \vect{A} \boldsymbol{\Psi} \vect{A}^H )} = \mathcal{O}(N^{-1})$.

Next, the case $l\in \mathcal{U}_{\perp}$ in \eqref{eq:det-equiv-inter-user-interference} follows from $\mathbb{E}\left\{ | \vect{h}_l^H \vect{v}^{\mathrm{UL}} |^2 \right\} =
\mathbb{E}\left\{ (\vect{v}^{\mathrm{UL}})^H \vect{R}_l \vect{v}^{\mathrm{UL}} \right\} \leq \| \vect{R}_l \|_2 = \mathcal{O}(1)$ since $\vect{h}_l$ and $\vect{v}^{\mathrm{UL}}$ are independent.

Finally, we note that the noise term in the denominator of \eqref{eq:capacity-lower-equivalent-uplink} would be
\begin{equation}
\begin{split}
\frac{\mathbb{E}\{ \vect{v}^H \vect{Q}_{\mathcal{H}} \vect{v} \} }{ p^{\mathrm{UE}} \tr(\vect{R}-\vect{C})} = \sum_{l \in  \mathcal{U}_{\parallel} }
\frac{ p^{\mathrm{UE}} \mathbb{E}\left\{ | \vect{h}_l^H \vect{v} |^2  \right\} }{ p^{\mathrm{UE}} \tr(\vect{R}-\vect{C})} + \mathcal{O}\left( \sqrt{\frac{1}{N}} \right )
\end{split}
\end{equation}
where the first term is equal to \eqref{eq:contamination-term} by exploiting \eqref{eq:det-equiv-inter-user-interference} and $\tr(\vect{R}-\vect{C}) = \sqrt{p^{\mathrm{UE}}} \tr(\vect{A} \vect{R})$.

\section*{Acknowledgment}

The authors would like to thank Romain Couillet and the anonymous reviewers for indispensable feedback on this paper.

\bibliographystyle{IEEEtran}
\bibliography{IEEEabrv,refs}

\end{document}